\documentclass[twocolumn,rmp,aps,nofootinbib]{revtex4}
\usepackage{amsmath} \usepackage{graphicx} \usepackage{dcolumn}
\usepackage{bm} \usepackage{epsf} \usepackage{amssymb}

\begin{document}
\ifx\href\undefined\else\hypersetup{linktocpage=true}\fi

\newcommand{\prtl}{\partial}
\newcommand{\la}{\left\langle}
\newcommand{\ra}{\right\rangle}
\newcommand{\dla}{\la \! \! \! \la}
\newcommand{\dra}{\ra \! \! \! \ra}
\newcommand{\we}{\widetilde}
\newcommand{\smfp}{{\mbox{\scriptsize mfp}}}
\newcommand{\sph}{{\mbox{\scriptsize ph}}}
\newcommand{\sinhom}{{\mbox{\scriptsize inhom}}}
\newcommand{\sneigh}{{\mbox{\scriptsize neigh}}}
\newcommand{\srlxn}{{\mbox{\scriptsize rlxn}}}
\newcommand{\svibr}{{\mbox{\scriptsize vibr}}}
\newcommand{\smicro}{{\mbox{\scriptsize micro}}}
\newcommand{\seq}{{\mbox{\scriptsize eq}}}
\newcommand{\teq}{{\mbox{\tiny eq}}}
\newcommand{\sinn}{{\mbox{\scriptsize in}}}
\newcommand{\tin}{{\mbox{\tiny in}}}
\newcommand{\scr}{{\mbox{\scriptsize cr}}}
\newcommand{\stheor}{{\mbox{\scriptsize theor}}}
\newcommand{\sGS}{{\mbox{\scriptsize GS}}}
\newcommand{\sNMT}{{\mbox{\scriptsize NMT}}}
\def\Xint#1{\mathchoice
   {\XXint\displaystyle\textstyle{#1}}%
   {\XXint\textstyle\scriptstyle{#1}}%
   {\XXint\scriptstyle\scriptscriptstyle{#1}}%
   {\XXint\scriptscriptstyle\scriptscriptstyle{#1}}%
   \!\int}
\def\XXint#1#2#3{{\setbox0=\hbox{$#1{#2#3}{\int}$}
     \vcenter{\hbox{$#2#3$}}\kern-.5\wd0}}
\def\ddashint{\Xint=}
\def\dashint{\Xint-}
\title{The Microscopic Quantum Theory of Low Temperature Amorphous
Solids}

\author{Vassiliy Lubchenko} \altaffiliation[Current Address:
]{Department of Chemistry, Massachusetts Institute of Technology,
Cambridge, MA 02139.}  \affiliation{ Department of Chemistry and
Biochemistry and Department of Physics, University of California at
San Diego, La Jolla, CA 92093-0371} \author{Peter G. Wolynes}
\affiliation{ Department of Chemistry and Biochemistry and Department
of Physics, University of California at San Diego, La Jolla, CA
92093-0371}


\begin{abstract}

The quantum excitations in glasses have long presented a set of
puzzles for condensed matter physicists. A common view is that they
are largely disordered analogs of elementary excitations in crystals,
supplemented by two level systems which are chemically local entities
coming from disorder. A radical revision of this picture argues that
the excitations in low temperature glasses are deeply connected to the
energy landscape of the glass when it vitrifies: the excitations are
not low excited states built on a single ground state but locally
defined resonances, high in the energy spectrum of a solid. According
to a semiclassical analysis, the two level systems involve resonant
collective tunneling motions of around two hundred molecular units
which are relics of the mosaic of cooperative motions at the glass
transition temperature $T_g$. The density of states of the TLS is
determined by $T_g$ and the mosaic's length scale, which is a weak
function of the cooling rate. The universality of phonon scattering in
insulating glasses is explained. The Boson Peak and the plateau in
thermal conductivity, observed at higher temperatures, are also
quantitatively understood within the picture as arising from the same
cooperative motions, but now accompanied by thermal activation of the
mosaic's vibrational modes. The dynamics of some of the local
structural transitions have significant quantum corrections to the
semiclassical picture. These corrections lead to a deviation of the
heat capacity and conductivity from the standard tunneling model
results and explain the anomalous time dependence of the heat
capacity. Interaction between tunneling centers contributes to the
large and negative value of the Gr\"{u}neisen parameter often observed
in glasses.

\end{abstract}

\date{December 17, 2004}

\maketitle

\tableofcontents

\section{Introduction}

During the past several decades, it has been gradually recognized in
the condensed matter and materials science community that amorphous
materials, while sharing many characteristics with the more common
crystalline solids, represent a distinct solid state of matter.  On
the one hand, glasses exhibit rigidity and elastic response on humanly
relevant time scales, thus qualifying them as solids for many
practical purposes. In fact, until the relatively recent advent of
systematic studies of the materials' response to mechanical and
electromagnetic perturbation, as well as of their detailed microscopic
structure, the only commonly known distinct attributes of amorphous
substances had been their optical properties and the low magnitude and
isotropic character of their thermal expansion. Those properties still
undergird the main technological importance of amorphous materials. On
the other hand, there are many ways in which glasses are fundamentally
different from crystals. This is most noticeable in their properties
at cryogenic temperatures.

We presently know very well that an amorphous solid is in reality a
liquid caught {\em locally} in a small set of metastable free energy
minima \cite{XW}, each of which are separated from the much lower free
energy crystalline arrangement by high barriers. Therefore, the glass
transition, as manifest in the laboratory, is not strictly speaking a
phase transition in the regular thermodynamic sense and is not
accompanied by a symmetry change or appearance of a free energy
singularity. In contrast, a liquid that was cooled below its melting
point fast enough so as to avoid crystallization - i.e. has become
supercooled - experiences a crossover to (highly viscous) activated
transport. As the temperature is lowered further, the relaxation
barriers grow in a very dramatic fashion thus confining the molecules
in their metastable arrangements long enough to give the appearance of
shear elasticity in the sample on the technologically relevant
frequency scales. A quantitative understanding of the physics behind
the glass transition has recently been achieved with the random first
order transition (RFOT) theory of glasses \cite{KTW,XW}. This theory
has provided a microscopic picture of molecular motions in supercooled
liquids, such as first principle predictions of the length scales of
these motions and the cooperativity lengths and the barrier heights of
the activated transport. At any given time, a supercooled liquid is a
mosaic of cooperatively rearranging regions, whose size becomes larger
as the temperature is lowered. This article describes how the RFOT
theory also provides the necessary microscopic input to understand the
cryogenic anomalies observed in glasses.

In spite of the absence of periodicity, glasses exhibit, among other
things, a specific volume, interatomic distances, coordination number
and local elastic modulus comparable to those of crystals. Therefore
it has been considered natural to consider amorphous lattices as
nearly periodic with the disorder treated as a perturbation,
often-times in form of defects, so such a study is not futile.  This
is indeed a sensible approach, as even the crystals themselves are
rarely perfect, and many of their useful mechanical and other
properties are determined by the existence and mobility of some sort
of defects as well by interaction between those defects. Nevertheless,
a number of low temperature phenomena in glasses have persistently
evaded a microscopic model-free description along those lines. A more
radical revision of the concept of an elementary excitation on top of
a unique ground state is necessary \cite{LW,LW_BP,LW_thesis}.

\begin{figure}[htb]
\includegraphics[width=.95\columnwidth]{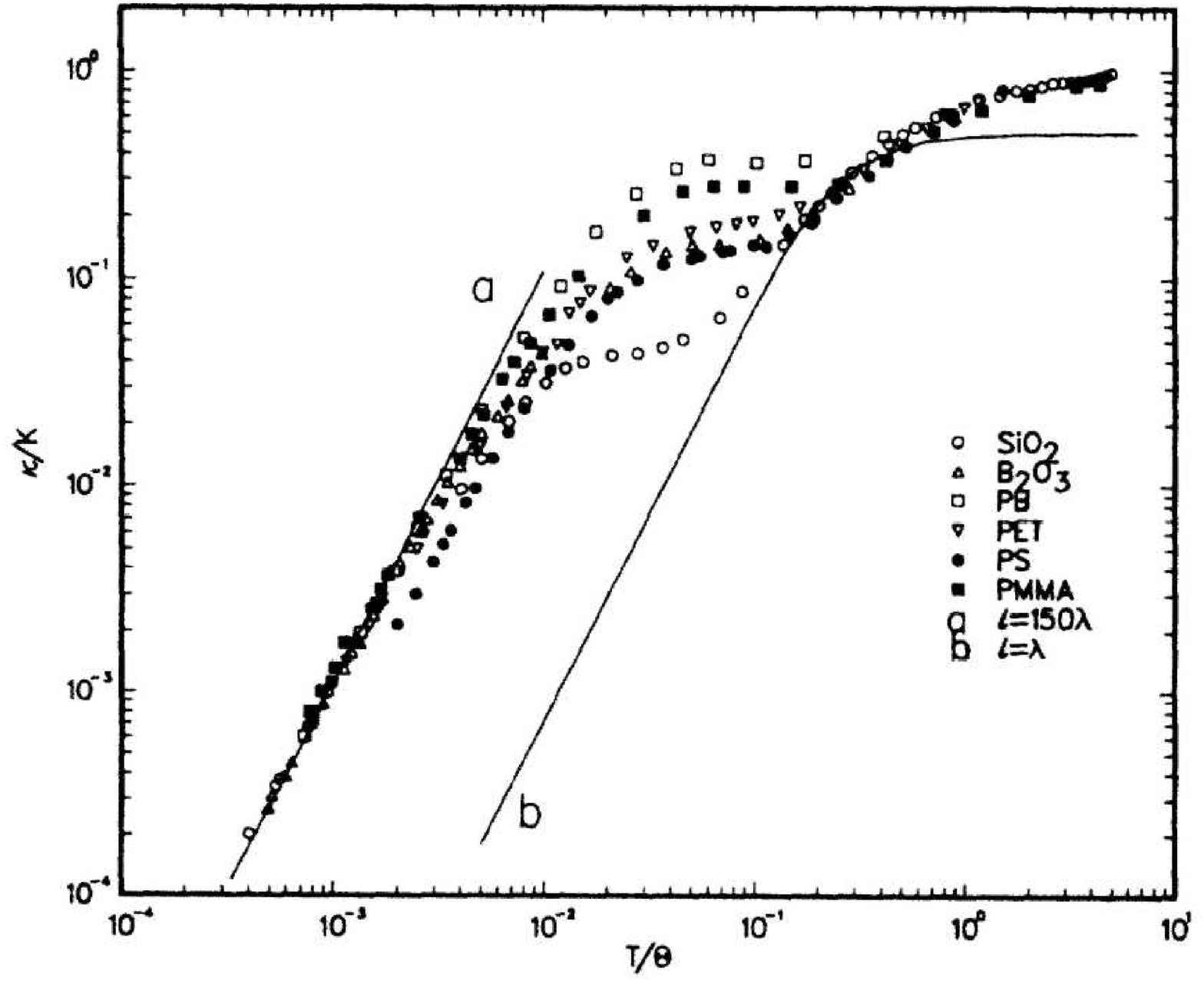}
\caption{\label{l_lambda} Scaled thermal conductivity ($\kappa$) data
for several amorphous materials is shown. The horisontal axis is
temperature in units the Debye temperature $T_D$. The vertical axis
scale $K \equiv \frac{k_B^3 T_D^2}{\pi \hbar c_s}$.  The value of
$T_D$ is somewhat uncertain, but its choice made in
\cite{FreemanAnderson} is strongly supported by that it yields
universality in the phonon localization region. The solid lines are
calculated using $\kappa \simeq \frac{1}{3} \sum_{\omega}
C\sph(\omega) l_\smfp(\omega) c_S$ with $l_\smfp/\lambda = 150$ and
$l_\smfp/\lambda = 1$ respectively \cite{FreemanAnderson}.}
\end{figure}

Let us give a brief historical overview of some of the most
outstanding issues in low temperature amorphous state physics.  It was
already noted in the 1960s that the thermal conductivity of amorphous
solids is significantly lower than that of crystals. A low-temperature
experimentalist using epoxy in his apparatus knew that its thermal
conductivity at liquid helium temperatures went roughly as
\verb+constant+$\times T^2$, where the \verb+constant+ was practically
the same for other amorphous substances as well \cite{ACAnderson_pc}.
Surprisingly, this had not particularly alarmed anyone, even though
one would not {\em \`{a} priori} expect low temperature properties of
disordered solids to be different from crystals, as the appropriate
thermal phonon length is much larger than the molecular scale which
was presumed to characterize the relevant heterogeneity scale. It was
not until Zeller and Pohl published their classic paper
\cite{ZellerPohl} that it became generally known that both the heat
capacity and thermal conductivity of glasses were significantly
different from those of crystals, and that these anomalies were
correlated. The heat capacity turned out to be approximately linear in
temperature and larger than the $T^3$ phononic contribution up to
temperatures $\sim 10$ K.  The challenge to the theorists was soon met
by the so called Standard Tunneling Model (STM) \cite{AHV,Phillips},
in which one assumes that due to a disordered pattern of molecular
bonds in glasses, there are a number of defects in the lattice
(something like ``loose'' atoms or ``dangling bonds''), which have two
alternative positions in space separated by a sufficiently low
tunneling barrier. At low temperatures, the dynamics of such a system
is described well by a two-level system (TLS) hamiltonian.  If one
assumes that the spectral density of these TLS's is flat, one recovers
the linear heat capacity. One also finds that the inverse mean free
path of a thermal phonon due to resonant scattering off the TLS's is
equal to $l^{-1}_\smfp \propto T$, which implies thermal conductivity
$\kappa \simeq \frac{1}{3} \sum_{\omega} C\sph(\omega) \,
l_\smfp(\omega) \, c_s \propto T^2$. Here, $C\sph(\omega)$ is the heat
capacity of a phonon mode of frequency $\omega$ and $c_s$ is the speed
of sound (one assumes here that heat is carried primarily by phonons,
which was experimentally demonstrated explicitly four years later by
Zaitlin and Anderson \cite{ZaitlinAnderson}).  Note that the resonant
character of phonon scattering implies that the scattering
cross-section of low-frequency phonons would be independent of the
scatterer size, but would scale with the phonon wavelength (squared)
itself.  Therefore no knowledge of scatterer's microscopic details are
needed. Rather, only a single coupling parameter is needed to estimate
the magnitude of scattering at low temperatures.  The STM did prove to
be very successful \cite{LowTProp}, as it predicted, among other
things, nonlinear sound absorption due to the saturation of the
resonant absorption and the phonon echo, both of which were later
observed \cite{satur,GG}. In spite of these successes, the microscopic
nature of these defects had remained unknown, although there later
appeared several indications in the literature that the tunneling
centers are not single atom entities but rather involve motions within
larger groups of atoms \cite{MonAshcroft,GuttmanRahman}. On the
experimental front, there had been a growing amount of evidence that
the number of these additional excitations and their coupling to the
phonons are correlated and also depend on $T_g$
\cite{Reynolds,Reynolds1, RaychaudhuriPohl}, which culminated in the
observation made by Freeman and Anderson \cite{FreemanAnderson}, that
the heat conductivities of all studied insulating glasses, if scaled
by elastic constants, fall onto the same line in two regions,
connected by a non-universal flat piece corresponding to the so called
``plateau''. In Fig.\ref{l_lambda} we show a facsimile of Fig.2 from
\cite{FreemanAnderson} that demonstrates this heat conductivity
universality.
The lower temperature straight line corresponds to the value $\simeq
150$ of the ratio of the thermal phonon mean free path $l_\smfp$ to
the thermal Debye wave-length $\lambda \equiv \hbar c_s/k_B T$. This
region spans roughly 1.5 decades in temperature between several mK
(lowest $T$ accessed so far for the heat conductivity measurements)
and to 1-10 K, depending on the substance. The short linear region at
higher temperatures (20-60 K) corresponds to $l_\smfp/\lambda \simeq
1$, which actually implies complete phonon localization \cite{ph_loc}
according to the heuristic Ioffe-Riegel criterion.  This implies,
among other things, that one can no longer use kinetic theory
expressions for heat transfer at these temperatures, as a diffusive
mechanism must prevail \footnote{This idea that the heat was
transfered by a random walk was used early on by Einstein
\cite{Einstein} to calculate the thermal conductance of crystals but,
of course, he obtained numbers much lower than those measured in the
experiment. As we now know, crystals at low enough $T$ support well
defined quasiparticles - the phonons - which happen to carry heat at
these temperatures. Ironically, Einstein never tried his model on the
amorphous solids, where it would be applicable in the $l_\smfp/\lambda
\sim 1$ regime.}.  The intermediate region (``plateau'') is usually
observed between 1 and 30 K, and does not scale with the Debye
temperature and speed of sound.  The standard tunneling model of
non-interacting two-level systems mentioned above is normally applied
to the region where $l_\smfp/\lambda \simeq 150$, that is generically
below 1 K. The universality of $l_\smfp/\lambda$ can be boiled down
\cite{LowTProp} to the universality of the following combination of
parameters: $\bar{P} \frac{g^2}{\rho c_s^2}$, where $\bar{P}$ is the
spectral and spatial density of the TLS's (empirically $\sim 10^{45
\pm 1} J^{-1} m^{-3}$), $g$ is coupling to the elastic strain on the
order of eV, $\rho$ is the mass density (for reference, $\sim
\frac{g^2} {\rho c_s^2 r^3}$ would be the interaction strength between
such TLS's at distance $r$ from each other).  Now, if the defects
involved the motion of only a single atom, one would reasonably assume
that the value of their spectral density and coupling to the lattice
or their combination would be very strongly material dependent. Even
though $\bar{P}$ and $g^2$ vary within almost two orders of magnitude
(still surprisingly little), the combination $\bar{P} \frac{g^2}{\rho
c_s^2}$ is constant within 50\% for different materials ($\rho$ and
$c_s^2$ vary considerably as well). It certainly takes a stretch of
imagination to think that this is merely a coincidence, as pointed out
in \cite{Leggett}. In 1988, Yu and Leggett proposed \cite{YuLeggett}
that the density of states of the TLS might itself be a result of
dipole-dipole interactions between some original non-renormalized
excitations. In short, this idea is motivated by the observation that
for TLS coupled to the phonons with strength $g$, the coefficient at
the dipole-dipole interaction term $g^2/\rho c_s^2$ has dimensions
energy times volume. Therefore the interaction induced renormalized
density of states $\bar{P}$ has to be the inverse of $g^2/\rho c_s^2$
with a coefficient, hence the universality of $\bar{P} g^2/\rho c_s^2$
for different materials. However, it so far has not proved possible to
use their approach to justify the value of that coefficient to yield
the experimental $l_\smfp/\lambda \simeq 150$. This is surprising,
since one expects such a simple dimensional argument to be very
robust.  (Several other studies of the universality
\cite{MeissnerSpitzmann,Coppersmith} were undertaken at the time, that
followed the paper by Freeman and Anderson \cite{FreemanAnderson}.)
There has been subsequent work applying a renormalization group style
calculation to a system of interacting TLS \cite{BurinKagan}, but it
seems from the results that renormalizations are relevant only at
ultra-low temperatures ($\mu$K's and below) \cite{Silbey}.  In spite
of the difficulties in justifying the strong interaction scenario
\cite{Carruzzo,LWunp}, the works \cite{YuLeggett, Leggett} that first
challenged the standard TLS paradigm remain the main conceptual
motivation behind the present paper.  One stresses however that the
idea that the observed coupling constant, which is quite small, could
be a result of some original ``bare'' strong interaction, is
consistent with the microscopic theory if we argue it is the molecular
interactions behind the glass transition itself which become
``renormalized'' in a somewhat unexpected fashion. This microscopic
theory suggests \cite{LW} that the phenomenological two-level systems
are discrete energy levels representing resonantly accessible local
degrees of freedom that exist in glasses due to the possibility of
collective transitions between alternative structural configurations
of compact regions encompassing roughly 200 molecular units.  The
theory of glassy ergodicity breaking shows the spectrum of these
excitations is nearly flat and the density of states scales with the
inverse glass transition temperature $T_g$, echoing the excitation
spectrum of a random energy model (REM) with that glass transition
temperature. Furthermore, the transitions are an alternative mode of
motion that must be in equilibrium with phononic excitations at
$T_g$. This equilibrium requirement makes one realize that TLS-phonon
coupling $g$, $T_g$ and the material's elastic constants are
intrinsically related. The universality of the $l_\smfp/\lambda$ ratio
is a consequence of this relationship reflecting the non-equilibrium
character of the glassy state. The structural transitions, that become
tunneling two-level systems at cryogenic temperatures, exist because a
glassy sample, when it falls out of equilibrium, resides in a
metastable configuration chosen from a very high density of
states. The sample is broken up into a mosaic of dynamically
cooperative regions. Alternatively speaking, the energy landscape is
local in nature; that is rearrangements of compact regions will not
change the structural state of the rest of the sample, but only deform
the surrounding regions weakly and purely elastically. A (small)
fraction of these rearrangements requires overcoming only a very low
barrier and can therefore occur even down to sub-Kelvin
temperatures. The tunneling occurs by consecutive molecular
displacement within the cooperativity length established at $T_g$. The
consecutive motion of atoms is conveniently visualized as a domain
wall separating the two alternative local structural states, moving
through the local region.

The thermal conductivity plateau has traditionally been considered by
most workers a separate issue from the TLS. In addition to the rapidly
growing magnitude of phonon scattering at the plateau, an excess of
density of states is observed in the form of the so called ``bump'' in
the heat capacity temperature dependence divided by $T^3$. The plateau
is interesting from several perspectives. For one thing, it is
non-universal if scaled by the elastic constants (say $\omega_D$ and
$c_s$).  It is, however, located between two universal regions and it
is important to understand which {\em other} scales in the problem
determine its location and shape. The excitations that give rise to
the dramatically increased phonon absorption at the corresponding
frequencies have been circumstantially associated with the excitations
observed as the so called Boson Peak (BP), directly seen in the
inelastic X-ray and neutron scattering experiments, also observed in
the optical Brillouin and Raman scattering measurements.  These
experimental developments date well into 90-s and became possible, in
the neutron spectroscopy case, due to the improved resolution in the
neutrons' velocity detection, combined with the ability to generate
higher energy incident beams \cite{neutron}.  Similarly, meV
resolution was needed to utilize the X-ray scattering technics to
discern the small inelastic wings on the sides of the strong elastic
peak \cite{Xray}.  The term ``Boson Peak'' comes from the fact that
its intensity scales roughly according to the Bose-Einstein
statistics.  The extraction of the density of states from the spectra
is unfortunately model dependent, and those models can be roughly
divided \cite{Pilla} into the ones where the Boson peak signifies the
energy scale on the edge of phonon localization, as promoted in
\cite{neutron}, and those following the other school of thought which
asserts that these modes are propagating even well above the frequency
of the BP, as supported by the interpretation in \cite{Pilla}. As far
as theoretical interpretation is concerned, it is our impression that
most of theories of the Boson Peak, existing until recently, have
postulated a sort of spatial heterogeneity in an otherwise perfectly
elastic medium (see a partial list of references in \cite{Parisi1}),
with the notable exception of the soft-potential model (SPM)
\cite{soft_pot,soft_potBuchenau}. It is, of course, always possible to
recover the observed magnitude of the heat capacity excess at the BP
temperatures by a particular choice of parameters. While a
contribution of the lattice disorder to the density of states
undoubtedly exists and can be very significant (see, for example,
simulations of silica's heat capacity by Horbach at
el. \cite{Horbach}), we must note that if amorphous lattices were
purely harmonic, the phonon absorption at the BP frequencies would be
of the Rayleigh type and should be significantly lower than observed
in the experiment \cite{ACAnd_Phi,Joshi}. There must be internal
resonances present in the bulk, that scatter phonons
inelastically. Though phenomenologically introduced, this feature is
present, for example, in the soft-potential model. An analysis of the
higher temperature behavior of the tunneling transitions that give
rise to the TLS at subKelvin energies was provided in the RFOT
approach in \cite{LW_BP}. When these transitions occur at high enough
temperature, the domain wall separating the two alternative states can
have its surface vibrations thermally excited. The large degeneracy of
these vibrational states, characteristic of a two dimensional
membrane, that accompany the underlying structural transition, is
sufficient to account for the enhancement of phonon scattering at the
plateau, as compared to the TLS regime. Finally, the superposition of
the domain wall vibrations on the underlying tunneling transition
leads to an excess of density of states that reproduces well the bump
in the heat capacity (these compound excitations we call
``ripplons''). We therefore arrive at a unified physical picture that
allows a unified quantitative explanation of previously seemingly
unrelated mysteries in the TLS regime and at the higher, plateau
energies.

The paper is organized as follows: the first section outlines the
basics of the RFOT theory and then proceeds in applying that theory to
understanding the origin of the tunneling centers in amorphous
solids. The spectrum of the two-level systems, their coupling to the
phonons and the origin of the universality of phonon scattering are
then discussed. Additionally, we show how details of the derived TLS'
tunneling amplitude distribution lead to a deviation of $T$ dependence
of the heat capacity from a strict linear form.  The second section
explains how the high energy vibrational excitations (ripplons) of the
tunneling interfaces gives rise to an excess of states which exhibits
itself as the heat capacity bump and yields the rapidly rising phonon
scattering at these higher energies.  A short discussion of the {\em
relaxational} absorption from these excitations is given and its
frequency dependent part is derived. The contents of these first two
sections are, for the most part, a detailed account of the
calculations underlying two earlier brief letters \cite{LW,LW_BP} that
have reported our explanation of the low temperature anomalies in
glasses within a semiclassical approach. The third and fourth chapters
are comprised of new results. There, we establish that, while not
altering the main conclusions of the semiclassical picture, a purely
quantum phenomenon of level mixing and repulsion has an observable
effect on the density of states of the tunneling centers at low $T$.
Finally, the interaction between tunneling centers, mediated by
phonons, is estimated and this is argued to make a significant
contribution to the negative thermal expansivity (and thus a negative
Gr\"{u}neisen parameter) observed in many amorphous materials.

\section{Overview of the Classical Theory of the Structural Glass 
Transition}
\label{RFOT}

From a physicist's perspective, a theory of the glass transition
describes what happens to a liquid when it is cooled down sufficiently
but is not observed to crystallize. To a mathematician, this is a
generalized problem of packing compact interacting objects of
comparable size given a specific constraint on the density
distribution (it is not periodic) and total energy of the system. A
nearly complete conceptual, microscopic picture of the amorphous state
has emerged in the course of the two last decades \cite{dens_F2}
\cite{dens_F1} \cite{MCT,MCT1} \cite{KT_PRL87,KT_PRB87,KTW}
\cite{XW,XWbeta,XWhydro} \cite{LW,LW_BP,LW_soft,LW_aging}. This
framework has lead to a unified, quantitative understanding of many
seemingly unrelated phenomena in supercooled liquids above and below
the glass transition. The glasses we consider form at temperatures
where quantum effects are small so classical statistical mechanics is
used. We review such a classical glass transition in what follows.

First, we make several comments on the phenomenology of supercooled
liquids. Strictly speaking, these are nonequilibrium systems: When
cooled sufficiently slowly, most simple liquids will crystallize at a
temperature just below the melting temperature $T_m$.  Randomly
atactic polymers become glassy but presumably never crystallize.  The
melting point is defined as the temperature at which the liquid and
crystal free energies are equal. Cooling the liquid at least a bit
below $T_m$ is necessary to create a free energy driving force so as
to make the nucleation barrier finite and to allow the system to
equilibrate.  The crystal, once formed is different from the liquid in
several ways, e.g. it scatters X-rays at precise angles and it is
anisotropic.  Crucially for us, a crystal supports transverse sound
waves, at {\em all} frequencies (including $\omega=0$, thence the
crystal retains its shape). In contrast, the supercooled liquid is a
finite lifetime state since crystallization will eventually occur by
nucleation.  However, the growth of crystalline nuclei, inside the
liquid, is subject to the slowing of all motions in liquids. Owing to
this dramatic slowing of liquid motions upon lowering the temperature,
one can supercool the liquid substantially below its melting point,
which is the key to forming glasses. The extra nucleation barrier
ensures there is adequate time to study the properties of the
supecooled noncrystalline state.  Local structures in supecooled
liquids persist for some time, call it $1/\omega_c$. This time is
longer than the time it takes to establish a Maxwell distribution of
velocity, which is at most a few vibrational periods. Such an
amorphous system will support transverse waves at frequencies $\omega
> \omega_c$, just as a crystal would, but will in contrast exhibit a
liquid like, equilibrium response to time dependent perturbations at
frequencies $\omega < \omega_c$. As we have said, $\omega_c$ drops
rapidly upon cooling.  If one is intent on observing equilibrium
response at {\em some} frequency range, one must prepare the sample by
cooling it more slowly than $\omega_c$.  Conversely, for any given
cooling rate, no matter how slow, the liquid will fall out of
equilibrium on {\em all} time scales and the sample will appear to be
mechanically solid. We say the liquid has undergone the glass
transition. (The corresponding $\omega_c$ usually ranges between
$10^2$ and $10^5$ sec, depending on the experimenter's patience.)  The
liquid just below the glass transition temperature $T_g$ is only
subtly different from the liquid just above $T_g$. Structurally, first
of all, the two are nearly identical. Even dynamically, both can flow,
although the $T$-dependences of the corresponding transport
coefficients are distinct in the two forms of the ``equilibrium''
supercooled liquid and the nonequilibrium glassy state
\cite{LW_aging}. The residual dynamics below $T_g$ is referred to as
``aging''. Aging is at least as slow as the motions just above $T_g$,
but can be much slower when the sample is studied well below $T_g$.
This requires a greater amount of the experimenter's patience in
studying system properties than even needed for sample
preparation. Finally, when the sample falls out of equilibrium at
$T_g$, a jump in the heat capacity is measured by differential
calorimetry, thus resembling, crudely, a phase transition.

The dramatic slowing down of molecular motions is explicitly seen in a
vast area of different probes of liquid local structures. Slow motion
is evident in viscosity, dielectric relaxation, frequency dependent
ionic conductance, as well as in the speed of crystallization
itself. In all cases, the temperature dependence of the generic
relaxation time obeys to a reasonable, but not perfect approximation
the empirical Vogel-Fulcher law:
\begin{equation}
\tau_{\srlxn} \propto e^{D
T_0/(T-T_0)}
\label{VF}
\end{equation}
For a review, see \cite{app_phys_rev,Bohmer}. A specific example of a
$\tau(T)$ dependence is shown in the l.h.s. panel of Fig.\ref{TNB}. In
the expression above, $T_0$ is a material dependent temperature at
which the relaxation times would presumably diverge, if the
experimenter had the patience to equilibrate the liquid at the
corresponding temperatures. Needless to say, measurements of
equilibrium dynamics near $T_0$ are essentially nonexistent.  The
coefficient $D$ is often called ``fragility'', with larger values of
$D$ corresponding to ``stronger'' substances, while smaller values are
associated with ``fragile'' liquids.  This terminology apparently
refers to the degree of covalent networking in the material
\cite{Angell_fragility}, a qualitative trend later rationalized by a
density-functional study of \cite{HallWolynes}. Fragility appears to
correlate with the Poisson ratio, at least for non-polymeric glasses
\cite{PoissonFrag}. At any rate, the value of coefficient $D$ is
directly related to what glassblowers refer to as ``short glasses''
and ``long glasses'', \cite{Schott}: (molten) glass can be worked or
shaped in the range of viscosities $10^4 - 10^9$ Poise. If the
corresponding temperature range is short, the glass is called
``short'', and vice versa for the ``long''glass. The former and the
latter obviously correspond to a small and large value of the
parameter $D$ respectively.

The non-equilibrium character of a supercooled liquid is exhibited in
the entropy of the liquid which is considerably larger at $T_g$ than
that of the corresponding crystall at this temperature.  This
additional entropy corresponds to all the molecular translations, that
would have otherwise frozen out at crystallization. In
crystallization, this would appear as the latent heat of the
liquid-to-crystal transition. In a supercooled liquid, the molecular
structure is dense enough to define a lattice locally. Vibrations
around lattice sites are small. The excess entropy associated with the
locations of these lattice sites has traditionally been designated as
the ``configurational'' entropy. This excess entropy, $s_c$, is
temperature dependent. It refers to all possible liquid configurations
that could be surveyed by the liquid if we wait long enough for
molecular translations to occur. Experimentally, we determine the
configurational entropy by relying on the third law of thermodynamics.
Using the third law, we know the total entropy of the liquid at $T_m$
by integrating the crystal's heat capacity (over $T$) and adding the
entropy of melting.  Now for the supercooled liquid, we integrate the
heat capacity difference between the liquid and the crystal. To do
this we, of course, assume the vibrational entropies of the ordered
and aperiodic lattices are close.  The heat capacity measured by
differential calorimetry above the glass transition depends on the
rate of the configurational and vibrational entropy decrease with
temperature right above $T_g$. Below $T_g$ the structure of the liquid
remains the same as of the moment of vitrification, apart from some
(normally insignificant) aging. The vibrational entropy decreases as
it did above $T_g$, but there is no component from configurational
change. Thus one observes a non-zero heat capacity jump at
$T_g$. Above $T_g$, the $s_c$ decreases and the density increases with
lowering the temperature. This is expected because there are fewer
ways to mutually arrange the molecules at higher densities. When
extrapolated past $T_g$, as was done by Simon \cite{Simon} and notably
by Kauzmann in his review \cite{Kauzmann}, the configurational entropy
vanishes at a temperature $T_K$, which is securely above the absolute
zero. This suggests that only a non-extensive number of low energy
aperiodic, liquid arrangements could be found at $T_K$ and the entropy
of the liquid is thus equal to the corresponding crystal (correcting
for differences in their vibrational spectrum).  This phenomenon is
sometimes referred to as the ``entropy crisis'', which, again, would
presumably occur only under completely equilibrium cooling.  Such an
entropy crisis strictly occurs in several mean-field spin glass models
with infinite interactions \cite{GrossMezard,GrossKanterSomp,MCT1}.
There are many sound arguments suggesting a strict singular vanishing
of configurational entropy at $T_K$ is unlikely for real liquids
\cite{EastwoodW, Stillinger88}. Nevertheless, $T_K$ is a useful
fiducial point for the analysis.  None of the results of the present
theory in the experimentally accessible regime depend on the
configurational entropy truly vanishing at any point.  As we shall
see, the configurational entropy is macroscopic but decreases with
temperature.  $s_c$ is typically $\sim .8 k_B$ per movable unit at the
conventional glass transition temperature corresponding to cooling
rate of inverse hour and decreases at a rate proportional to $\Delta
c_p/T_g$. For simplicity, we will assume $s_c$ extrapolates so as to
scale linearly with the proximity to the entropy crisis (see
\cite{RichertAngell}): $s_c = \Delta c_p (T-T_K)/T_K$.

Before our formal discussion, let us make several qualitative
statements about molecular transport above $T_g$.  The motions of a
supercooled liquid are much slower than the local vibrations.  The
potential felt by an individual molecule comforms to a local
``cage''. This local ``cage'' is formed by the neighboring molecules,
of course. In order to translate irreversibly a given molecule, as
opposed to vibrating about the current position, the cage must be
destroyed.  In other words, a number of surrounding molecues must be
translated as well. Upon lowering the temperature, the density
increases and $s_c$ decreases, therefore fewer alternative states are
available to any given group of molecules. Thus it is clear that
conforming the liquid to an arbitrary translation of a given molecular
unit will require readjusting the positions of more and more
surrounding molecules at the same time. This leads to a larger
cooperative region size, leading in turn to higher barriers for
relaxation processes and higher viscosity. At a crude level, this
picture underlies the arguments from \cite{AdamGibbs}, but those
arguments fail to relate the size of the moving regions to the energy
landscape itself.  In contrast, the Random First Order Transition
(RFOT) theory \cite{KTW,XW} explicitly shows how these
reconfigurational motions occur and thus establishes intrinsic
connection between the kinetic properties and the thermodynamics of
supercooled liquids. Our account is based on \cite{LW_aging} which
also discusses the intrinsic connection between cooperative, activated
motions in the supercooled liquid both above and the classical aging
dynamics below the glass transition. These arguments also pave the way
for understanding the quantum dynamics at cryogenic temperatures.

\begin{figure}[htbp!]
\includegraphics[width=.55\columnwidth]{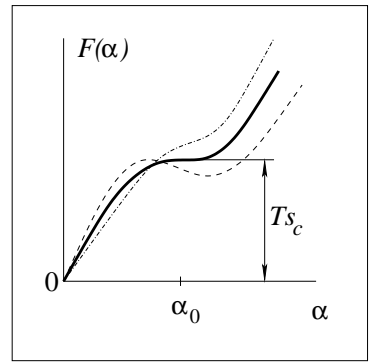}
\caption{\label{F_alpha} This is a schematic of the free energy
density of an aperiodic lattice as a function of the effective
Einstein oscillator force constant $\alpha$ ($\alpha$ is also an
inverse square of the localization length used as input in the density
functional of the liquid. Specifically, the curves shown characterize
the system near the dynamical transition at $T_A$, when a secondary,
metastable minimum in $f(\alpha)$ begins to appear as the temperature
is lowered. This figure is taken from \cite{LW_soft}.}
\end{figure}

The main prerequisite of the RFOT theory is the existence of time
scale separation between vibrational thermalization and equilibrating
structural degrees of freedom that involve crossing saddle points on
the free energy surface. This only occurs below a crossover
temperature $T_A$ which is predicted by the theory itself.  The
existence of local trapping in cages is well established by
experiment: there is a long plateau in the time dependent structure
factor as measured by the inelastic neutron scattering \cite{Mezei}.
In RFOT, such trapping was first established theoretically using a
density functional theory (DFT) in \cite{dens_F1}: This paper shows
there are aperiodic free energy minima by computing the free energy of
an aperiodic variational density distribution function: $\rho({\bf r})
\equiv \rho({\bf r},\{{\bf r}_i\}) = \sum_i
\left(\frac{\alpha}{\pi}\right)^{3/2} e^{-\alpha ({\bf r}-{\bf
r}_i)^2}$.  The set of coordinates $\{{\bf r}_i\}$ denotes a
particular aperiodic lattice. The typical lattice spacing is $a$. A
zero value for the parameter $\alpha$ would correspond to a completely
delocalized, uniform liquid state, such as just below the liquid-vapor
transition. $\alpha \rightarrow \infty$ would imply freezing into an
infinitely rigid lattice. $\alpha$ can also be interpreted as the
spring constant of an equivalent Einstein oscillator forcing each
molecule to remain near its proper location in the aperiodic
lattice. $F(\alpha)$ develops a metastable minimum, at non-zero
$\alpha = \alpha_0 \ne 0$, only below some temperature $T_A$.  This
minimum has higher free energy than than the lowest minimum at $\alpha
= 0$ (see Fig.\ref{F_alpha}).
In the mean field limit, the appearance of such minimum would lead a
lattice stiffness and would represent a state with a divergent
viscosity. This localization transition and the viscosity catastrophe
of mode-coupling theories are essentially identical as was established
in \cite{MCT}. A single such high lying free energy minimum would be
thermodynamically irrelevant, but one must recall that this
$F(\alpha)$ is computed for a {\em single}, particular aperiodic
lattice, which is actually only one of many possibilities. Taking into
acount the thermodynamically large number of alternative aperiodic
packings increases the entropy of the (set of) localized, aperiodic
state(s) and thus lowers the metastable free energy minima just the
right amount to make them competitive with the mean-field uniform,
delocalized state. The correspondence between the free energy
difference in mean field theory and the configurational entropy was
rigorously shown for the Potts Glass by \cite{MCT1} who argued such
systems have similar symmetry properties to structural glasses.  For
structural glasses this correspondnce may also be shown more formally
using a replica formalism \cite{repl_Lind}.  The localization
transition at $T_A$ is accompanied by a discontinuous change in the
order parameter $\alpha$. This is why the transition is called
``Random First Order''.  Although there is a discontinuity in
$\alpha$, the actual structure in which the system freezes is chosen
at random out of a multitude of possibilities (given by the
configurational entropy) At the same time, such an ordered phase will
persist only for finite times, therefore this is a true transition
only for high-frequency motions, comparable at first to the
vibrational time scale. This transition at $T_A$ only signifies a soft
cross-over, as far as the {\em whole} dynamical range is concerned. We
emphasize, there are many different ``phases'' below $T_A$, all of
which are random packings. The number of random packings, thermally
available to a region of size $N$, $e^{s_c N}$, decreases {\em
gradually} with temperature. (This corresponds to gradual freezing out
the translational degrees of freedom with lowering the temperature, as
signified by the decreasing $\omega_c$.) Because the decrease is
gradual, the {\em random} first order transition does not exhibit a
latent heat.  In a finite range system, different minima can
interconnect by barrier crossing. (Such barriers would be infinite in
mean field.) Even though the transition at $T_A$ is a crossover, the
temperature $T_A$ itself is a useful parameter characterizing material
properties.

The resulting time scale separation at and below $T_A$ has two
important consequences. First, one may perform canonical averaging
over the vibrations within a particular structural state. This gives a
free energy of a particular structural state: $\Phi = E - T S_\svibr$,
where $S_\svibr$ is the vibrational entropy. Note the vibrations are
not necessarily harmonic. To define $\Phi$, all that matters is that
the local vibrations equilibrate much faster than the structural
degrees of freedom.  As a consequence, $\Phi$ can be termed the bulk,
{\em microcanonical} energy of a given {\em structural} state. To any
value of this energy one may associate a bulk, microcanonical {\em
entropy} $S_c(\Phi)$ counting states with similar contributions from
energy and vibratrional entropy; both $\Phi$ and $S_c(\Phi)$ scale
linearly with the size of the system. One may thus to work with
morphologically distinct, globally defined aperiodic phases without
actually specifying their precise molecular constitution, so long as
we know their spectrum, i.e. their number as a function of the
microcanonical free energy.  These statistics are directly measurable
by calorimetry just as in our discussion of the Kauzmann paradox.

\begin{figure}[htbp!]
 \includegraphics[width=\columnwidth]{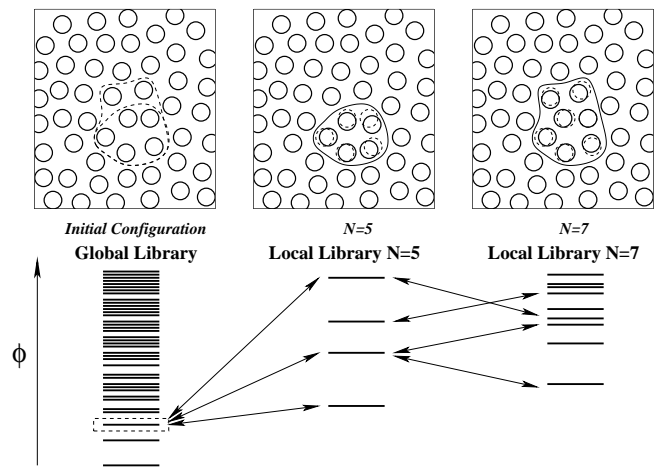}
 \caption{\label{library} This figure is taken from
\cite{LW_aging}. In the upper panel on the left a global configuration
is shown, chosen out of a global energy landscape. A region of $N=5$
particles in this configuration is rearranged in the center
illustration. The original particle positions are indicated with
dashed lines. A larger rearranged region involving $N=7$ particles is
connected dynamically to these states and is shown on the right. In
the lower panel, the left most figure shows the huge density of states
that is possible initially. The density of states found in the local
library originating from a given initial state with 5 particles being
allowed to move locally is shown in the second diagram. These energies
are generally higher than the original state owing to the mismatch at
the borders. The larger density of states where 7 particles are
allowed to move is shown in the right most part of this panel. As the
library grows in size, the states as a whole are still found at higher
energies but the width of the distribution grows.  Eventually with
growing $N$, a state within thermal reach of the initial state will be
found.  At this value of $N^*$ we expect a region to be able to
equilibrate. } 
\end{figure}

Having established the transitory existence of a global aperiodic
structure, we may next enquire into how molecular motions allow the
system to escape such a phase \footnote{Of course, the issue of
producing the aperiodic state in the {\em laboratory} would also
involve estimating whether corresponding quenching rates can be
experimentally achieved.}. This occurs by replacing locally one part
of the aperiodic packing by a different local packing. This will be an
activated event.  The RFOT theory allows one to compute the mean
activation barrier and its distribution. Also, the theory determines
critical region size and the spatial extent of the excitations
corresponding to the cooperative rearrangement. The magnitude of an
individual molecular displacement during the transition is determined
by $\alpha$. To estimate the activation free energy, let us make the
following construct.  Considering a library of possible local
aperiodic arrangements at a particular location, as illustrated in
Fig.\ref{library}.
This {\em local} library of states can be constructed based on the
existence of the {\em global} library of states introduced earlier
that we described by the energy variable $\Phi$ and the corresponding
entropy $S_c(\Phi)$ reflecting the spectrum. Clearly, the energy
density $\exp[S_c(\Phi)]$ is extremely high and grows rapidly with
$\Phi$. We might perform a full survey of local states by mentally
carving out a small region of size $N$, while freezing in place the
lattice sites surrounding the region. One can then heat the local
region and then allow that region to equilibrate.  Unless the new
local arrangement is exactly the same as the original one, its energy
will likely be significantly higher: A local substitution
statistically must cost free energy, stemming from a structural
mismatch between two randomly chosen aperiodic packings of a given
energy $\Phi$. This mismatch energy corresponds to the usual surface
energy, such as that between two different crystal forms or at a
liquid-crystal interface.  The free energy cost of locally replacing
the initial phase (labelled as ``in'') by another phase, call it $j$,
can therefore be written as
\begin{equation} 
\phi_j^{lib}({\bf R}) - \phi_\sinn^{lib}({\bf R}) = \Phi_j(N) -
\Phi_\sinn(N) + \Gamma_{j,in},
\label{phi_j}
\end{equation}
where $\Gamma_{j,in}$ is the mismatch energy and ${\bf R}$ is the
location of the local region. As before, the capital $\Phi$ denotes
the {\em bulk} energy, corresponding to a distinct aperiodic packing,
with the vibrational entropy already included. To compute the
likelihood of such a local rearrangement, substitute for the specific
surface energy $\Gamma_{j,in}$ its average value which should scale
with size: $\gamma N^x$.  $\gamma$ depends on the material and on
temperature. Naively, the usual surface energy scaling is
$N^{(d-1)/d}$, expected in $d$ dimensions. One can argue however that
$x$ will actually turn out to equal 1/2. Such a surface tension
renormalization was first conceived by Villain \cite{Villain}, in the
context of the random field Ising model (RFIM). In RFIM, the Ising
spins, in addition to their coupling, are subjected to a random static
magnetic field obeying certain fluctuation statistics. A flat
interface, or domain wall, between spin-up and spin-down domains will
distort so as to conform to the local variation of the field. An RG
argument incarnating this distortion on a hierarchy of length scales
yields a scale dependent renormalization of the surface tension,
giving a surface free energy exponent $x=1/2$ \cite{Villain}.  The
structure-structure interface in a supercooled liquid resembles the
RFIM, owing to the fluctuations of local energies of the various
aperiodic packings. The statistics of these fluctuating local energies
require that $\delta \Phi(\delta N) \sim \Phi_0 \sqrt{\delta N}$,
where $\Phi_0$ is $\delta N$-independent, echoing the fluctuation
statistics of the frozen random field of the RFIM. Thus, as \cite{KTW}
suggest, the originally thin flat interface will become diffuse
yielding $x=1/2$. In the liquid case, a vivid interpretation of the
surface energy renormalization is possible: Since the interface is
distorted down to the smallest scale (allowed by the material's
discretness), the region occupied by the now diffuse wall is neither
of the two original structures it separates. Instead it may be
interpreted as accommodating {\em other} structures. These
intermediate packings interpolate structurally two randomly chosen,
and thus {\it \`{a} priori} energetically disagreeable packings. In
other words, the original thin interface separating two given
packings, is ``wetted'' by other packings thus lowering the overall
interface energy. As we shall see, real liquids have only modest size
regions of rearrangement, so it is hard to argue about the exact value
of the exponent. Nevertheless, we note two felicitous observations:
With $x=1/2$, the usual scaling argument will give precisely a
discontinuity in $\Delta c_p$ at any ideal transition, to be seen at
$T_K$. Also, while the RFIM itself remains the subject of discussion,
Villain's argument does give a length scale exponent agreeing with the
majority of experiments and numerical studies \cite{Nattermann,
Belanger}.

The role of the interface mismatch energy in the reconfiguration
process can be beneficially understood from a statistical point of
view, as illustrated in Fig.\ref{library}. It costs free energy to
reconfigure a small number $N$ of molecules because considering a
small region severely limits the number of available liquid
configurations. The interface energy grows with $N$, however the
available density of states, too, grows with $N$, both in terms of its
absolute value and the distribution's {\em width}. At some size $N^*$,
that will be computed shortly, all relevant liquid states become
available.  The rate of escape of a group of $N$ molecules to another
structural state can be determined by a canonical type sum accounting
for the multiplicity of the final states at energy $\phi_j$:
\begin{eqnarray} 
k & = & \tau^{-1}_{\smicro} \int (d \phi_j^{lib}/c_\phi)
e^{S_c(\Phi_j)/k_B} e^{-(\phi_j^{lib}-\phi_\tin^{lib})/k_B T}
\nonumber \\ & \simeq & \tau^{-1}_{\smicro} e^{S_c(\Phi_{\teq})/k_B}
e^{-(\phi_{\teq}-\phi_\tin^{lib})/k_B T}.
\label{k}
\end{eqnarray}
In the second step, a steepest descent evaluation is made where
$\phi_{\seq}$ maximizes the integrand. $c_\phi$ is some constant of
units energy that reflects the local curvatures of the energy
landscape. The quantities $\phi_j^{lib}$ and $\Phi_j$ are related
through Eq.(\ref{phi_j}). The time scale $\tau_\smicro$ is the time
scale of a molecular scale non-activated process, typically of the
order a picosecond. The value $\phi_{\seq}$ that maximizes the
integrand above will be the internal (equilibrium) free energy
characteristic of the system at the ambient (i.e. vibrational)
temperature $T$. In other words, the greatest kinetically
accessibility of a state, as embodied in the optimization in
Eq.(\ref{k}), implies that the state will be most frequently visited
by the system, therefore it must be the {\em equilibrium} state.  The
integration in Eq.(\ref{k}) is similar to a canonical sum; yet it is
different in an important way: The summation in Eq.(\ref{k}) is far
more general than the usual expression for the partition function
because when relaxation times are continuously distributed, one must
{\em explicitly} weigh the contribution of a state (to the canonical
sum) by its kinetic accessibility. The latter, in general, will depend
on the spatial extent of the excitation corresponding to a transition
between two states; in this regard, the integration variable $\phi_j$
is, in a sense, a local {\em microcanonical} energy. Consequently, the
energy $\phi_\seq$ corresponds to a {\em canonical} energy. Yet,
$\phi_j$ and $\phi_\seq$ would strictly become a microcanonical and
canonical energy, in their conventional sense, only in the large $N$
limit, when the boundary effects are small. In contrast, the very
thermodynamic relevance of the glassy state is due to the locality of
the landscape and non-smallness of the surface term.  Finally, since
the bulk entropy $S_c(\Phi_{\seq})$ corresponds to the equilibrium
energy, it will be given by the equilibrium configurational entropy
$S_c(T)$, measured by calorimetry. Thus given $\phi_{\seq}$, one can
compute the value of the typical escape rate to a structure where $N$
particles have moved. This gives:
\begin{equation}
k(N) = \tau^{-1}_{\smicro} \exp\left\{S_c(N,T) -
\frac{\phi_{\teq}-\phi_\tin^{lib}}{k_B T} \right\}.
\end{equation}
The number of particles that must be moved for complete equilibration
is determined by the minimum of this expression over $N$.  We thus
determine an activation free energy profile
\begin{eqnarray}
F^\ddagger(N) &=& \phi_{\seq} - \phi_\sinn^{lib} - T S_c(N,T)
\nonumber \\ &=& \Phi_\seq (N) - \Phi_\sinn(N) + \gamma \sqrt{N} - T
S_c(N,T), \hspace{2mm}
\label{F_phi}
\end{eqnarray}
where we used Eq.(\ref{phi_j}) in the second equality.  The maximum of
the $F(N)$ curve defines the bottleneck location. This equation is
suitable for finding the rate of structural rearrangement both in the
equilibrated supercooled liquid (before it crystallizes!) and in the
nonequilibrium glass, which ages below $T_g$.

\begin{figure}[htb]
\includegraphics[width=.6\columnwidth]{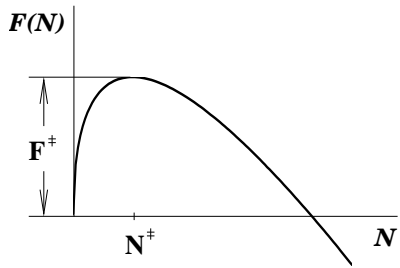}
\caption{\label{F_Ngraph}  The droplet growth free energy profile from
Eq.(\ref{F_N}) is shown.}
\end{figure}

Let us first consider equilibrium liquid rearrangements.  In this case
typically $\Phi_{\seq} = \Phi_\sinn$, apart from fluctuations.  Thus
one arrives at the following simple expression,
\begin{equation}
F(N)=\gamma \sqrt{N} - T s_c \, N,
\label{F_N}
\end{equation}
where we have used the thermodynamic scaling of the configurational
entropy, $S_c(N) = s_c N$. In the supercooled equilibrated liquid,
molecular transport is driven by only the multiplicity of mutual
molecular arrangements. For this reason, the reconfigurations
following the activation profile from Eq.(\ref{F_N}) have been called
``entropic droplets''.  The graph of the function in Eq.(\ref{F_N}) is
shown in Fig.\ref{F_Ngraph}.
The transition state configuration will satisfy $\prtl F/\prtl N = 0$,
corresponding to an unstable saddle point of this free energy. This
gives for fixed $\gamma$ a rearranging region size $N^\ddagger$ that
grows as $s_c$ diminishes: $N^\ddagger = (\gamma/2 s_c T)^2$. The
resulting barrier also scales inversely proportionally to $s_c$:
\begin{equation}
F^\ddagger = \frac{\gamma^2}{4 s_c T}.
\end{equation}
An inverse scaling of the barrier with the configurational entropy was
arrived at by Adam and Gibbs \cite{AdamGibbs} in a different (and
inequivalent) way. Notice if $\gamma$, as function of temperature, is
smooth around $T_K$ and $s_c$ is described by the linear law $s_c
\propto (T-T_K)$, the resulting activation barrier is exactly of the
Vogel-Fulcher law form (\ref{VF}), which, as we have said, fits data
well. Many arguments can lead to increasing relaxation times at low
temperatures and with enough adjustable parameters, can fit data. What
is different about the RFOT theory is that it establishes an intrinsic
link between the rate law and the entropy crisis. In addition, if the
entropy of the equilibrated fluid can be estimated, the density
functional theory allows the vibrational entropy and thus, by
substraction, the configurational entropy to be determined. Therefore
$T_K$ can be estimated from the microscopic force laws. This has been
done for simple soft spheres by Mezard and Parisi \cite{repl_Lind},
giving reasonable results.  Hall and Wolynes \cite{HallWolynes} have
also calculated $T_0$ and $T_A$ for a simplified model of a network
fluid. Their study is consistent with known chemical trends for $T_A$
and $T_K$ as the network becomes more thoroughly crosslinked.

The idea of the configurational entropy itself driving liquid
rearrangements still appears to generate some confusion. One possible
reason for this is that $s_c$ is totally unambiguously defined only in
the mean field limit. In the latter limit, rearrangements are infinite
so dynamics driven by $s_c$ do not arise.  This is a good place to
emphasize that the RFOT theory is not mean-field! Only the local
landscape, within an entropic droplet, is actually well described by a
mean-field, Random Energy Model like approximation. We took advantage
of this in extracting the energy spectrum of low energy structural
excitations in a frozen glass \cite{LW}, as explained in detail in the
following Section. We wish to point the reader to the recent elegant
treatment of \cite{BouchaudBiroli} re-analyzing the RFOT conclusions
for rearrangements in an equilibrated fluid from the viewpoint of
Derrida's Random Energy Model (REM) \cite{Derrida}.

Now, calculations of $T_A$ and $T_K$ are plagued by the usual
difficulties of liquid state structure theory and the accuracy of
approximations some of which are hard to control. Still, even in the
face of such approximations, such microscopic considerations lead us
to expect a universal value of $\gamma/T_g$ at $T_g$ as we shall
discuss below.

The RFOT theory allows the coefficient $\gamma$ in the mismatch energy
to be estimated from a microscopic argument. It turns out to be
proportional to $T_K$ and to depend logarithmically on the inverse
square of the so called Lindemann ratio. Early in the 20th century,
Lindemann argued that the thermal fluctuations of an atom's position
could only be a fraction of the lattice spacing $a$ in a solid, if the
packing is to be mechanically stable \cite{Lindemann}.  Since the
threshold value of the vibrational amplitude of an atom in the lattice
is finite, the transition in which the lattice disintegrates must be
first order. For crystals, the Lindemann ratio of this threshold
displacement $d_L$ to the lattice spacing is about $1/10$.  For
amorphous materials, the $d_L/a$ ratio can be obtained from the
plateau in the self correlation functions measured by neutron
scattering experiments \cite{Mezei}. Again, this ratio turns out to be
approximately one-tenth (universally!). This number is reproduced in
several microscopic calculations consistent with the RFOT theory, such
as the self-consistent phonon theory and density functional theories
\cite{dens_F1,dens_F2}, and dynamical mode coupling theory
\cite{Gotze_MCT,MCT,MCT1,MCT2}, with modest quantitative
variations. The meaning of $\alpha \simeq 0.1$ as a mechanical
stability criterion has been also corroborated within the replica
formalism \cite{repl_Lind}. In terms of the DFT calculation dicussed
earlier, $\alpha_L$ corresponds with the metastable minimum that the
free energy $F(\alpha)$ develops below the dynamical transition
temperature $T_A$ (see Fig.\ref{F_alpha}). It has a relatively weak
temperature dependence.  The logarithmic scaling of the surface
tension coefficient with the Lindemann length follows from a detailed
calculation by \cite{XW}, but can be rationalized in a simple way:
Below $T_A$, motions span only the length $d_L$, while in the liquids,
they can move a distance $a$ before losing their identity with a
neighboring molecule. The entropy of the ``caged'' fluid is less and
thus the free energy cost of confining a molecule within length $d_L$,
as opposed to $a$, can be assessed by recalling the free energy
expression for an ideal monatomic gas: $- f = \frac{3}{2} k_B T \ln
\left[ \left(\frac{eV}{N}\right)^{2/3} \frac{mT}{2\pi
\hbar^2}\right]$, written deliberately here so as to have a length
scale squared in the logarithm.

\begin{figure}[htbp!]
\includegraphics[width=.75\columnwidth]{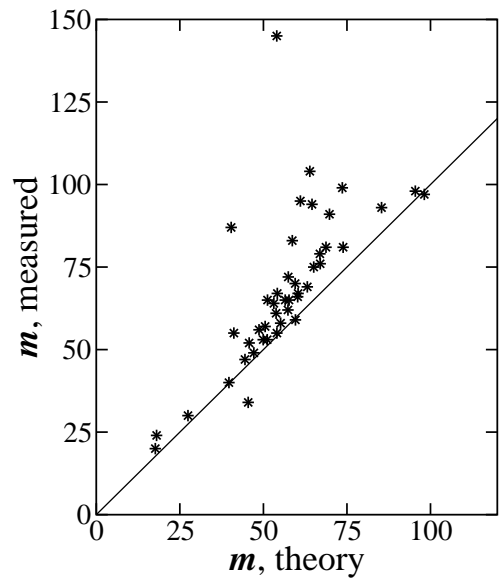}
\caption{\label{jake} The horisontal axis shows the value of fragility
as computed from the thermodynamics by the RFOT theory, and the
vertical axis contains the fragility directly measured in kinetics
exoeriments. Here $m$ is the so called fragility index, defined
according to $m=T[d \log_{10} \tau(T)/d(1/T)]$. $m$ is somewhat more
useful than the fragility $D$, because deviations from the strict
Vogel-Fulcher law, $\tau = \tau_0 e^{D T_K/(T-T_K)}$, are often
observed, see text. $m$ essentially gives the apparent activation
energy of relaxations at $T_g$, in units of $T_g$, it is roughly an
inverse of $D$. This figure is taken from \cite{StevensonWunp}.  In
evaluating $m$ theoretically, one needs to know the size of the moving
unit, or ``bead'', in each particular liquid. The latter can be
estimated using the entropy loss at crystallization \cite{LW_soft},
resulting in $m_\stheor \propto \frac{\Delta c_p(T_g) T_g}{\Delta H_m
s_c^2(T_g)} \propto \frac{\Delta c_p(T_g) T_g}{\Delta H_m}$, in view
of the near universality of $s_c(T_g)$ (see text).}
\end{figure}

$\gamma$ is proportional to $T_K$ and only logarithmically depends on
a nearly universal quantity, the Lindemann ratio. If $T_g$ is near
$T_K$, i.e.  for slow quenches, $\gamma/T_g$ is thus nearly material
independent and calculable: $\gamma = \frac{3}{2} \sqrt{3 \pi} k_B T_g
\ln(\alpha_L a^2/\pi e)$. Quantifying the mismatch energy this
specifically leads to many predictions about the dynamics near $T_g$,
for a range of substances. First, the coefficient in the Vogel-Fulcher
law $D$ is predicted to follow from the measured thermodynamics.
Using the $\gamma$ value above, we find not only the VF dependence of
the relaxation times on the temperature, $\propto e^{D T_0/(T-T_0)}$,
but also a remarkably simple formula relating $D$ and the heat
capacity jump: $D = 32.  R/\Delta c_p$ \cite{XW}. The coefficient
$32.$ is nearly universal and, as we see, follows numerically from the
microscopic theory since the universal value of the Lindemann ratio
enters only logarithmically in the localization entropy cost. The
numerical relation between $D$ and $\Delta c_p$ from this simple
explicit calculation is in rather remarkable agreement with
experiment. In Fig.\ref{jake}, we plot the so called fragility index
$m$, as computed from calorimetry and extracted from direct relaxation
measurements. $m$ is proportional to the slope of the $\log \tau$
vs. $1/T$ relation at $T_g$ and thus is directly related to $D$ if the
VF law is valid. ($D$ values in the literature are obtained from
global fits of $\log \tau$ vs. $1/T$ and depend somewhat on the
fitting procedure.)

Two other remarkable universalities emerge from the value of $\gamma$.
First, at a reference laboratory time scale of 1 hr $\sim 10^{17}
\tau_0$ we have a universal value of $s_c \simeq 0.8 k_B$. This
implies $s_c(T_g)/s_c(T_m) \simeq 0.7$, where $s_c(T_m)$ is, of
course, also the fusion entropy. This relation is independent of
question of what is the moving subunit. The relation holds very
well. A second important universal feature emerges from the universal
value of $\gamma/T_g$: the cooperative size at $T_g$ is nearly
universal.

Let us now consider in greater detail the pattern of cooperative
structural rearrangements in a supercooled liquid. These turn out to
presage the existence of the residual degrees of freedom in a glass
below $T_g$.  Within a period of time shorter than the typical
relaxation time $\tau$, the molecular motions within regions of size
$\xi^3$ will be highly correlated and, at the same time, approximately
decoupled from the surrounding. That is, the liquid is broken up in to
a (flickering) mosaic pattern of cooperative regions. This mosaic
structure is directly manifested in the dynamical heterogeneity
recently observed in supercooled liquids using single molecule
experiments \cite{RusselIsraeloff}, nonlinear relaxation experiments
\cite{Silescu} and non-linear NMR experiments \cite{Spiess}. (These
experimental tools became available only a decade after the RFOT
theory was first formulated.) The size of a typical mosaic cell is
found from the thermodynamic condition $F(N^*)=0$. Unlike the regular
nucleation of one distinct phase within another (as in crystal growth
in the liquid), by crossing the barrier from Eq.(\ref{F_N}) the local
region arrives at a statistically similar but an alternative solution
of the free energy functional, thus that solution still represents a
typical liquid state! An informal analogy here is that distinct low
energy dense local liquid packings are like the fingerprints of
different individuals - different in detail, yet generically
equivalent liquid states. Since we have agreed that $F=0$ is the
liquid {\em equilibrium} free energy at this temperature (the
crystalline state is assumed to be hidden behind a high enough crystal
nucleation barrier), the condition $F(N^*)=0$ specifies the size of
region to which an arbitrary liquid configuration is
available. Therefore, a region of size $N^*$ is able to reconfigure on
the experimental time scale characterized by $F^\ddagger$. In terms of
physical length, $F(N^*)=0$ implies $\xi \equiv {N^*}^{1/3} a = a
\left[\frac{8}{3 \sqrt{3 \pi}} \ln\left(\frac{\tau}{\tau_0}\right)/
\ln\left(\frac{\alpha_L a^2}{\pi e}\right) \right]^{2/3} \simeq 5.8 \,
a$ ($N^* \simeq 190$). The critical radius $r^*$ at $T_g$ is a
multiple of $\xi$.  Droplets of size $N>N^*$ are thermodynamically
unstable and will break up into smaller droplets, in contrast to what
prescribed by $F(N)$, if used naively beyond size $N^*$. This is
because $N=0$ and $N=N^*$ represent thermodynamically equivalent
states of the liquid in which every packing typical of the temperature
$T$ is accessible to the liquid on the experimental time scale, as
already mentioned.  In view of this ``symmetry'' between points $N=0$
and $N^*$, it may seem somewhat odd that $F(N)$ profile is not
symmetric about $N^\ddagger$. Droplet size $N$, as a one dimensional
order parameter, is not a complete description.  The profile $F(N)$ is
a projection onto a single coordinate of a transition that must be
described by $e^{s_c N^*}$ order parameters - the effective number of
distinct aperiodic packings explored by the liquid. At the point
$N^*$, the free energy functional actually has a minimum as a function
of the (multi-component) order parameter.  A more detailed discussion
of this can be found in Ref.\cite{LW_soft}, where we compute the
softening of the barrier $F^\ddagger$ near $T_A$ due to order
parameter magnitude fluctuations that are important near $T_A$.

We thus see that the length scale of the mosaic and number density of
the mosaic domain walls is determined by the competition between the
energy cost of a domain wall and the entropic advantage of using the
large number of configurations. We emphasize again, the relative
domain size $\xi/a$ depends only on the logarithms of the relaxation
rate and the Lindemann ratio, nearly universal parameters themselves,
and is therefore the same for different substances. This high
temperature phenomenon of universality at $T_g$ has direct
consequences for the universality of the {\em ultra-low} temperature
glassy anomalies.

\begin{figure}[htbp!]
\includegraphics[width=.95\columnwidth]{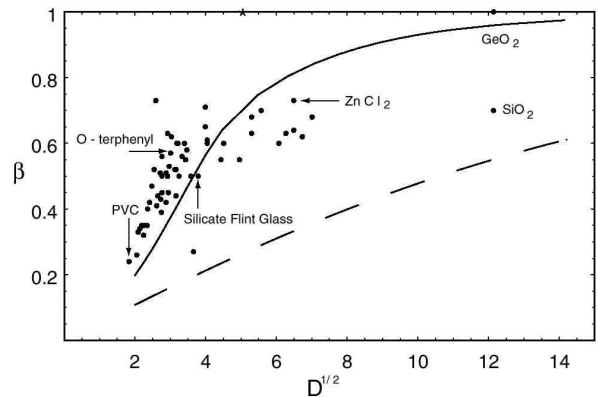}
\caption{\label{beta_D} Shown is the correlation between the liquid's
fragility and the exponent $\beta$ of the stretched exponential
relaxations, as predicted by the RFOT theory, superimposed on the
measured values in many liquids taken from the compilation of 
\cite{Bohmer}. The dashed line assumed a simple
gaussian distribution with the width mentioned in the text. The solid
line takes into account the existence of the highest barrier by
replacing the barrier distribution to the right of the most probable
value by a narrow peak of the same area; the peak is located at that
most probable value.  This figure is taken from \cite{XWbeta}.}
\end{figure}

\begin{figure}[htbp!]
\includegraphics[width=.95\columnwidth]{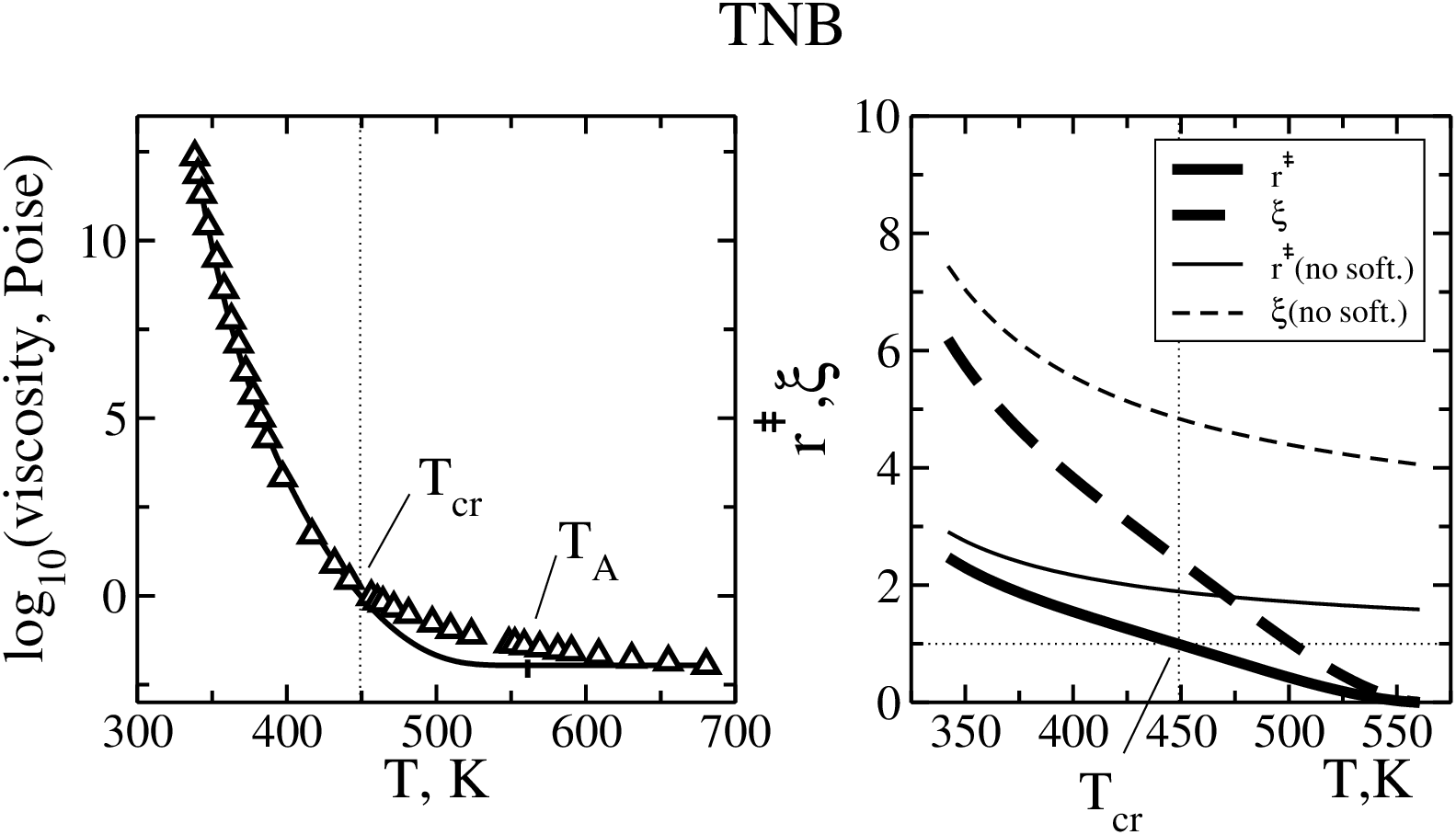}
\caption{\label{TNB} Experimental data (symbols) for TNB's viscosity
\cite{TNBexp}, superimposed on the results of the fitting procedure
(line) from \cite{LW_soft} are shown. $T_A$ is shown by a tickmark.
The temperature $T_\scr$ signifies a cross-over from activated to
collisional viscosity, dominant at the lower and higher temperatures
respectively (see text).  The temperature is varied between the
boiling point and the glass transition.  The r.h.s.  pane depicts the
temperature dependence of the length scales of cooperative motions in
the liquid. The thick solid and dashed lines are $r^\ddagger$ and
$\xi$ respectively. This figure is taken from \cite{LW_soft}.}
\end{figure}

We have seen that the cooperative region, which represents a nominal
dynamical unit of liquid, is of rather modest size, resulting in
observable fluctuation effects. Xia and Wolynes computed the
relaxation barrier distribution \cite{XWbeta}.  The configurational
entropy must fluctuate, with the variance given by the usual
expression: $\la (\delta S_c)^2 \ra = C_p \propto 1/D$
\cite{LLstat}. The barrier height for a particular region is directly
related to the local density of states, and hence to the
configurational entropy itself by Eq.(\ref{F_N}), $F^\ddagger \propto
1/s_c$. As a result, the barrier distribution width must correlate
with the fragility. A gaussian approximation leads to a simple formula
$\delta F^\ddagger/F^\ddagger = 1/2 \sqrt{D}$ \cite{XWbeta}.  There
are also calculable deviations from gaussianity.  The barrier
distribution gives rise to non-exponentiality of relaxations.  These
are well fitted by a {\em stretched} exponential
$e^{-(t/t_0)\beta}$. The measured $\beta$ correlates with the
fragility, in good agreement with the theory, see Fig.\ref{beta_D}.

We have so far presented a simplified picture of activated relaxation
in liquids, which is more accurate at temperatures close to $T_K$, and
thus sufficiently lower than $T_A$ - the temperature at which
activated processes emerge. The transition at $T_A$ where metastable
minima emerge, along with a mosaic structure with intermediate tense
regions, i.e. domain walls, is in many respects similar to a spinodal
for an ordinary first order transition, except that the number of
alternative phases is very large ($e^{s_c N}$ for a region of size
$N$). The proper treatment of this transition must include
fluctuations of the order parameter and consequent softening of the
droplet surface tension at temperatures close to $T_A$. As a result of
this, closer to $T_A$ the structural relaxation barriers are lowered
from what would be expected extrapolating from near $T_K$ - this gives
deviations from the VF law. The corresponding length scales
$r^\ddagger$ and $\xi$ also should be smaller than would be predicted
by the ``vanilla'', $T_K$-asymptotic version of the RFOT theory. These
barrier ``softening'' effects were quantitatively estimated in
\cite{LW_soft}.  They demonstrated that softening effects do vary
between different substances and are more pronounced for fragile
liquids. As a result, the value of the configurational entropy at
$T_g$, as predicted by the RFOT theory with softening varies somewhat,
within a factor of two or so among different substances. This is in
contrast to the universal $s_c(T_g) = .82$ of the vanilla
RFOT. Nonetheless, the value of $\xi$ at $T_g$ is much less sensitive
and seems to be always within 5\% of the simple estimate above. This
is shown in the r.h.s. panel of Fig.\ref{TNB}.
Understanding of the softening effect has allowed us to compute the
activation barrier for liquid rearrangements in the full temperature
range, including the high $T$ part near $T_A$, where the barriers
become low, and the transport is dominated by activationless,
collisional phenomena.  Consistent with this predicted softening, the
$T$ dependence of relaxation times, $\tau = \tau_\smicro e^{\gamma^2/4
T^2 s_c(T)}$, as predicted by the RFOT (see Eq.(\ref{F_N})), fits well
the experimental dependences in the low frequency range, but
underestimates the viscosity near boiling. After softening is
included, one can compute the activation component of the molecular
transport, with the temperature $T_A$ as a fitting parameter of the
theory \cite{LW_soft}.  Fitting the viscosity was performed using the
following obvious constraints: (a) at low temperatures, the order
parameter $\alpha$ fluctuations are negligible, the barriers are fully
established and high, and the transport is thus fully activated; (b)
near boiling, the barrier vanishes, and the viscosity (known to be
around a centipoise for all liquids) gives the value of
$\tau_\smicro$. The fit, shown in Fig.\ref{TNB}, demonstrates that of
the 16-17 orders of the total dynamical range, about three orders, on
the low viscosity side, are dominated by collisions. The experimental
and activation-only theoretical curve differ from each other above a
temperature $T_\scr$. The three order of magnitude time scale
separation, arising {\em internally} in the theory, is indeed
consistent with the prerequisite of the transport being fully
activated at $T_\scr$ and below. The discussion above indicates that
samples quenched (sufficiently fast) from a temperature $T > T_\scr$
may exhibit somewhat distinct detailed molecular motions, also
implying quantitative deviations form the RFOT predictions.  At any
rate, these sample, being caught in a very high energy state, are
expected to have small cooperative regions, and also be very brittle
and in general mechanically unstable.  Such rapid quenches would be
extremely difficult to produce in a lab, because $T_\scr$ corresponds
to relaxation times of the order $10^{-8 \ldots 9}$ sec. On the other
hand, it is these ultra fast quenches, that must be currently employed
by simulations owing to the limitations of computer power.  We
speculate that the thin ``amorphous'' films made by vapor deposition
on a cold substrate also may sometimes correspond to such
ultra-quenches. While one may expect a number of behaviors in the {\em
bulk} that are qualitatively distinct from what we have discussed
here, various {\em surface} effects are likely to be important too:
For one thing, such films are thin, have a large free surface, and
strongly interact with the substrate. Further, there is a good reason
to believe these films undergo local cracking, and spontaneous
crystallization \cite{Perry_private}.

\begin{figure}[htbp!]
 \includegraphics[width=.6\columnwidth]{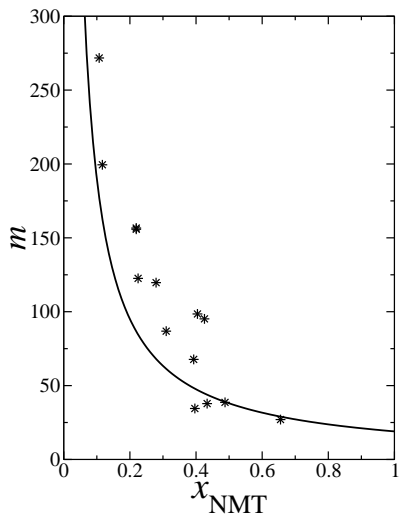}
 \caption{\label{m_x} The fragility parameter $m$ is plotted as a
function of the NMT nonlinearity parameter $x_{\sNMT}$. The curve is
predicted by the RFOT theory when the temperature variation of
$\gamma_0$ is neglected. The data are taken from
Ref. \cite{app_phys_rev}.  The disagreement may reflect a breakdown of
phenomenology for the history dependence of sample preparation. The
more fragile substances consistently lie above the prediction, which
has no adjustable parameters. This discrepancy may be due to softening
effects.} \end{figure}

The present article deals with phenomena in glasses at temperatures
much much lower than the temperatures at which the samples form. If a
sample, upon vitrification, is cooled significantly below $T_g$, its
lattice remains practically the same as of the moment of
freezing. Indeed, the typical reconfiguration barrier is at least
$\ln(10^{15}) \sim 35 k_B T_g$, as already mentioned. If, on the other
hand, the sample in maintained at some temperature $T$ close enough to
$T_g$, exceedingly slow structural relaxations take place. These
attempts of the sample to equilibrate to a structure characteristic of
temperature $T$ can be detected. Achieving quantitative accuracy in
such experiments is difficult.  Consistent with the notion that the
lattice, and the barrier distribution, freeze in at the glass
transition, the relaxation below $T_g$, obeys approximately the
following temperature dependence:
\begin{equation}
k_{\mbox{\scriptsize n.e.}} = k_0 \exp\left\{-x_{\sNMT}\frac{\Delta
E^*}{k_B T} - (1-x_{\sNMT}) \frac{\Delta E^*}{k_B T_g} \right\},
\hspace{4mm}
\end{equation}
where $E^*$ is the equilibrated apparent activation energy at $T_g$
and $x_{\sNMT}$ lies between 0 and 1. This equation is part of the
Nayaranaswany-Moynihan-Tool (NMT) empirical description of aging
\cite{Tool,Narayanaswamy,Moynihan}. The difference in the apparent
activation energy above and below $T_g$, as expressed by the parameter
$x_{\sNMT}$, will depend on how fast the barrier itself was changing,
with cooling, above $T_g$, under ``equilibrium'' cooling
conditions. Since the rate of that change depends on the fragility, $m
= \left. \frac{1}{T_g} \frac{\partial \log_{10}
\tau}{\partial(1/T)}\right|_{T_g} = \frac{\Delta E^*}{k_B
T_g}\log_{10}e$, one expects that $x_\sNMT$ and $m$ are
correlated. The RFOT based theory of aging in \cite{LW_aging} analyzes
structural rearrangements in a non-equibrium glassy sample by means of
Eq.(\ref{F_phi}), where the initial state is not equilibrium, but
instead corresponds to the structure frozen-in at $T_g$. The predicted
correlation between $x_\sNMT$ and $m$ is very simple: $m \simeq
19/x_\sNMT$, and is consistent with experiment, see Fig.\ref{m_x}.

For some of the comparison of theory and experiment it is necessary to
be specific about the molecular length scale $a$ (a very detailed
discussion of this quantity can be found in \cite{LW_soft}). The
molecular scale denotes the lattice spacing between molecular units
(or ``beads'') that act as idealized spherical objects at the ideal
glass transition at $T_K$. The determination of $a$, though
approximate, is rather unambiguous and can be done using the knowledge
of chemistry to give values accurate within 15\%. For example, the
number of beads in a chain molecule, that interacts with the
surrounding only weakly, is always close to the number of
monomers. Highly networked substances, such as amorphous silica,
present a more difficult case, because it is not clear how covalent
the intermolecular bonds in these substances are. Since melting also
involves freeing up molecules, with encreased entropy, an independent
check on the soundness of a particular bead number assignment can be
done by comparing the fusion entropy of the substance (if it exists in
crystalline form) with the known entropy of fusion of a hard sphere
liquid or Lennard-Jones liquid, equal to $1.16 k_B$ and $1.68 k_B$
respectively \cite{Hansen}. Note, however, the knowledge of the
absolute value of $a$ is not required for most of the numerical
predictions the theory will make in the quantum regime.
 
We thus see that the RFOT theory provides a rather complete picture of
vitrification and the microscopics of the molecular motions in
glasses. The possibility of having a complete chart of allowed degrees
of freedom is very important, because it puts strict limitations on
the range of {\it \`{a} priori} scenarios of structural excitations
that can take place in amorphous lattices. This will be of great help
in the assessment of the family of strong interaction hypotheses
mentioned in the introduction.  To summarize, the present theory
should apply to all amorphous materials produced by routine quenching,
with quantitative deviations expected when the sample is partially
crystalline. The presence and amount of crystallinity can be checked
independently by X-ray. It is also likely that other classes of
disordered materials, such as disordered crystals, will exhibit many
similar traits, but of less universal character.

\section{The Intrinsic Excitations of Amorphous Solids}
\label{Intrinsic}

\subsection{The Origin of the Two Level Systems}
\label{TLS}

In this section we discuss how phenomena near the glass transition
temperature, described in the previous subsection, dictate the
existence and character of the quantum excitations in glasses at
liquid helium temperatures and below. As mentioned earlier, a
dynamical pattern of cooperative regions forms in a supercooled liquid
below $T_A$. Each cooperative region is defined by the existence of at
least {\em two} distinct configurations mutually accessible within the
time scale $\tau$, which chatacterizes the life-time of the local
mosaic pattern.  Conversely, a molecular transport event is made
possible by rearranging molecules within the cooperative length scale.
The mosaic pattern ``flickers'' on the time scale $\tau$; this process
slows down dramatically upon vitrification and, below $T_g$ is
referred to as ``aging'', as it corresponds to (very slow) structural
changes.  At $T_g$, the existent pattern of transitions (with
distributed energy changes and reconfiguration barriers) freezes in
because each cell is now surrounded by a rigid lattice (this is
because the rearrangements of the neighboring domains were
uncorrelated at $T_g$). Each region of the material can now explore
the phase space as prescribed by the environment at the time of
freezing. Below $T_g$, the mosaic is spatially defined by the
molecular motions that were not arrested at $T_g$, and is thus
strictly speaking only {\em dynamically} detectable. It is true that
the weaker walls will probably be the site of (unstable) instantaneous
normal modes in the fluid state with imaginary frequencies.  This
dynamical correlation pattern does not necessarily imply any easily
discernible spatial heterogeneity in the atomic locations.  In fact,
there has been no direct evidence for any static type of heterogeneity
of the appropriate scale in glasses so far, which definitely
contributes to the (underappreciated) mystery of glasses
\footnote{We note, however, that there have been instances of
mistaking polycrystalline samples for truly amorphous ones.}.  But can
the {\em dynamical} heterogeneity be seen directly?  We will claim
later that this is done for us by thermal phonons: the magnitude of
scattering at the plateau can only be explained by presence of {\em
dynamical} heterogeneities. The latter are signified by structural
transitions that scatter the phonons {\em inelastically}. Apart from
aging (which we will ignore in the rest of the work), a particular
pattern of flipping regions, as frozen-in at $T_g$, will persist down
to the lowest temperature. The apparent size of each cell in this
mosaic of flippable regions will depend on the observation time. The
longer this time is, the more structural relaxation degrees of freedom
(from the high barrier tail of the barrier distribution) one should
observe. Eventually, in fact, the glass should
crystallize\footnote{Note that there are, in principle, other ways to
move molecules in a glass, in addition to the cooperative
rearrangements: for example by creating defects such as vacancies (the
corresponding barriers are prohibitively high, of course)}. In order
to estimate the number of tunneling centers that are thermally active
at low temperatures we will have to find the size of the regions that
allow for a rearrangement accompanied by a small energy change and, at
the same time with a low barrier. It may reasonably seem that
typically such barriers for multiparticle events would be very
high. Nevertheless, the lattice is arrested in a high energy state. We
can thus foresee the possibility of stagewise barrier crossing (or
tunneling) events, when the width of the barrier for each consecutive
atomic movement is only a small fraction of a typical interatomic
distance, thus rendering individual atomic movements nearly
barrierless. This is as if one could define an instantaneous mode of
nearly zero frequency, at each point along the tunneling
trajectory. (Yet at no point is the motion harmonic per se!)  The
presence of such low frequency modes should be expected given the high
number of configurational states available to the sample as the moment
of freezing, as reflected in the high value of the configurational
entropy at $T_g$. After all, the material is unstable, both globally
and locally!  (Note, the extent of bond deformation during an
individual atomic movement is low - within the Lindemann length -
actually affording a few ``hard'' places along the tunneling
trajectory, where the ``instantaneous'' frequencies are not
necessarily low.) One may contrast the situation above with, say,
tunneling of a substitutional impurity in a crystal, a system which is
indeed near its true ground state. Such tunneling would not contribute
to the very low $T$ thermal properties owing to a large barrier. Also,
we note that multiparticle barrier crossing events {\em have} been
seen in computer simulations of amorphous systems
\cite{GuttmanRahman}, anticipated theoretically
\cite{MonAshcroft,HeuerSilbey}, and recently inferred from simulations
of dislocation motions in copper \cite{Sethna}.

\begin{figure}[htb]
\includegraphics[width=.8\columnwidth]{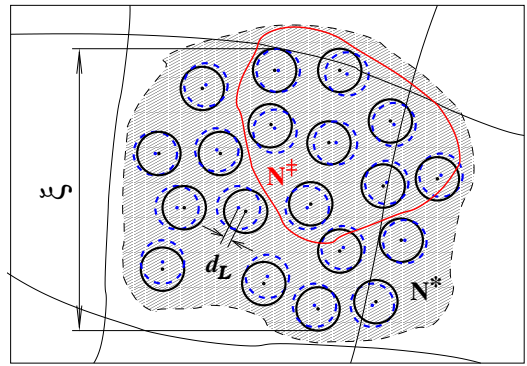}
\caption{\label{coop_reg} A schematic of a tunneling center is
shown. $\xi$ is its typical size. $d_L$ is a typical displacement of
the order of the Lindemann distance. The doubled circles symbolize the
atomic positions corresponding to the alternative internal states.
The internal contour, encompassing $N^*$ beads, illustrates a
transition state size, to be explained later in the text.}
\end{figure}

We summarize the discussion so far by noting that the preceding
Section has demonstrated that the possible atomic motions in a
supercooled liquids are either purely vibrational excitations or
structural rearrangements. Any possible motions {\em below} $T_g$, in
terms of the {\em classical} basis set must be a subset of the motions
above $T_g$, although the dynamics of these events become
quantum-mechanical at low enough temperatures.  Even after the system
is cooled to an arbitrarily low temperature, it remains essentially in
the configuration in which it got stuck at the glass transition. The
density of directly accessible states at that high energy
configuration is rather high; the {\em total} density of states is, of
course, exponentially larger, but inaccessible on realistic time scale
without other regions of the glass rearranging.  Since the {\em
typical} rearrangements near $T_g$ span about a length $\xi$ across,
we may make the following, preliminarily conclusion: The
non-equilibrium character of the glass transition necessarily dictates
the existence of intrinsic additional non-elastic degrees of freedom
in a glass, tentatively one per region of roughly size $\xi$, in
addition to the usual vibrations of a stable lattice.  The
universality of $\xi$, in a sense, is the main clue to the cryogenic
universality that is observed.  A schematic of a cooperative region is
shown in Fig.\ref{coop_reg}.

Note that showing the existence of low energy tunneling paths is
really a mathematically problem of finding hyper-lines, connecting two
points of particular latitude on a high-dimensional surface, that
meander within a certain latitude range. Visualizing high-dimensional
surfaces is prohibitively difficult, while the field of topology, at
its present stage of development, is of little help. Yet, a completely
general argument is not required here: We only need to consider a very
small subset of all surfaces, such that they satisfy the (very severe)
constraint on the liquid density distribution above $T_g$, namely such
that conforms to an ``equilibrated'' liquid at $T_g$.  Because (and
with the help) of this constraint, it is possible to put forth a
formal argument showing that there are indeed enough low energy
structural transitions in a frozen glass: This argument will follow
(albeit in the reverse order, in a sense) the argument from the
preceding Section, where we found the {\em typical} trajectory for
rearrangement. The key point of the microcanonical-like library
construction from Section \ref{RFOT} is that the distribution width of
energies of a region increases with region size.  A region is
guaranteed to have a state at some low energy, call it $E_\sGS(N)$, as
found by integration in Eq.(\ref{k}). Past a certain critical size,
this energy decreases as $N$ grows larger, giving rise to the
existence of a resonant state at a large enough $N$. One must bear in
mind, however, that $E_\sGS(N)$ reflects the typical freezing energy.
It really gives an {\em upper bound} on the lowest energy level. The
{\em actual} lowest energy state fluctuates and always lies {\em
below} $E_\sGS(N)$, although most likely not much below. Here we will
look in detail the statistics of these energy states below the typical
reconfiguration profile, with the aim to find the probability of a low
energy trajectory for reconfiguring a region size $N$.

We will make several preliminary, quite general notions that will
guide us in constructing an adequate approximation for the local
statistics of the energy landscape of a frozen lattice.  First, we
give a general argument of the density of frozen-in excitations,
valid, as we will see shortly, in the limit of infinitely slow aging:
Since the atomic arrangement does not change upon freezing, the {\em
classical} density of states of a frozen glass is that of the
supercooled liquid at $T_g$. Those states correspond to configurations
in which the system could have frozen at $T_g$ and in principle can
explore, provided they are thermally accessible and have a
sufficiently low barrier separating them from a given
configuration. Take a generic liquid state at $T_g$ as the reference
state. Then the Boltzmann probability to switch to a conformation
higher in energy by amount $\epsilon$ is $\propto
e^{-\epsilon/T_g}$. That a configuration with that energy was one of
the allowed configurations upon freezing means there must have been
$n(\epsilon) = \frac{1}{T_g} e^{\epsilon/T_g}$ of them. The factor
$1/T_g$ arises because the energy spectrum by construction is
continuous, while the actual {\em local} spectrum is discrete and
$\epsilon = 0$ gives the upper bound on the location of the actual
ground state. The latter must be somewhere between $0$ and $-\infty$:
$\int_{-\infty}^0 d\epsilon \, n(\epsilon) = 1$. This argument,
however, is silent as to what the {\em spatial} characteristics of
such excitations or their time scale are.

This inverse ``Boltzmann'' density of states has been computed
explicitly in frustrated mean-field spin systems \cite{MPV}, but is of
more general nature. Indeed, such distributions arise {\em
universally} when describing the statistics of the lowest energy state
of a wide class of energy distributions \cite{BouchaudMezard},
including the random energy model \cite{Derrida}, that will be used
later on.  Kinetic considerations did not explicitly enter our
heuristic derivation above (or, the mean-field estimates in
\cite{MPV,BouchaudMezard}). This is directly seen by differentiating
$\prtl \log n(\epsilon)/ \prtl \epsilon = 1/T_g$. Clearly,
$n(\epsilon)$ is the microcanonical density of states corresponding to
the translational (liquid-like) degrees of freedom, and the system is
assumed to be completely ergodic within that set of states. This
corresponds to an approximation where we consider all degrees of
freedom which are faster than a given time scale as {\em very} fast,
and everything slower than that chosen time scale is regarded to be
{\em much} slower than can be detected in the experiment.  By using
this same density $n(\epsilon)$, as it was at $T_g$, also at $T <
T_g$, we formally express the fact that this subset of the total
density of states no longer thermally equilibrates but stays put where
it was at $T_g$ - the subsystem of the translational degrees of
freedom has undergone an entropy crisis, a glass
transition. Everywhere in the discussion above, we have been ignoring
the contribution of the purely vibrational excitations to the total
free energy. We thus assume that the spectrum of those elastic
excitations is independent of precisely where on the glassy landscape
the liquid is.

We now give the argument, first laid out in \cite{LW}, that allows one
to estimate the classical density of states and will also
simultaneously yield the size of the region where the excitation takes
place. First we address the question of how many structural states are
available to a compact fragment of lattice of size $N$, regardless of
the barrier that separates those alternative states from the initial
ones. This corresponds to the assumption of time scale separation
mentioned just above. Within this assumption, the low energy limit of
the spectrum must obey $e^{E/T_g}$ so as to give a glass transition at
$T_g$. Next, the spectrum, when integrated, must give $e^{s_c N}$ for
the total number of states available to the region. Notice further
that we expect the reconfiguring regions to be relatively small.  The
atomic motions within these small regions are directly coupled and so
a mean-field, gaussian density of states, that only describes lowest
order fluctuations around the mean, should be accurate enough.  An
energy density satisfying the requirements above actually corresponds
to the well known Random Energy Model (REM) \cite{Derrida}, which also
describes the pure state free energy in mean field frustrated spin
models:
\begin{equation}
\Omega_N(E) \sim \exp \left\{s_c N - \frac{[E - (N \Delta \epsilon +
\gamma \sqrt{N})]^2}{2 \delta E^2 N}\right\},
\label{O_E}
\end{equation}
where $\delta E^2$ is the variance to be determined shortly.  Here,
the factor $e^{s_c N}$ gives the correct total number of states, the
term $\gamma \sqrt{N}$ takes into account the interface energy cost of
considering distinct atomic arrangements with the region.  Note the
fluctuations in the surface term are expected but are automatically
included in the fluctuation of the microcanonical energy $E$
itself.The term $N \Delta \epsilon$ is a bulk energy necessary to
account for the observed excess energy of the frozen structure
relative to the energy of the ideal structure at $T_K$. It is easy to
relate to measured quantities: To do this, recall that the system
freezes in its ``ground state'', with energy $E_\sGS$, when its
entropy becomes non-extensive:
\begin{equation}
\Omega_N(E_\sGS) = 1.
\label{O_N_1}
\end{equation}
We take the energy $E_g$ of the liquid state at $T_g$ as the {\em
reference} energy.  Next note that in the {\em absence} of the surface
energy term $\gamma \sqrt{N}$, the lowest available energy state is
that of the liquid at $T_K$: $(E_K - E_g)/N = - \int_{T_K}^{T_g} dT
\Delta c_p(T) \simeq - \Delta c_p (T_g - T_K) \simeq - T_g s_c$. (The
two latter equalities are accurate for $T_g$ close to $T_K$. The
corrections would be observable \cite{LW_aging,LW_soft}, but small.)
One immediately gets from Eqs.(\ref{O_E}) and (\ref{O_N_1}) that
$\Delta \epsilon = \sqrt{2\delta E^2 s_c} - T_g s_c$ ($\gamma=0$ must
be used in this estimate, but nowhere else!).  Further, using the
microcanonical $\prtl \ln \Omega_N(E)/\prtl E|_{E=E_\sGS}=1/T_g$ fixes
the value of the variance $\delta E^2 = 2 T_g^2 s_c$.  The resultant
density of states is proportional to $e^{(E-E_\sGS)/T_g}$ at $T_g$, as
already shown above by a general argument. Now that we have determined
the thermodynamical quantities entering Eq.(\ref{O_E}), we can find
how the excess energy of an alternate ground state depends on the size
$N$:
\begin{equation}
E_\sGS(N) = \gamma \sqrt{N} - T_g s_c N,
\label{EGS_N}
\end{equation}  
where $E_\sGS$ is defined by Eq.(\ref{O_N_1}).  Only low energy
excitations will be thermally active at the lowest
temperatures. Therefore, we are looking for excitations that are
nearly isoenergetic with the reference state. This imposes an
additional condition $E_\sGS (N) = 0$ thus prescribing the minimal
size $N_0$ of a region such that has $\Omega_N(0) \ge 1$ for $N \ge
N_0$. A region of this size has at least one alternative structural
state at the same energy.  One obtains from Eq.(\ref{EGS_N}) that $N_0
= (\gamma/T_g s_c)^2 = N^*$, consistent with our previous argument
that any region of size $N^*$ has a spectral density of states equal
to $\frac{1}{T_g} e^{E/T_g}$.  Note that Eq.(\ref{EGS_N}) echoes the
free energy profile of droplet growth from Eq.(\ref{F_N}), but unlike
Eq.(\ref{F_N}), it can be used for $N > N^*$ as well. Eq.(\ref{EGS_N})
explicitly shows that a droplet of size larger than $N^*$ has an
exponentially increasing number of available configurations
corresponding to lattices typical of $T_g$, consistent with the
instability of droplets larger than $N^*$ at temeperatures above $T_g$
mentioned earlier.

The microcanonical argument above is basically a gedanken experiment
in which we had the demon-like ability to browse through all possible
atomic arrangements, given the total number of allowed states equal to
$e^{N s_c}$.  The total sample is thus comprised of regions of the
type considered in the argument (the interface energy has been taken
into account by the term $\gamma \sqrt{N}$, this energy may be viewed
as the penalty for considering the states of a given compact region as
if this region were totally independent from the rest of the sample;
c.f. our earlier comments on the locality of the liquid's energy
landscape). Therefore, if the rate of conversion between the
alternative glassy states can be ignored, the argument immediately
yields the density of residual excitations in a frozen glass: $\simeq
\frac{1}{N^* a^3 T_g} e^{\epsilon/T_g}$ ($N^* a^3 \equiv \xi^3$, of
course).  However, even though each of these imaginary regions has an
alternative resonant state, there is so far only an undetermined
chance to reach it within any particular time.  In fact, the typical
{\em classical} barrier for the excitations available to the regions
of size $\le N^* \simeq 190$ is $F^\ddagger \simeq 39 k_B T_g$. Such a
barrier would seem to exclude the possibility for tunneling for a
typical domain of size $N^*$. But to account for kinetics issues, we
should repeat the argument for the critical size, but also
simultaneously include the life-time of each considered configuration
as a selection criterion. In other words, one should compute the {\em
combined} distribution of the excitation energies, their spatial
extent and the corresponding tunneling amplitudes. Later in the paper,
we will discuss one source of correlation between the excitation
energy and the tunneling amplitude owing to level repulsion effects.
Nonetheless here, we present a simpler argument, given in \cite{LW},
in which the tunneling rate distribution is assumed to be independent
from that of $\epsilon$. Simpler yet, we will look for the density of
regions that allow for a rearrangement with a {\em zero}-height
barrier. As vindicated {\it post factum}, all of these simplifications
can lead only to at most a 10\% error in the resulting density of
states.
 
\begin{figure}[htb]
\includegraphics[width=.8\columnwidth]{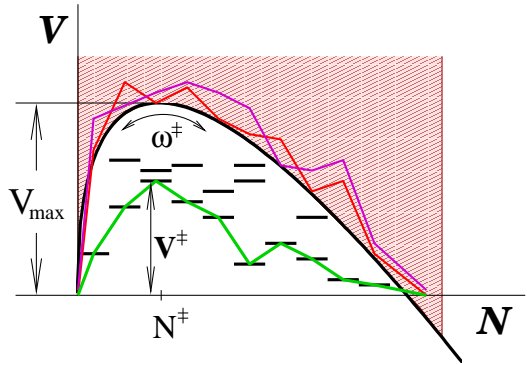}
\caption{\label{V_N} The black solid line shows the barrier along the most
probable path. Thick horizontal lines at low energies and the shaded
area at energies above the threshold represent energy levels available
at size $N$.  The red and purple line demonstrate generic paths, green
line shows the actual (lowest barrier) path, which whould be followed
if $\hbar \omega^\ddagger < k_B T/2\pi$.}
\end{figure}

Imagine the process of conversion to another state as a step-wise
process where the ``nucleus'' of this new state is increased by adding
one atom at a time, as signified by the horisontal axis in
Fig.\ref{V_N}.
Such addition involves moving the atom a distance of the order of the
Lindemann distance $d_L$. It follows then that the path connecting the
two states is likely to encounter a high barrier of the order
$F^\ddagger$, which effectively disconnects those two states. However,
the possible configurations through which one can pass and therefore
the barrier heights are {\em distributed} and there is a chance even
for a region of size $N^*$ to have an arbitrarily low barrier. What
would such a distribution be? We first have to decide whether the
tunneling probability is a sum of contributions of many (interfering)
paths or, whether it is dominated by a single path, which has the
lowest barrier. The first scenario would be realized in a highly
quantum glass, where Debye temperature rivals or exceeds the glass
transition temperatures \cite{SchmalianWolynes}. Such a highly quantum
glass could in fact melt due to quantum fluctuations. In our case,
freezing is a completely classical process, which is signified by the
fact that the barriers are proportional to a classical energy scale
$T_g$. We now assume more specifically that the contribution of a
tunneling path is proportional to $e^{-\pi V^\ddagger/\hbar
\omega^\ddagger}$, where $\omega^\ddagger$ is a quantum frequency
scale, a multiple of $\omega_D$, and barrier $V^\ddagger$ scales with
$T_g$, as mentioned earlier. This would be an accurate assignment in
the case of a parabolic barrier. The form of the tunneling amplitude
$e^{-\pi V^\ddagger/\hbar \omega^\ddagger}$ conforms to our
expectation that the tunneling trajectory is dominated by a single
path with the lowest barrier, as $V^\ddagger$ and $\hbar
\omega^\ddagger$ are taken from distributions characterized by scales
$k_B T_g$ and \verb+const+$\times \hbar \omega_D$ respectively, the
former one generally much larger than the latter (the \verb+const+
self-consistently will turn out to be less than one in subsection
\ref{multilevel}). Since the energy profile along the tunneling
trajectory has a complicated shape formed by {\em many} intermediate
states separated by {\em small intermediate} barriers (see below), it
is fair to say that the state of the system at the highest barrier
corresponds to the highest energy intermediate state (the
``transition'' state).  The statistics of energy states have already
been found earlier. We therefore use distribution (\ref{O_E}) with
only one difference: we must double the variance because the barrier
height is actually the difference between two fluctuating quantities:
the energy of the final (or initial) energy and the highest energy
along the path. As a result,
\begin{equation}
\Omega_N(V) \sim \exp \left\{s_c N - 
\frac{[V - (T_g s_c N + \gamma \sqrt{N})]^2}{4 \delta E^2 N}\right\}.
\label{O_V}
\end{equation}
Distribution (\ref{O_V}) thus gives the typical value of the barrier
for the (quantum) growth of a droplet. It is easy to see from
(\ref{O_V}) that the highest barrier corresponds to rearranging a
region of size $N^\ddagger \simeq$ 14 and is equal to
\begin{equation}
V_{max} = F^\ddagger/(2\sqrt{2}-1) \simeq 26 T_g s_c.
\end{equation}
Since this is the hardest place to get through, we must take it as the
transition state. Hence, the final distribution of (transition state)
barriers is the density of pure states corresponding to Eq.(\ref{O_V})
with $N=N^\ddagger$ (similar to the $e^{\epsilon/T_g}$, obtained
above).  Thus,
\begin{equation}
\Omega(V^\ddagger) \sim
\exp\left\{\frac{V^\ddagger-V_{max}}{\sqrt{2} T_g} \right\}
= \exp\left\{-18 \cdot s_c + \frac{V^\ddagger}{\sqrt{2} T_g} \right\}.
\label{O_V1}
\end{equation}
As one can see, the probability to have a small barrier path is
exponentially suppressed. Nevertheless, owing to the large value of
the energy parameter in distribution (\ref{O_V}) the fraction of zero
barrier paths per mosaic cell $\sim e^{-18 s_c} \simeq 3 \cdot
10^{-7}$ is actually not prohibitively small. A region larger by only
18 molecules (less than a single layer) will have $e^{18 s_c}$ more
final states (and therefore paths) to go to. We therefore conclude,
any region of size $\simeq 200$ molecules will have an accessible
alternative state with spectral density $1/T_g$. Finally, we stress a
remarkable feature of the tunneling paths statistics in glass. Mark
the very rapid - exponential - scaling of the number of paths leading
out of a particular local structural state on the size of the
respective region. This means that the final estimate of the density
of structural transitions that have low enough barriers to be
thermally relevant is rather insensitive to the details of correlation
between the energy of the transition and its tunneling
amplitude. Consequently, even a very simple estimate of this density,
such as the one above, is very robust.  Finally, note that the
tunneling argument above is, again, a microcanonical argument, such as
the one leading to Eq.(\ref{O_E}), that also takes into account (in a
rather crude manner) the mutual accessibility between alternative
energy states.

As we will see later, the tunneling barriers, and hence the relaxation
times of the tunneling centers, are distributed. This would lead to a
time dependent heat capacity. Ignoring this complication for now, the
{\em classical}, long time heat capacity is easy to estimate already
(assuming it exists): Since our degrees of freedom span a volume
$\xi^3$ and their spectral density is $1/T_g$ at low energy, one
obtains for the low $T$ heat capacity per unit volume: $T/T_g \xi^3$,
up to an insignificant coefficient. The coefficient at the linear heat
capacity dependence is often denoted $\bar{P}$.  For silica, $T_g
=$1500 K and realistic $\xi = 20 \AA$ yields $\bar{P} \simeq
6\cdot10^{45}$m$^{-3}$J$^{-1}$ in agreement with the experiment (we
took $a = 3.5 \AA$ - a length scale appropriate for a tetrahedron
formed by four oxygens around a silicon atom.  These tetrahedra appear
to be moving units in a-Silica \cite{Trachenko}). The assumption of
the existence of the long-time heat capacity is empirically justified
(within logarithmic accuracy), but is also consistent with the present
theory, see Subsection \ref{Barr_Dist} and Section
\ref{LevelRepulsion}.

In conclusion, the main result of this Subsection is that the
non-equilibrium nature of the glass transition results in the
existence of residual motional degrees of freedom, a significant
fraction of which remain thermally active down to the lowest
temperatures. These degrees of freedom are collective highly
anharmonic atomic motions within compact regions of size $\xi^3$,
determined mainly by the length scale of the entropic droplet mosaic
determined at $T_g$. The energy scale in the spectrum of these
excitations is set by the glass temperature $T_g$ itself. We now turn
to the question of what determines the {\em strength} with which these
entropic droplet excitations couple to the phonons. This will explain
the universality in the heat conductivity at temperatures below
$\sim$1 K.

\subsection{The Universality of Phonon Scattering}
\label{universality_sec}

First of all, what do we mean by ``phonons'' in amorphous materials?
There is no periodicity, therefore one can only strictly speak of
elastic {\em strain}, even if the structure is completely stable.  In
the latter case, low gradient strains $\nabla \phi$ are still
described by a simple bi-linear form:
\begin{equation}
H\sph \equiv \int d^3 {\bf r} \frac{\rho c_s^2}{2} (\nabla \phi)^2.
\label{H_ph}
\end{equation} 
The $\nabla \phi$ field is defined on a isotropic, translationally
invariant flat metric, as in continuum mechanics \cite{LLelast}, and
so a wave-vector $k$ is an operational concept. It is easily seen, by
dimensional analysis, that strains arising specifically due to
disorder will be of higher order in $k$ than the term in
Eq.(\ref{H_ph}): The corresponding energy terms should scale with some
positive power of the lattice inhomogeneity lengthscale(s),
$l_\sinhom^{\zeta}$ ($\zeta > 0$), so as to vanish upon ``zooming
out''. The terms will subsequently go as $k^2 (l_\sinhom k)^\zeta$.
But, as we have already seen, the amorphous lattice is not stable,
that is there are anharmonic transitions with arbitrarily small energy
and barrier. Still, the regions encompassing the transitions are quite
small, at most 6 lattice spacings across, which is much less than the
thermal de Broglie wave-length at 1 K (about $10^3 a$ in
a-silica). The unstable regions interact with the strains of the
otherwise stable lattice. We conceptualize this interaction by
approximating the strain with pure phonons and computing the phonon
mean-free path. The latter will turn out to be about 150 times longer
than the phonon wave-length, so that the phonon approximation is
internally consistent in that the phonons are indeed reasonably good
quasi-particles, at the plateau temperatures and below. Finally, note
in Eq.(\ref{H_ph}), we have used only one phonon polarization for
simplicity (it will be usually obvious how to account approximately
for all three acoustic phononic branches at the end of a calculation;
using this ``scalar'' version of the lattice dynamics, for the
purposes of this paper, boils down to neglecting the difference
between the longitudinal and transverse sound, except in the later
discussion of the Gr\"{u}neisen parameter).

The structural transitions interact with the phonons because the
energy of the transition changes in the presence of a strain: To see
this exlicitly, consider the elastic energy within a droplet-sized
region capable of undergoing a {\em low} energy transition, as
relevant at low $T$. For the sake of argument, assume there is no
sheer deformation; a similar argument applies to the transverse phonon
branches. The stress energy is then $\int_{V_\xi} d^3 {\bf r} K
u_{ii}^2/2$, where $K$ is a compressibility constant on the order of
$\rho c_s^2$, which we are allowed to treat as a constant, with the
error contributing in a higher order in $k$, as already mentioned.
$u_{ii}$ is the trace of the elastic tensor \cite{LLelast}, which has
the same meaning as $\Delta \phi$. Since the transition energy is low,
the lowest order, quadratic expansion suffices. This implies that all
individual bonds within the region distort by a very small amount
(already shown not to exceed $d_L$, even at the transition's
bottle-neck). We have demonstrated that such regions do indeed exist
and found their density in the previous section.  We separate the
total elastic tensor $u_{ij}$ into a contribution $\phi_{ij}$ due to
the elastic stress and $d_{ij}$ due to the tunneling displacement. The
$d_{ii} \phi_{ii}$ cross term represents the coupling between the
transition and the strain. If the phonon wave-length is larger than
$\xi$, $\phi_{ii}$ is constant within the integration boundaries and
can be taken out of the integral.  One consequently arrives at the
following {\em energy difference} for the defect states in the
presence of a phonon: $\rho c_s^2 \phi_{ii} \int_{V_\xi} d^3 {\bf r}
\, d_{ii}$. Here, $d_{ii}$ corresponds to the transition induced
displacement between two given structural states. Each tunneling
center is a multilevel system. However, since we are presently
interested in the coupling of the lowest energy transition to the
strain, we consider here the set of $d_{ii}$ corresponding to the {\em
two lowest} energy states of the region. (Higher energy transitions
turn out to be intimately related to the lowest energy transition, and
are discussed later, in Section \ref{plateau_chapter}.)  We therefore
conclude that a tunnelling transition, active at low temperatures, is
linearly coupled to a lattice strain with the strength defined as $g =
\rho c_s^2 \int_{V_\xi} d^3 {\bf r} \, d_{ii}/2$; the corresponding
term in the Hamiltonian reads:
\begin{equation}
H_{int} \equiv {\bf g} \nabla \phi \sigma_z.
\label{H_int}
\end{equation}

We present next two independent ways to estimate the coupling constant
$g$. The first one is based on the realization that at the glass
transition, purely phononic excitations and a frozen-in structural
transition must coexist, that is they are are of marginal stability
with respect to each other. On the one hand, a local region posed to
harbor a structural transition below $T_g$, must not be ``crumpled''
by a passing phonon. On the other hand the energy of the transition
can only be so high, as to be sustainable by the lattice stiffness. In
other words, an atom will be part a of frozen-in transition, if that
atom is roughly in equilibrium between the transition driving forces
and the ambient lattice strain. This stability condition gives at the
molecular scale $a$, by Eqs.(\ref{H_ph}) and (\ref{H_int}): ${\bf g}
\sigma_z = - \rho c_s^2 a^3 \nabla \phi$. The lattice strain will be
distributed throughout the material in the usual manner, subject, of
course, to the equipartition that fixes the variance of the strain so
as to conform to the thermally availabe energy. We take advantage of
this by multiplying the equilibrium condition by $\nabla \phi$ and
noting that thermal averaging is also ensemble averaging. This yields
that for an atom posed to be part of low energy structural transitions
below $T_g$, it is generically true that
\begin{equation}
|\la {\bf g \nabla} \! \phi \: \sigma_z \ra| \simeq \rho c_s^2 \la
({\bf \nabla} \! \phi)^2 \ra a^3 \simeq k_B T_g.
\end{equation} 
Noting that $\la {\bf g \nabla} \! \phi \: \sigma_z \ra \simeq \la
|{\bf g \nabla} \phi| \ra$, one arrives at a simple relation:
\begin{equation}
g = \sqrt{\rho c_s^2 a^3 k_B T_g},
\label{g_est_orgnl}
\end{equation}
which is the main result of this Subsection. It is understood that
this estimate is accurate up to a number of order one, and $g$'s are
likely distributed. At any rate, we observe that the TLS-phonon
coupling is the geometric average between the glass transition
temperature and an energy parameter $\rho a^3 c_s^2$ ($\sim$ 10$^1$eV)
related to the cohesion energy of the lattice (note the quantity $\rho
c_s^2$ is a multiple of the Young's modulus). We point out the
estimate above applies quantitatively to low barrier transitions
only. The mechanical stability criterion is a zero frequency, static
condition. However, it takes a finite time for a structural transition
to respond to an external stress. In other words, a region harboring a
{\em slow} transition will likely appear perfectly elastic to a high
frequency phonon. We thus arrive at the conclusion that TLS-phonon
coupling must be frequency dependent, however deviations from the
result obtained above will enter in a higher order in $\omega, k$.

Alternatively, one may attempt to estimate the integral over the
derivative of the displacement field that entered in the expression
for the coupling constant $g = \rho c_s^2 \int_{V_\xi} d^3 {\bf r} \,
d_{ii}/2$.  Since $d_{ii}$ is the divergence of a vector, the integral
is reduced to that over a surface within the droplet's boundary:
$\int_{S_\xi} d{\bf s} \, {\bf d}({\bf r})$, where ${\bf d}({\bf r})$
is the tunneling displacement itself, near the boundary.  How near?
The Gauss theorem applies so long as the field ${\bf d}$ is
continuous. This field is roughly $d_L$ in magnitude close to the
droplet's center and is zero outside of the region. A function defined
on a discrete lattice is expressly discontinuous. Requiring that one
be able to cast a continuous tunneling displacement field on a
discrete manifold, so that the field interpolates smoothly between
$d_L$ in the middle and zero outside, imposes constraints on the ${\bf
d}$ values at the droplet's boundary. We argue this value should
generically go as $(a/\xi) d_L$, up to a constant, in order to realize
the interpolation and spread evenly the tensile field of the domain
wall throughout the droplet. In other words, $\sim (a/\xi) d_L$ is
quite obviously the lower bound, while higher values statistically
imply a higher stress, and higher transition energy.  Now, since ${\bf
d}(r)$ is randomly oriented, the integral over the droplet's border is
of the order $a^2 \sqrt{{N^*}^{2/3}} (a/\xi) d_L$, where ${N^*}^{2/3}$
is the number of molecules at the boundary.  The Lindemann distance at
$T_g$ is equal to the magnitude of a thermal fluctuation, hence $d_L/a
\simeq |\nabla \phi|$
\footnote{For reference, $|{\bf \nabla} \! \phi| \sim (T_g/\rho c_s^2
a^3 )^{1/2}$ at $T_g$ is about 0.05 for SiO$_2$, 0.06 for B$_2$O$_3$
(oxide glasses), 0.03 for PS and PMMA (polymer glasses), in agreement
with the Lindemann criterion. We stress the sensitivity to the value
of the molecular size $a$, which is somewhat arbitrary. Here, we have
not calculated the bead size based on chemistry, but instead used the
values of the speed of sound, as employed in the scaling procedure of
Freeman and Anderson \cite{FreemanAnderson}. According to the
definition of the Debye frequency, $a =
(c_s/\omega_D)(6\pi^2)^{1/3}$.}.  As a result, the coupling to the
extended defects is still about $g \simeq \sqrt{\rho c_s^2 a^3 T_g}$,
again within a factor of two or so.

Note that, when considering a particular value of lattice distortion
$\nabla \! \phi$ in the discussion above, we did not specify the
wave-length(s) of the phonons that contributed to this distortion,
therefore the estimate of $g$ in Eq.(\ref{g_est_orgnl}) is correct as
long as the form of the interaction term (Eq.(\ref{H_int})) is
adequate.  This surely holds for long-wave phonons relevant at the TLS
temperatures.

With the knowledge of $g$, we can estimate the inverse mean free path
of a phonon with frequency $\omega$.  As done originally within the
TLS model, the quantum dynamics of the two lowest energies of each
tunneling center are described by the Hamiltonian $H_{TLS} = \epsilon
\sigma_z/2 + \Delta \sigma_x/2$. This expression, together with
Eqs.(\ref{H_ph}) and (\ref{H_int}), form a complete (approximate)
Hamiltonian of the TLS plus the lattice vibrations. The phonon inverse
mean free path is then calculated in a standard fashion
\cite{Jackle,LowTProp}:
\begin{equation}
l_\smfp^{-1}(\omega) = \pi \frac{\bar{P} g^2}{\rho c_s^3} \omega
\tanh\left(\frac{\hbar \omega}{2 k_B T}\right).
\label{l_l1}
\end{equation}
This  yields 
\begin{equation}
\lambda_{dB}/l_\smfp = \frac{2\pi^2}{3} \tanh(1/2) 
\left( \frac{a}{\xi} \right)^3,
\label{l_l2}
\end{equation}
where factor $1/3$ comes from the averaging with respect to different
orientations of the defects and we used $\bar{P} \simeq 1/T_g
\xi^3$. It follows that $l_\smfp/\lambda_{dB} \sim (\xi/a)^3 \simeq
200$ up to a constant of order one. Hence, the analysis above explains
the universality of the combination of parameters $\bar{P} g^2/\rho
c^2_s$, and relates it to the geometrical factor $(a/\xi)^3 \simeq
10^{-2}$, which is the relative concentration of cooperative regions
in a supercooled liquid, an almost universal number within the random
first order glass transition theory \cite{XW}, depending only
logarithmically on the speed of quenching. Strictly speaking, our
argument predicts the universality in $l_\smfp/\lambda$ only within
10\%-20\% or so. This is a consequence of indeterminacy of $\epsilon$
vs. $\Delta$ correlation that may be system specific, or could be due
to deviations of the $\xi/a$ ratio from the universal 5.8 at $T_g$.
Since the latter ratio depends on the glass preparation time, the
corresponding experimental study seems worthwhile.

Numerically, Eq.(\ref{l_l2}) yields $l_\smfp/\lambda \simeq 70$, a
factor of $2$ less than the empirical $150$ \cite{FreemanAnderson}.
We could not have expected much better accuracy from our estimates,
that used no adjustable parameters.  Although it may seem that we have
slightly over-estimated the number of scatterers, the size of the
error is too small to reliably support this suggestion. However, this
is a good place to make a few comments on the distribution of the
tunneling matrix elements $\Delta$, which will also prove useful for
the discussion of the phonon scattering at higher frequencies in
Chapter \ref{plateau_chapter}. The estimate for the phonon mean free
path in Eq.(\ref{l_l1}) is not terribly sensitive to the form of
tunneling amplitude distribution \cite{LowTProp} (within reasonable
limits).  This is because the contribution of an individual TLS to the
total scattering cross-section is proportional to $\Delta^2/E^2$,
where $E \equiv \sqrt{\Delta^2 + \epsilon^2}$.  Two common
distributions have been used in the literature.  One distribution
simply assumed $(\Delta^2/E^2) \sim 1$ in the absence of a more
specific knowledge and a flat distribution of the total energy
splitting $E$ (this is actually the original TLS model). The earlier
Standard Tunneling Model (STM) \cite{AHV,Phillips}, on the other hand,
postulates $P(\Delta) \propto \frac{1}{\Delta}$ (approximately
supported by our own conclusions too), which predicts a nearly flat
$E$ distribution as well. In the end, both models differ only in that
the TLS model has to postulate the average $(\Delta^2/E^2)$ value when
calculating the scattering cross-section (this number is absorbed into
the TLS-phonon coupling constant $g$).  On the other hand, the STM
allows for somewhat more closed-form derivations, however it still has
to introduce the cut-off values $\Delta_{min}$ and $\Delta_{max}$ as
parameters (fortunately, many measured quantities depend on these
parameters only logarithmically). (The distinction between the two
models is described in detail in the front article by Phillips in
\cite{LowTProp}.) One point in favor of the STM is that it necessarily
predicts time dependence of the specific heat. While a time dependence
has been observed, its specific functional form has not been
unambiguously established in the experiment (see
\cite{Pohl,HunklingerRaychaudhuri}). We also mention, for
completeness, there is a different way to parametrize the two-level
system motions within the more general, soft-potential model
\cite{soft_pot,soft_potBuchenau}. At any rate, we see that while a
two-level system's contribution to the total phonon scattering depends
on the value of its $\Delta \sim E$ ratio, the precise form of the
$\Delta$ distribution will change the answer quantitatively, but not
qualitatively.  We note, that in the context of the present
calculation it is preferable to consider the simpler, TLS setup that
does not specify the $\Delta$ distribution, because our argument has
so far been only {\em semi}-classical.  Indeed, so far the tunneling
amplitudes have only interested us from the perspective of the volume
{\em density} of {\em allowed} transitions. We saw that an
indeterminacy of the density could not exceed a 10\% or so due to a
weak (logarithmic) dependence of that density on a specific $\Delta$
distribution. Therefore, we are more confident in the numerical
estimates using the TLS-model setup that does not require introducing
additional parameters (such as $\Delta_{min}$) explicitly.
Nevertheless, the special role of $\Delta \sim E$ defects in
scattering is worth noting. These defects have low classical energy
splittings $\epsilon < \Delta$ and their dynamics are mostly
determined by the {\em quantum} energy scale. These are the ``fast'',
or ``zero-barrier'' excitations discussed earlier in the literature
\cite{BlackHalperin,Geszti}, whose tunneling matrix element probably
can not be directly estimated by WKB, but we can still guess that it
scales with $\omega_D$. This suggests that using the same distribution
function $P(\Delta)$ for {\em all} thermally defects may not be
justified, as circumstantially supported by results of Black and
Halperin \cite{BlackHalperin} who found that the density of TLS
derived from the heat capacity and conductivity measurements
respectively are not exactly equal to each other.  While this
indeterminacy in the exact barrier distribution introduces only an
error of order one in quantitative estimates \cite{BlackHalperin} of
the density of states and is not of special concern here, we note that
the present theory, upon inclusion of the effects beyond the strict
semi-classical picture, {\em does} in fact provide a mechanism for the
excess of the ``fast'' two-level systems, as will be explained in
Section \ref{LevelRepulsion}.

{\em Strong Interaction Scenarios.} By deriving the density of states
of structural transitions, and their coupling to the phonons based on
the known properties of the amorphous lattice, we have constructively
established the microscopics of glassy excitations in excess to the
purely elastic excitations. It follows from the discussion that {\em
no other} excitations are present in glasses at 1 K and below (see
also the discussion on the exhaustive classification of excitations in
amorphous lattices in Subsection \ref{classific}).  Importantly, the
density of states (DOS) in excess of the phonons, is due to {\em
local} motions.  This is in contrast with Strong Interaction Scenarios
(SIS) \cite{YuLeggett, Leggett, BurinKagan, Coppersmith} that posit
that {\em any} local excitations (other than pure strain) would give
rise to a ``universal'' density of states. Such universal density of
states arises in SIS as a consequence of long range, $1/r^3$,
interaction mediated by the phonons, so that the actual observed DOS
is a highly renormalized entity. The corresponding excitations are
expressly non-local, possibly infinite in extent. The idea is very
attractive because of its generality but remains a pure abstraction,
until those bare excitations are constructively shown to {\em exist}
in the first place. Additionally, even upon assuming some bare
excitations are present, demonstrating the quantitative relationship
between the effective density of states and the phonon coupling that
conforms to the experimental $\bar{P} g^2/\rho c_s^2 \sim 10^{-2}$ has
so far proven elusive \cite{Leggett_private,Carruzzo,LWunp}. On the
other hand, we have shown that local structural transitions, that
interact with phonons with a particular strength, must indeed take
place in amorphous solids. In order to determine where the current
theory stands in relation to the SIS, one may inquire whether the
phonon-mediated interaction leads to the emergence of some collective
density of states. It should be immediately clear that no such
additional, collective DOS appears at $T_g$, because the argument in
the previous subsection has already included {\em all} the effects of
the surrounding of a structural transition, which simply amounted to
the thermal noise at $T_g$ delivered to the transition by the elastic
waves. It, of course, does not matter what the phonon source is.  What
happens at low temperatures should be considered separately. The
effects of interaction turn out to be small in the TLS regime and are
discussed in detail in the final Section of this paper. Here, we give
several qualitative estimates for the sake of completeness, both at
high and low temperatures. The phonon-mediated interaction goes
roughly as $\frac{g^2}{\rho s_c^2} \frac{1}{r^3}$, with a numerical
factor less than one (see Section \ref{Gruneisen}). Ignoring the
factor, the interaction is expressed, with the help of
Eq.(\ref{g_est_orgnl}), via the glass transition temperature according
to $k_B T_g \frac{a^3}{r^3}$. Two neighbouring domains, a distance
$\xi$ apart, would thus couple with strength $J_\sneigh = T_g
(a/\xi)^3$. In order to assess the effects of interaction on the
effective energies of individual transitions, or whether it even makes
sense to talk about on-site energies after the interaction is turned
on, one must compare the interaction strength to the width of the
distribution of the on-site energies as derived in the absence of
interaction, exactly the same way it is done in the context of
Anderson localization. According to the previous Subsection the latter
width, call it $\Delta E$, is of the order $T_g$. The ratio
$J_\sneigh/\Delta E \simeq (a/\xi)^3$ is a small number, as
expected. There will be no long range effects, due to resonant
interactions, at high temperatures near $T_g$. At very low
temperatures, only tunneling centers (TC) with energy splitting $\sim
k_B T$ or less are thermally active. While the relevant spread of the
on-site energies $E_T \sim k_B T$ is now down by a factor $T/T_g$
compared to the glass transition temperature, the {\em concentration}
of active TC is also down by the same factor, namely $(T/T_g \xi^3)$,
thus increasing the mutual separation between the regions of mobile
transitions. The total dipole-dipole induced static field due to all
those thermally active two-level systems at a given spot is simply
$\frac{g^2}{\rho c_s^2} (T/T_g \xi^3) \simeq k_B T (a/\xi)^3$, again
much smaller than the relevant on-site energy range $E_T \sim k_B
T$. The motions within the tunneling centers are quantum-mechanical at
these low temperatures, and so one may consider possible effects of
{\em resonant} interactions between distinct TC's, as in the
Burin-Kagan scenario \cite{BurinKagan}. These effects have been shown
to become important only at ultra-cold temperatures of $\mu$K and
below \cite{Silbey}, as already mentioned in the introduction.

Besides having explained the origin of the universality of combination
$\bar{P}\frac{g^2}{\rho c_s^2}$ ubiquitous in the STM, we have also
seen why the value of this parameter is different from $1$, suggested
by the strong defect-defect interaction universality scenario.  This
value can be traced to the relative concentration of the domains
$(a/\xi)^3 < 0.01$, as just mentioned.  It is curious that the
defect-phonon interaction in the long wave-length limit can be
expressed as a surface integral.  Besides supporting our picture of
the residual excitations as motions of domain walls, it points at a
connection with string theories, where the elementary particles
exhibit internal structure at high enough energies, which is also true
in our case. In fact, this internal structure is ultimately the cause
of the phenomena observed in glasses at higher temperatures, namely
the so called bump in the specific heat and the thermal conductivity
plateau, which are dealt with in Chapter \ref{plateau_chapter}.

\subsection{Distribution of Barriers and the Time Dependence of the
Heat Capacity} \label{Barr_Dist}

The STM postulated tunneling matrix element distribution $P(\Delta)
\propto 1/\Delta$ implies a weakly (logarithmically) time dependent
heat capacity. This was pointed out early on by \cite{AHV}, while the
first specific estimate appeared soon afterwards in \cite{Jackle}. The
heat capacity did indeed turn out time-dependent, however its
experimental measures are indirect, and so a detailed comparison with
theory is difficult. Reviews on the subject can be found in
\cite{Nittke,Pohl}. Here, we discuss the $\Delta$ distribution
dictated by the present theory, in the semi-classical limit, and
evaluate the resulting time dependence of the specific heat. While
this limit is adequate at long times, quantum effects are important at
short times (this concerns the heat condictivity as well). The latter
are discussed in Subsection \ref{Mixing}.

In the tunneling argument from Section \ref{TLS}, we have suggested a
WKB type expression for the tunneling amplitude: 
\begin{equation}
\Delta = \Delta_0
e^{-\frac{\pi V^\ddagger}{\hbar \omega^\ddagger}},
\label{D_D0}
\end{equation} 
which would be completely correct in the case of a parabolic barrier
with frequency $\omega^\ddagger$ and height $V^\ddagger$ and was
motivated by the necessity to maintain the proper scaling with $\hbar$
in the denominator of the exponent, given that the typical barrier
height is determined by the {\em classical} landscape characteristics
and should scale with $T_g$.  According to Eq.(\ref{O_V1}),
$P(V^\ddagger) \propto e^{\frac{V^\ddagger}{\sqrt{2} T_g}}$.  It
follows then that
\begin{equation}
P(\Delta) d\Delta = A \left( \frac{\Delta_0}{\Delta} \right)^c 
\frac{d \Delta}{\Delta},
\label{P_D}
\end{equation}
where $c = \hbar \omega^\ddagger/\sqrt{2} T_g$ should be less then
$0.1$ according to our estimates of $\omega^\ddagger$ (see section
\ref{multilevel}). $A$ is a constant, to be commented on very shortly.
The distribution in Eq.(\ref{P_D}) becomes $P_{STM}(\Delta) d\Delta
\propto d\Delta/\Delta$ postulated in the standard tunneling model, if
$c \rightarrow 0$.  As shown next, the non-zero $c$ gives rise to an
anomalous exponent $\alpha = c/2$ in the heat capacity $C \propto
T^{1+\alpha}$ and a power law $t^{c/2}$ for the specific heat time
dependence at long times, as opposed to a logarithmic one, predicted
by the STM. While $c \simeq 0.1$ implies $\alpha \simeq 0.05$,
experimentally, $\alpha$ seems to vary from $0.1$ to $0.5$. This
larger value is consistent with quantum mixing effects that go beyond
the semiclassical analysis as we will discuss later.

Scaling $\hbar \omega^\ddagger$ in the denominator of the tunneling
exponent implies that $\omega^\ddagger$ must be a quantum energy scale
and it is indeed shown in Section \ref{multilevel} that
$\omega^\ddagger$ is proportional to the Debye frequency
$\omega_D$. While the tunneling argument from Section \ref{TLS} only
explicitly considered the statistics of the {\em highest} energy state
along the tunneling trajectory, the expression in Eq.(\ref{D_D0})
actually does not use such a simplified picture but considers a {\em
finite} vicinity of the barrier top. The conclusion of Section
\ref{multilevel} that the barrier heights are distributed
exponentially, such as in in Eq.(\ref{O_V1}), remains true in either
case. The leads to a non-zero value of $c$, and here we explore what
consequences this has for the low temperature heat capacity and
conductivity.  As follows from the discussions in Section \ref{TLS},
constant $A$ in Eq.(\ref{P_D}) is of order one.

Since the temperatures in question here are so low ($1 K$ and below),
we will ignore the energy dependence of $n(\epsilon)$ in this section
and take $n(\epsilon) = \bar{P}$.  In order to see the time dependence
of the heat capacity we obtain the combined distribution of the TLS
energy splittings $E$ and relaxation rates $\tau^{-1}$ -
$P(E,\tau^{-1})$, much as was done in \cite{Jackle}, - and then count
in only those TLS whose relaxation time $\tau$ is shorter than a
particular experimental observation time $t$.

The (phonon irradiation induced) relaxation rate of a TLS is
\cite{Jackle}:
\begin{equation}
\tau^{-1} \simeq \frac{3 g^2 \Delta^2 E}{2 \pi \rho c_s^5} \coth(\beta
E/2).
\label{tau_TLS1}
\end{equation}

It follows from Eqs. (\ref{P_D}) and (\ref{tau_TLS1}) that
\begin{widetext}
\begin{equation}
P(E,\tau^{-1}) = \bar{P} A 
\left(\frac{\tau}{\tau_{min}(E)}\right)^{c/2}
\left(\frac{\Delta_0}{E}\right)^c 
\frac{\tau}{2 \sqrt{1-\tau_{min}(E)/\tau}},
\end{equation}
\end{widetext}
where 
\begin{equation}
\tau_{min}^{-1}(E) \equiv \frac{3 g^2 E^3}{2 \pi \rho c_s^5}
\coth(\beta E/2)
\label{tau_min}
\end{equation}
is the fastest relaxation rate of a TLS with energy splitting $E$,
achieved at $\Delta = E$, of course. As follows from
Eq.(\ref{tau_min}), the rate scales roughly as the cube of temperature
and is empirically of order an inverse millisecond at $10^{-2} K$.
The resultant sample's heat capacity per unit volume is then:
\begin{equation}
C(t) = \int_0^\infty dE \left(\frac{\beta E}{2 \cosh(\beta E/2)}
\right)^2 \int_{t^{-1}}^{\tau_{min}^{-1}} d \tau^{-1} \,
P(E,\tau^{-1}),
\label{hcap1}
\end{equation}
where $[\beta E/2 \cosh(\beta E/2)]^2$ is the TLS heat capacity.
With a change of variables, Eq.(\ref{hcap1}) reads:
\begin{widetext}
\begin{equation}
C(t) = \int_0^\infty dE \left(\frac{\beta E}{2 \cosh(\beta E/2)}
\right)^2 \int_0^{\log(t/\tau_{min}(E))} d z \, \frac{A}{2} \left(
\frac{\Delta_0}{E} \right)^c \, \frac{e^{\frac{c}{2} z}}
{\sqrt{1-e^{-z}}}.
\label{hcap2}
\end{equation}
\end{widetext}

At long times the expression Eq.(\ref{hcap2}) yields a power law for
both time and temperature dependence of the specific heat:
\begin{equation}
\lim_{t\rightarrow \infty} C(t) \propto t^{c/2} T^{1+c/2},
\label{power_asympt}
\end{equation}
where, note, the temperature dependence also comes from the energy
dependence of $\tau_{min}(E) \propto E^{-3}$ in Eq.(\ref{hcap2}).  The
value $c \sim 0.1$ implies the long time heat capacity should obey $C
\propto T^{1+\alpha}$ at low $T$, where $\alpha \sim 0.05$, a smaller
number than observed in amorphous materials. We must bear in mind that
the issue of the exact form of the time dependence in the laboratory
still appears to be unresolved, as there is no definite agreement
between different experiments; for references, see
\cite{Pohl,HunklingerRaychaudhuri,Nittke,Sahling2002}. While there is
no doubt that the specific heat {\em is} time-dependent, some
experiments agree with the logarithmic time profile, as predicted by
the STM, others give a lower or higher speed of variation with time.
The present semiclassical prediction with $c=0.1$ would be hard to
distinguish from a logarithmic law. Finally, even though a correction
to the linear temperature heat capacity dependence with $c=0.1$ is
most likely smaller than what is experimentally observed, the value of
this correction is non-universal, consistent with empirical data.

\begin{figure}[htb]
\includegraphics[width=.8\columnwidth]{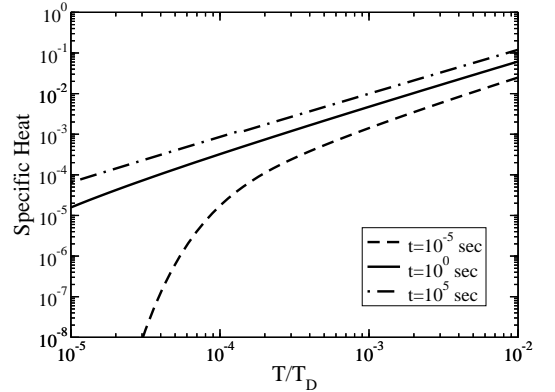}
\caption{\label{CT_t1} Displayed are the TLS heat capacities as
computed from Eq.(\ref{hcap2}) appropriate to the experiment time
scales of order a few microseconds, seconds and hours.  Value of
$c=0.1$ was used here. If one makes an assumption on the specific
value of $\Delta_0$, it is possible to superimpose the Debye
contribution on this graph, which would serve as the lowest bound on
the total heat capacity. As checked for $\Delta_0 = \omega_D$, the
phonon contribution is negligible at these temperatures.}
\end{figure}

The expression in Eq.(\ref{hcap2}) can be evaluated numerically for
all values of $t$, and the results for three different waiting times
are shown in Fig.\ref{CT_t1} for $c=0.1$.  The value of $\tau_{min} =
2.0 \mu sec$ at $E/T_D = 5.7 \cdot 10^{-4}$, derived from the present
theory (also consistent with \cite{GoubauTait}) was used. The results
for $t=10 \mu sec$ demonstrate that due to a lack of fast relaxing
systems at low energies, short time specific heat measurements can
exhibit an apparent gap in the TLS spectrum. Otherwise, it is evident
that the power-law asymptotics from Eq.(\ref{power_asympt}) describes
well Eq.(\ref{hcap2}) at the temperatures of a typical experiment.

\vspace{5mm}

As clear from the discussion above, the long-time power law behavior
of the heat capacity is determined by the ``slow'' two-level systems
corresponding to the higher barrier end of the tunneling amplitude
distribution, argued to be of the form shown in Eq.(\ref{P_D}). If one
assumes that this distribution is valid for the zero-barrier tail of
the $\log(\Delta)$ distribution as well, one would expect that the
heat conductivity would scale as $T^{2+c}$ at the TLS temperatures, in
contrast to an observed experimental sub-quadratic dependence
$T^{2-\alpha'}$. As we shall see in Section \ref{LevelRepulsion},
other quantum effects are indeed present in the theory and we will
discuss then how these contribute both to the deviation of the
conductivity from the $T^2$ law and the way the heat capacity differs
from the strict linear dependence, both contributions being in the
direction observed in experiment.  Finally, when there is significant
time dependence of $c_V$, the {\em kinematics} of the thermal
conductivity experiments are more complex and in need of
attention. When the time-dependent effects are included, both phonons
and two-level systems should ideally be treated by coupled kinetic
equations. Such kinetic analysis, in the context of the time dependent
heat capacity, has been conducted before by other workers
\cite{StrehlowMeissner}.

\section{The Plateau in Thermal Conductivity and the Boson Peak}
\label{plateau_chapter}

In this section we continue to explore the consequences of the
existence of the low temperature excitations in amorphous substances,
which, as argued in Chapter \ref{Intrinsic}, are really resonances
that arise from residual molecular motions otherwise representative of
the molecular rearrangements in the material at the temperature of
vitrification. We were able to see why these degrees of freedom should
exist in glasses and explain their number density and the nearly flat
energy spectrum, as well as the universal nature of phonon scattering
off these excitations at low T ($< 1$ K).

At higher temperatures ($K_B T \sim$10$^{-2.0}$ to 10$^{-0.5} \hbar
\omega_D$), an apparently different kind of excitations begins to
appear, leading to the so called bump in the heat capacity and plateau
in the thermal conductivity, as was discussed in the Introduction.  We
argue in this chapter that the transitions between the mutually
accessible frozen-in minima in the amorphous lattice that give rise to
the two level system behavior at the lowest $T$ also explain the
existence of the modes responsible for this ``Boson Peak'' and the
intense phonon scattering at the corresponding frequencies. This thus
removes the need to invoke theoretically any additional mechanisms,
although other contributions may well be present to some extent; we
will try to assess this possibility in the following Section.

\subsection{Introduction: Classification of Excitations in Glasses}
\label{classific}

While we believe to have mostly achieved a microscopic understanding
of the excitations that are specific to the amorphous solids and are
not present in other types of materials, this description is rather
new and, naturally, there is certain lack of established language that
could be efficiently used to characterize these excitations. In this
section, we will introduce some terminology that will be used in the
rest of the article. At the same time, we will provide a brief general
analysis of what possible qualitatively distinct types of molecular
motions can exist in glasses.

Any atomic motions that take place in a frozen glass, obviously may
also be present in the liquid above $T_g$. For example, those high $T$
motions that correspond to shear attain stiffness (on realistic time
scales) below freezing.  The motions in the liquid, apart from pure
volume change, corresponding to the longitudinal sound, are molecular
translations, or, informally speaking, jumps.  Above $T_A$, such jumps
are not accompanied by a noticeable volume change and bond stretching,
as no metastable structures form in the liquid at these
temperatures. The barriers are therefore largely entropic.  (It is
nice to compare Feynman's discussion of the absence of energy barriers
in superfluid He \cite{Feynman} in this regard.)  Below $T_A$, such
hopping already involves moving a number of molecules from one local
minimum of the free energy functional to another such minimum and thus
requires structural rearrangement within a certain cooperative length
owing to the formation of metastable local arrangements. Molecular
translations do not conserve momentum, which subsequently must be
provided by the rest of the bulk. We thus call these degrees of
freedom, which are relics of the translational motions in the liquid
above $T_g$, {\em inelastic} degrees of freedom. They are truly
inelastic also in the macroscopic sense of the word, because the
existence of alternative configurations in the solid bulk, which are
also coupled to the phonons, ultimately leads to irreversible
relaxation, if the sample is subjected to mechanical stress thus
causing a shift in the thermal population of the alternative internal,
structural states. This is the mechanism behind the so called bulk
viscosity \cite{LLelast} (incidentally, it also contributes to the so
called relaxational phonon absorption, which we discuss in subsection
\ref{rel_abs}).  The switching from one energy minimum to another is
accomplished by moving the domain wall - the interface between the two
alternative configurations - through the local region. As mentioned
earlier, this domain wall is something of an abstract entity, really a
quasiparticle of a sort. Yet it has many ponderable attributes. For
one thing, it has a mass (per unit area), which will be obtained in
this section. It also has surface tension, therefore it can support
surface vibrations, again, of a sort. Although these vibrations are
realized through real atomic motions, it is more beneficial to think
of them as vibrational modes of an imaginary membrane. In fact, as
will be argued later, the oscillations of this membrane correspond to
the indeterminacy in the exact boundary of the frozen-in domain that
has more than one kinetically accessible internal state.  Therefore,
highly anharmonic atomic motions in the real space correspond to {\em
harmonic} motions in the space where the domain walls are
defined. This mental construction does the trick of enabling us to
calculate the ripplon spectrum, as demonstrated in section
\ref{spectrum}.  Now, since it was shown in \cite{XW}, that the liquid
degrees of freedom below $T_A$ consist of switching to alternative
local energy minima; we can claim our assignment of different {\em
inelastic} modes is exhaustive (but not unique, of course!).  These
are, again, translations and vibrations of the domain walls.

On the other hand, any purely {\em elastic} motions in the glassy
lattice can be thought of as a sum of ordinary, affine, displacements
and non-affine displacements (see e.g. \cite{non-aff}). The affine
component would be the only one present in a perfectly isotropic
medium and would follow the stress pattern according to a Poisson
equation (the situation with a non-isotropic crystal is conceptually
the same). The non-affine displacements are a consequence of the
absence of periodicity.  They involve a small number of molecules and
are characterized by a non-zero circulation of the displacement
field. It is not clear at present whether the size of these non-affine
``islands'' could be inferred in present day computer simulations,
since the amorphous structures that can be currently generated on
modern computers still correspond to unrealistically rapid quenching
rates.  The resulting structure corresponds to a sample caught in a
very high energy state with extraordinarily low barriers. As is clear
by now from the random first order transition theory, such structures
correspond to temperatures close to $T_A$ and will have very small
cooperative regions approaching the molecular scale $a$.

We conclude this subsection by repeating ourselves that one important
difference between the elastic and the inelastic modes is in how they
absorb the phonons. While any static disorder can only provide
Rayleigh scattering with a characteristic length scale equal to the
size of the heterogeneity, the inelastic (resonant) absorption's
cross-section scales as the square of the phonon wave-length, it thus
will considerably dominate the Rayleigh mechanism for the longer
wave-length phonons (absorption saturation in the TLS's does not occur
at the sound intensities typical of heat transfer).

\subsection{The Multilevel Character of the Entropic Droplet Excitations}
\label{multilevel}

We hope to have convinced the reader by now that the tunneling centers
in glasses are complicated objects that would have to be described
using an enormously big Hilbert space, currently beyond our
computational capacity. This multilevel character can be anticipated
coming from the low temperature perspective in \cite{LW}.  Indeed, if
a defect has at least two alternative states between which it can
tunnel, this system is at least as complex as a double well potential
- clearly a multilevel system, reducing to a TLS at the lowest
temperatures. Deviations from a simple two-level behavior have been
seen directly in single-molecule experiments \cite{BTLBO}.  In order
to predict the energies at which this multilevel behavior would be
exhibited we first estimate the domain wall mass. Obviously, the total
mass of all the atoms in the droplet is so large that the possibility
of {\em simultaneous} tunneling of all atoms is completely
excluded. The tunneling, we argue, occurs stagewise; each individual
motion encounters a nearly flat potential, implying low frequency
instantaneous modes.

In addition, the effective mass of the domain wall turns out to be
low, {\em also} owing to the collective, barrierless character of the
tunneling events. This is because moving a domain wall over a
molecular distance $a$ involves displacing, at any one (imaginary!)
time, individual atoms only a Lindemann length $d_L$. Suppose this
occurs on the (imaginary) time scale $\tau$. The resulting kinetic
energy is $M_w (a/\tau)^2 = N_w m (d_L/\tau)^2$, where $N_w \simeq
(\xi/a)^2$ is the number of molecules in the wall and $m$ is the
molecular mass.  Thus the mass of the wall $M_w$ is only $ m (\xi/a)^2
(d_L/a)^2$.  Using $(\xi/a) \simeq 5.8$ and $(d_L/a)^2 \simeq 0.01$
gives $M_w \simeq m/3$. This implies the mass of the wall per {\em
atom} is very small - about a hundredth of a molecular mass,
consistent with the simulations of certain barrierless dislocation
motions in copper \cite{Sethna}.  Using $(d_L/a)^2 \simeq k_B T_g/\rho
c_s^2$, derived earlier, one can express the wall's mass through the
material constants as $M_w \simeq (\xi/a)^2 k_B T_g/c_s^2$.  The wall
mass estimate above, inspired by the Feynman's argument on the
effective mass in liquid helium \cite{Feynman_He}, is entirely
analogous to the well known estimate of the soliton mass in
polyacetylene, see e.g. \cite{Solitons_poly}. In the latter, the
soliton moves a large distance, while individual atoms undergo only
small displacements leading to a low soliton mass.

We can now use the typical value of the barrier curvature from our
tunneling argument in section \ref{TLS} (see Fig.\ref{V_N}) to
estimate the typical frequency $\omega^\ddagger$ of motion at the
tunneling barrier top.  We now express the barrier profile $V(N)$ as a
function of the droplet's radius $r \equiv a (3 N/4\pi)^{1/3}$ and
obtain
\begin{equation}
\omega^\ddagger = -\partial^2 V/\partial r^2/M_w 
\simeq 1.6 (a/\xi) \omega_D.
\end{equation}
According to the quantum transition state theory \cite{PGW_QTST}, and
ignoring damping, at a temperature $T' \simeq \hbar
\omega^\ddagger/2\pi k_B \simeq (a/\xi) T_D/2\pi$, the wall motion
will typically be classically activated.  This temperature lies within
the plateau in thermal conductivity \cite{FreemanAnderson}. This
estimate will be lowered if damping, which becomes considerable also
at these temperatures, is included in the treatment. Indeed, as shown
later in this section, interaction with phonons results in the usual
phenomena of frequency shift and level broadening in an internal
resonance.  Also, activated motion necessarily implies that the system
is multi-level. While a complete characterization of {\em all} the
states does not seem realistic at present, we can extract at least the
spectrum of their important subset, namely those that correspond to
the vibrational excitations of the mosaic, whose spectral/spatial
density will turn out to be sufficiently high to account for the
existence of the Boson Peak.

\subsection{The Vibrational Spectrum of the Domain Wall Mosaic
and the Boson Peak}
\label{spectrum}

\begin{figure}[htb]
\includegraphics[width=.6\columnwidth]{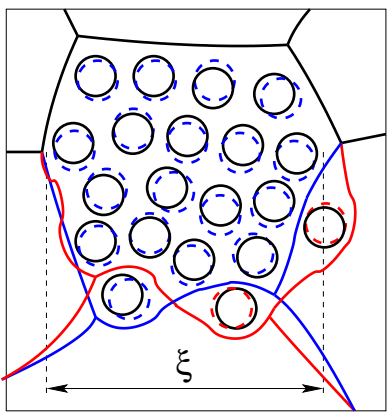}
\caption{\label{ripples} Tunneling to the alternative state at energy
$\epsilon$ can be accompanied by a distortion of the domain boundary
and thus populating the ripplon states. The doubled circles denote
atomic tunneling displacements. The blue line signifies, say, the
lowest energy state of the wall, and the blue circles correspond to
the respective atomic displacements. An alternative wall's state is
shown in red, the corresponding alternative sets of atomic motions are
coded using different colors. The domain boundary distortion is shown
in an exagerated fashion. The boundary does not have to lie {\em in
between} atoms and is drawn this way for the sake of argument; its
position in fact is not tied to the atomic locations in an {\it \`{a}
priori} obvious fashion.}
\end{figure}

At low temperatures the two-level system excitations involve tunneling
of the mosaic cells typically containing $N^* \simeq 200$ atoms. The
tunneling path involves stagewise motion of the wall separating the
distinct alternative configurations through the cell untill a near
resonant state is found. At higher temperatures, other final states
are possible since the exact number and identity of the atoms that
tunnel can vary. These new configurations typically will be like the
near resonant level but will also move a few atoms at the boundary,
i.e. at the interface to another domain.  This is schematically shown
in Fig.\ref{ripples}.
Alternatively, due to the quantum mechanical uncertainty of the exact
location of the domain wall, its shape is intrinsically subject to
fluctuations (these are zero-point vibrations of the domain wall). It
is thus not surprising that the ripplon's frequencies turn out to be
proportional to $\omega_D$, the basic quantum energy scale in the
system.  These fluctuations of the domain boundary shape can be
visualized as domain wall surface modes (``ripplons'').  A detailed
calculation of the ripplon spectrum would require a considerable
knowledge of the mosaic's geometry. At each temperature below $T_A$
the domain wall foam is an equilibrium structure made up of flat
patches of no tension (remember the renormalized $\sigma(r) \propto
r^{-1/2}$; however fluctuations will give rise to finite curvature and
tension).  To approximate the spectrum we notice that the ripples of
wave-length larger than the size of a patch will typically sense a
roughly spherical surface of radius $R = \xi (3/4\pi)^{1/3}$. The
surface tension of the mosaic has been calculated from the classical
microscopic theory and is given by $\sigma(R) = \frac{3}{4} (k_B
T_g/a^2) \, \log((a/d_L)^2/\pi e) (a/R)^{1/2}$ \cite{XW}, where
$d_L/a$ is the universal Lindemann ratio. It could appear that such
tension could collapse an individual fragment of the mosaic but this
tension is, of course, compensated by stretching the frozen-in outside
walls. We approximate the effect of this compensation by an isotropic
positive pressure of a ghost (i.e. vanishing density) gas on the
inside.

\begin{figure}[htb]
\includegraphics[width=.8\columnwidth]{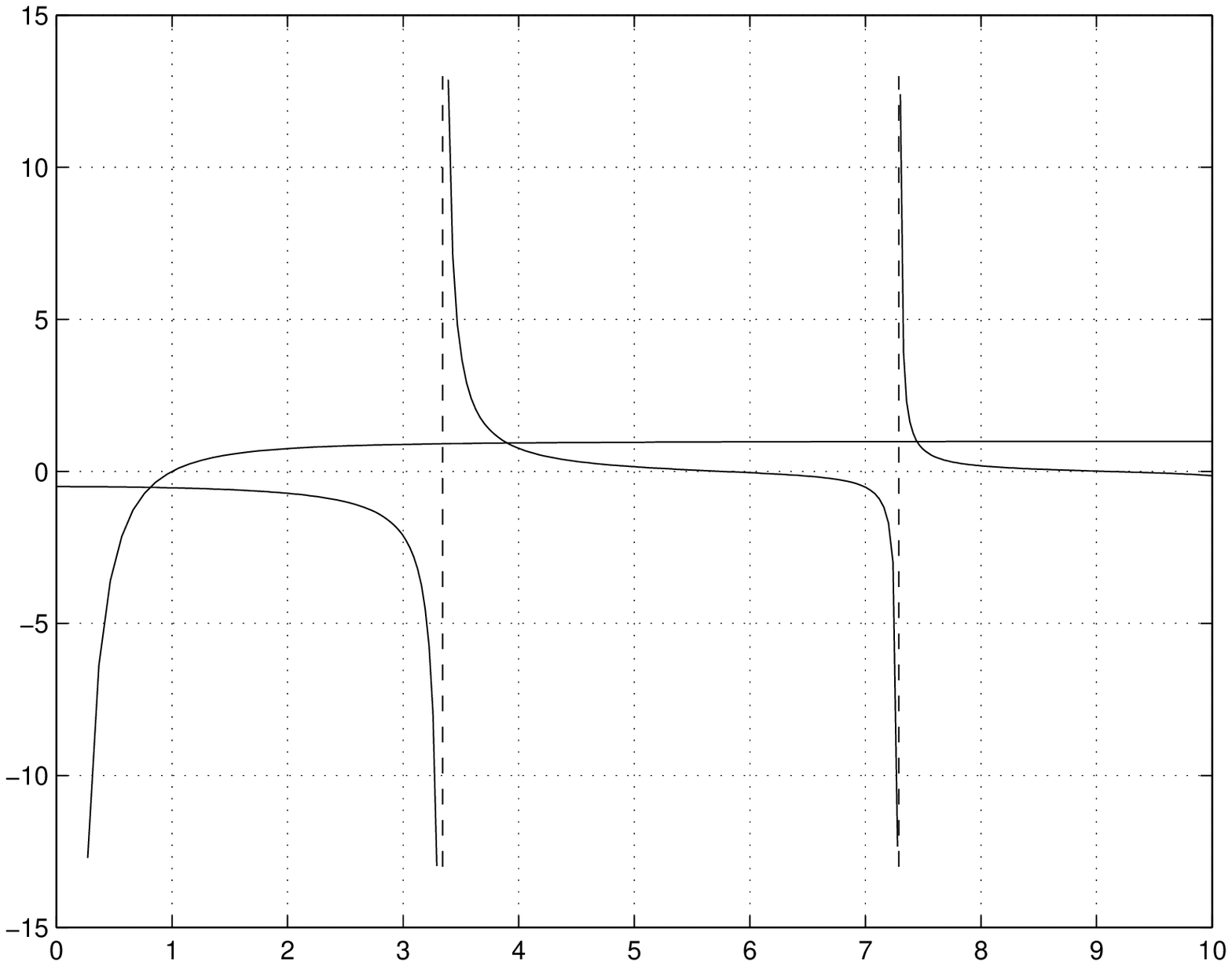}
\caption{\label{cotan} The functions entering Eq.(\ref{all_sltn})} are shown
for some arbitrary parameter values. Here, $l=2$.
\end{figure}

The eigen-frequency spectrum of the surface modes of a hollow sphere
with gas inside is well known (see e.g. \cite{Morse}, as well as our
appendix \ref{Rayleigh_app}).  If we pretend for a moment that the
surface tension coefficient $\sigma$ is curvature independent, the
possible values of the eigen-frequency $\omega$ are found by solving
the following equation:
\begin{equation}
\cot[\alpha_l(\omega R/c_g)] = \left(\frac{\rho_W}{\rho_g R}\right)
-\frac{(l-1)(l+2)}{(\omega R)^2}\left(\frac{\sigma}{\rho_g R}\right),
\label{all_sltn}
\end{equation}
where $\rho_W$ is the membrane's mass per unit area, $\rho_g$ and
$c_g$ are gas' mass density and sound speed respectively.  As stated
earlier, Eq.(\ref{all_sltn}) only applies for $l \ge 2$.  Finally,
function $\cot[\alpha_l(z)] \equiv \left(-l + z
\frac{j_{l+1}(z)}{j_l(z)} \right)^{-1}$, where $j_l(z)$ is the
spherical Bessel function of $l$-th order, does exhibit behavior
similar to that of the regular trigonometric cotangent for arguments
of the order unity and larger, going however to $-1/l$ as $z
\rightarrow 0$.  Its graph for $l=2$ is shown in Fig.\ref{cotan}.
An inspection shows that for each $l$, the smallest solution of
Eq.(\ref{all_sltn}) gives the frequency of the proper eigen-mode of
the shell itself (shifted due to the presence of the gas inside),
whereas the rest of the solutions represent the standing acoustic
waves in the gas. This is especially clear in the $\rho_g \rightarrow
0$ limit, when the lowest frequency does not even depend on the gas'
sound speed, whereas the rest of the solutions are obviously
determined by the inverse time it takes the sound in the gas to
traverse the sphere.

Since we are interested only in the wall's {\em proper} modes in the
limit $\rho_g \rightarrow 0$, we get unambiguously for the frequency
of an $l$-th harmonic:
\begin{equation}
\omega_l^2  = (l-1)(l+2) \left(\frac{\sigma}{\rho_W R^2}\right);
\: \: \: (l \ge 2). 
\label{stand_res}
\end{equation}

Accounting for the unusual $r$ dependence of the surface tension
$\sigma(r) \propto r^{-1/2}$ modifies the standard result from
Eq.(\ref{stand_res}) by a factor of $9/8$. The reason is, the peculiar
surface energy dependence $F_{surf}(R) = 4 \pi R^2 \sigma = 4 \pi
\sigma_0 R^{3/2} a^{1/2}$ calls for the following dependence of
pressure on the curvature: $p = \frac{1}{4 \pi R^2} \frac{\prtl
F_{surf}(R)}{\prtl R} = \frac{3}{2} \frac{\sigma}{R}$ (as compared to
the regular $p = 2 \frac{\sigma}{R}$).  The eigen-frequencies, in
their turn, are determined by calculating the (frequency dependent)
excess pressure due to a variation in curvature. Since now $p \propto
R^{-3/2}$, varying $p$ with respect to $R$ brings down another factor
of $3/2$, thus giving $9/4$ instead of the $2$ of the curvature
independent case. Hence the (barely significant, but curious)
correction factor of $9/8$ used in \cite{LW_BP}.  Since we have been
assuming that the amplitude is infinitesimally small, this factor is
the only consequence of having a curvature dependent $\sigma$, which
should have made the membrane oscillations even more non-linear (as
compared to $\sigma=$ const case) in the case of {\em finite}
displacements.  Pinpointing this effect, however, is clearly beyond
the accuracy attempted by the present model. Finally, one finds a
spectrum with
\begin{equation}
\omega_l^2 = \frac{9}{8} \frac{\sigma}{\rho_W R^2} 
(l-1)(l+2); \: \: \: (l \ge 2),
\label{spec_final}
\end{equation}
where each $l$-th mode of a sphere is $(2 l+ 1)$-fold degenerate.
Using $\rho_W = (d_L/a)^2 \rho a$, obtained earlier in the chapter and
$T_g \simeq \rho c_s^2 a^3 (d_L/a)^2$ (section
\ref{universality_sec}), one finds
\begin{eqnarray}
\omega_{l} & \simeq 1.34 & \, \omega_D (a/\xi)^{5/4} \sqrt{(l-1)(l+2)/4}
\nonumber \\
& \simeq & 0.15 \, \omega_D \, \sqrt{(l-1)(l+2)/4}.
\label{om_l_numer}
\end{eqnarray}
Because of the universality of the $(a/\xi)$ ratio \cite{LW},
$\omega_l$ is a multiple of the Debye frequency. Apart from the barely
significant $(a/\xi)^{1/4}$ factor, again, due to the $R$ dependent
$\sigma$, the ubiquitous scaling $\omega_l \sim (a/\xi) \omega_D$
stresses yet another time the significance of the scale $\xi$. Such a
scale has been previously empirically deduced by interpreting
inelastic scattering experiments but has been usually ascribed to the
static heterogeneity length scale, in contrast with the dynamical
nature of the mosaic in the present theory. We note, again, that this
``static heterogeneity'' has never been unambiguously seen in X-ray
diffraction.  Owing to the material's discreteness, one does not
expect harmonics of higher than $\pi \left(\frac{3}{4 \pi} N^*
\right)^{1/3} [(R-a/2)/R] \simeq 9..10$th order, a relatively large
number, which justifies the tacitly assumed continuum
approximation. The lowest allowed ripplon mode is $l=2$ (corresponding
frequency is $\sim 1$THz for silica, in remarkable agreement with the
inelastic neutron scatering data \cite{bos_peak}).

The requirement $l \ge 2$ can be understood from the symmetry
considerations. $l=1$, the case of no restoring force, corresponds to
a domain translation. Within our picture, this mode corresponds to the
tunneling transition itself.  The ``translation'' of the defects
center of mass violates momentum conservation and must be thus
accompanied by absorbing a phonon. Such resonant processes couple {\em
linearly} to the lattice strain and contribute the most to the phonon
absorption at the low temperatures, dominated by one-phonon processes.
$l=0$, on the other hand, corresponds to a uniform dilation of the
shell. This mode is formally related to the domain growth at $T >
T_g$, and is described by the theory in \cite{XW}. It is thus
possible, in principle, to interpret our formalism as a multi-pole
expansion of the interaction of the domain with the rest of the
sample. Harmonics with $l \ge 2$ correspond to pure shape modulations
of the membrane.

\begin{figure}[htb]
\includegraphics[width=.7\columnwidth]{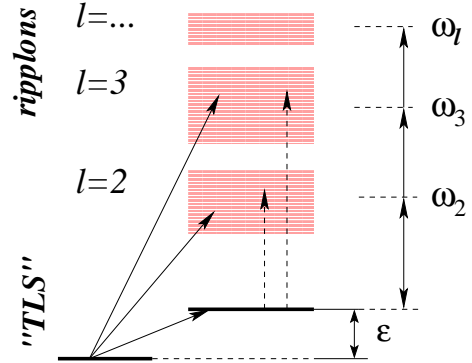}
\caption{\label{en_levels} Tunneling to the alternative state at energy
$\epsilon$ can be accompanied by a distortion of the domain boundary
and thus populatiting the ripplon states. All transitions exemplified
by solid lines involve tunneling between the intrinsic states and are
coupled linearly to the lattice distortion and contribute the
strongest to the phonon scattering. The ``vertical'' transitions,
denoted by the dashed line, are coupled to the higher order strain
(see Appendix \ref{Rayleigh_app}) and contribute only to Rayleigh type
scattering, which is much lower in strength than that due to the
resonant transitions.}
\end{figure}

The existence of the domain wall vibrations explains and allows us to
visualize, at least in part, the multilevel character of the tunneling
centers as exhibited at temperatures above the TLS regime.  Curiously,
the existence of TLS's, even though displayed at the lowest $T$, is
basically of classical origin due to the non-equilibrium nature of the
glassy state. Yet the ripplons, even though seen at higher $T$, are
mostly due to quantum effects and would not be predicted by a strictly
semi-classical theory, in which $\hbar \rightarrow 0$. A schematic of
the resultant droplet energy levels is shown in Fig.\ref{en_levels}.
The arrangement of the combined internal (configurational) and
ripplonic density of states, as depicted in Fig.\ref{en_levels}, has
the following motivation behind it. We include the possibility of
distorting the domain wall during the tunneling transition by
providing a set of vibrational states on top of the alternative
internal state.  The arrangement of the energy states as depicted in
Fig.\ref{en_levels} insures that only thermally active tunneling
centers have mobile (and thus vibrationally excitable boundaries).
The atomic motions at the {\em inactive} defects' sites (i.e. that
cannot tunnel or cross the barrier) would be indistinguishable from
the regular elastic lattice vibrations. Importantly, direct
transitions to the ripplonic state can occur from one of the two
lowest - ``TLS'' - energy states of a tunneling center.  This inherent
assymetry between the two structural states of a tunneling center
actually reflects the thermodynamical inequivalence of the two states
at the glass transition temperature.  While one of the states
represents the local structure in (meta-stable) equilibrium with the
current liquid arrangement around it, the other state is a
configuration that must only be regarded as one of the structures
along the many escape routes from the current equilibrium local state.
At $T_g$, most of those escape routes become too costly energetically.

This is a good place to remind the reader that existing explanations
of the large density of states at the BP energies have to do either
with purely harmonic excitations of disordered, but perfectly stable
lattice (see Introduction), or by generalizing the low energy
inelastic, two-level degrees of freedom to multilevel systems, as was
done e.g. by the soft potential model (SPM)
\cite{soft_pot,soft_potBuchenau}. Such generalizations imply a
connection between the anomalies seen in the TLS regime and at these
higher energies. Such a connection is strongly suggested by
experiment, most prominently by the strength of phonon scattering. The
latter is {\em inelastic} at the BP energies, as it was at the TLS
energies. We stress, the rate of increase of the ripplonic density of
states is much much higher than that empirically assumed in the purely
empirical SPM. Again, there is virtually no freedom to adjust the
numbers in our theory.

In order to compute the heat capacity of the ripplons on top of the
structural transitions we will need to consider the (classical)
density of the inelastic states in more detail than in the previous
section. The density of states $n(\epsilon) =\frac{1}{T_g}
e^{\epsilon/T_g}$ was derived earlier taking as the reference state
the generic global liquid state corresponding to the (high-energy)
configuration frozen-in at $T_g$. It turns out that only transitions
to states with $\epsilon < 0$ (relative to the liquid state!) will
contribute to the TLS density of states. Indeed, as we have shown, the
size of the region that permits a {\em low-barrier} rearrangement must
be slightly (by $\sim$18 molecules) larger than the generic
cooperative size at $T_g$. On the other hand, we know from the RFOT
theory that larger cooperative regions correspond to lower energy
liquid structures. Therefore one of the two alternative states must be
lower in energy than the generic liquid state at $T_g$.  As a result,
the negative $\epsilon$'s correspond to some of the very numerous but
mostly unavailable lower lying energy states, now accessed by
tunneling. Now, if each of those true {\em local} ground states is
taken as the reference one, the spectral density will be now
$n(\epsilon) =\frac{1}{T_g} e^{-\epsilon/T_g}$ ($\epsilon > 0$). We
consequently can let $\epsilon$ from Fig.\ref{en_levels} take both
positive and negative values by writing
\begin{equation}
n(\epsilon) =\frac{1}{T_g} e^{-|\epsilon|/T_g}.
\label{n_eps}
\end{equation} 
  
We can now calculate each domain's partition function by
including all possible ways to excite the system: 
\begin{equation}
Z_{\epsilon}=1+ \sum_{\{n_{lm}\}} e^{-\beta 
(\epsilon+\sum_{lm} n_{lm} \omega_{lm})} 
= 1+e^{-\beta \epsilon}
\prod_l Z_l^{2l+1},
\label{part_fun} 
\end{equation}
where $Z_l \equiv 1/(1-e^{-\beta \omega_l})$ is the partition function
of an $l$th order ripplon mode and we used $m=-l..1$.  Here we assume
each ripplon is a harmonic oscillator. Note that since the
``harmonic'' excitations of frequency $\omega_l$ are on top of another
(structural) excitation, we must consider the issue of the zero-point
energy of these ``harmonic'' excitations, that is no longer a matter
of simply choosing a convenient reference energy. Note that this
zero-point energy is actually several orders of magnitude higher than
the subKelvin energies that are sufficient to excite {\em some} of the
local structural transitions. And indeed, the energy that comprises
the ripplons' ground state energy is not extracted from the thermal
fluctuations of the medium, but, one may say, is simply ``converted''
from the zero-point energy of local elastic vibrations of the lattice.
At the site of a ``slow'' (or, thermally inactive) structural
transition, domain wall vibrations are indistinguishable from the
regular lattice phonons, as already mentioned.

The specific heat corresponding to the partition function in
Eq.(\ref{part_fun}) is found by computing $c_\epsilon = \beta^2
\frac{\prtl^2 log Z_\epsilon}{\prtl^2 \beta}$:
\begin{widetext}
\begin{equation}
c_\epsilon = \frac{\left[\beta \epsilon + \sum_l (2 l+1) 
\frac{\beta \omega_l}{e^{\beta \omega_l}-1} \right]^2}
{\left[2 \cosh \frac{\beta \epsilon + \sum_l (2 l+1) 
\log (1-e^{-\beta \omega_l})}{2}\right]^2} +
\frac{\sum_l (2 l+1) \left(\frac{\beta \omega_l}{2 \sinh 
\frac{\beta \omega_l}{2}}\right)^2}
{e^{\beta \epsilon + \sum_l (2 l+1) \log (1-e^{-\beta \omega_l})}+1}.
\label{c_eps}
\end{equation} 
\end{widetext}
Expression (\ref{c_eps}) clearly becomes the TLS specific heat
$c_{TLS} = \left( \frac{\beta \epsilon}
{2 \cosh(\beta \epsilon/2)} \right)^2$ for $T \ll \omega_l$. 

\begin{figure}[htb]
\includegraphics[width=.75\columnwidth]{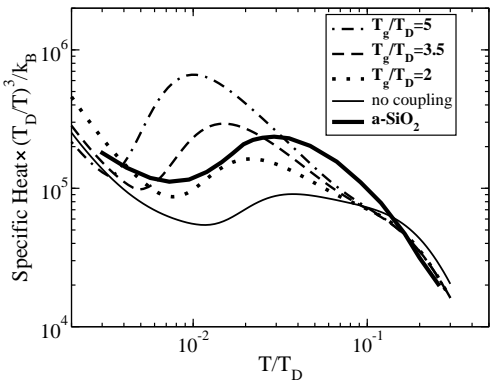}
\caption{\label{bump} The bump in the amorphous heat capacity, divided
by $T^3$, as follows from the derived TLS $+$ ripplon density of
states.  The thin curve corresponds to Eq.(\ref{c_eps}). The thick
solid line is experimental data for a-SiO$_2$ from \cite{Pohl}.  The
experimental curve, originally given in J/gK$^4$, was brought to our
scale by being multiplied by $\hbar^3 \rho c_s^3 (6 \pi^2)
(\xi/a)^3/k_B^4$, where we used $\omega_D = (c_s/a)(6\pi^2)^{1/3}$,
$(\xi/a)^3=200$, $\rho = 2.2 \mbox{g/cm}^3, c_s=4100 \mbox{m/sec}$ and
$T_D = 342 K$ \protect\cite{FreemanAnderson}.  The other curves take
into account effects of friction and frequency shift in the ripplon
frequencies. They will be explained later in subsection
\ref{friction}. The Debye contribution was included in our estimate of
the total specific heat; it was calculated according to (per particle)
$c_D = 9 (T/T_D)^3 \int_0^{T/T_D} dx \frac{x^4 e^x}{(e^x-1)^2}$
\cite{Kittel}.  This equals to $234 (T/T_D)^3$ at the low $T$. When
multiplied by $(\xi/a)^3 \simeq 200$, this gives a value $4.6 \cdot
10^4$ in good agreement with the experiment (see Fig.3.10 from
\cite{Pohl}); note, however that the amorphous $T_D$ is {\em lower}
than the corresponding crystalline one, still it seems $T_D^{amorph} >
\frac{1}{2} T_D^{cryst}$. We remind the reader that no adjustable
parameters have been used so far.}
\end{figure}

In order to obtain the amorphous heat capacity per domain, we
(numerically) average $c_\epsilon$ with respect to $n(\epsilon)$; the
result is shown in figure \ref{bump} with the thin solid line.

\vspace{4mm}

\subsection{The Density of Scatterers and the Plateau}
\label{plateau_sec}

In order to estimate the phonon scattering strength and thus the heat
conductivity, we need to know the effective scattering density of
states, the transition amplitudes and the coupling of these
transitions to the phonons.

Any transition in the domain accompanied by a change in its internal
state is coupled to the gradient of the elastic field with energy $g
\sim \rho c_s^2 \int d {\bf s \, d}({\bf r})$, where ${\bf d}({\bf
r})$ is the molecular displacement at the droplet edge due to the
transition (see section \ref{universality_sec}). An additional
modulation in the {\em domain wall} shape due to the current
vibrational state cancels out due to the high symmetry ($l \ge 2$), as
easily seen when computing the angular part of the surface integral.
We therefore conclude that any transitions between groups marked with
solid lines in Fig.\ref{en_levels} are coupled to the phonons {\em
with the same strength as the underlying (TLS-like) transition}.
(Notice this also implies inelastic scattering off those transitions!)
Incidentally, no selection rules apply for the change in the ripplon
quantum numbers, being essentially a consequence of strong
anharmonicity of the total transition.
  
We do not possess detailed information on the transition amplitudes,
however they should be on the order of the transition frequencies
themselves, just as is the case for those two level systems that are
primarily responsible for the phonon absorption at the lower $T$ which
also have their transition amplitudes comparable to the total energy
splitting.  The argument is thus essentially the same as proposed
earlier in section \ref{TLS}. It should be noted, however, that the
Hilbert spaces corresponding to the {\em quantum} in nature ripplons
and the {\em classical} inelastic states are quite distinct (although
overlapping); it thus should not be surprising that the matrix element
beween {\em superpositions} of these spaces is on the order of the
energy differences themselves. In what follows, we circumvent to an
extent the question of what the precise distribution of the tunneling
amplitudes of the TLS+ripplon transitions is and simply calculate the
{\em enhancement} of the bare TLS induced scattering due to the
presence of the ripplons. This is suggested by an earlier notion that
the structural transitions in glasses couple to the phonons with the
same strength even if accompanied by exciting vibrational modes of the
mosaic.

We now calculate the density of the phonon scattering states.  Since
we have effectively isolated the transition amplitide issue, the fact
of equally strong coupling of all transitions to the lattice means
that the scattering density should directly follow from the partition
function of a domain via the inverse Laplace transform. We will not
proceed this way for purely technical reasons. In addition, we will
separate the cases of positive and negative $\epsilon$ (see
Fig.\ref{en_levels}), corresponding to absorption from ground and
excited states respectively.

The phonon-ripplon interaction exhibits itself most explicitly through
the phonon scattering, which becomes so strong by the end of the
plateau as to cause complete phonon localization. This interaction
also results in other observable consequences, such as dispersion (or
frequency shift) of the ripplon frequencies, as well as rendering the
resonances finite width. Furthermore, we will argue, this interaction
suffices to account for the non-universality of the plateau. First,
however, we consider a simpler situation, where we assume the ripplon
spectrum itself is unaffected by coupling to the phonons.

\subsection{Phonon scattering off frictionless ripplons}

If $\epsilon > 0$, the phonon absorbing transition occurs from 
the ground state. The total number of ways to admit energy $\omega$
into the system is 
\begin{widetext}
\begin{equation}
\rho (\omega) = \int_0^{\infty} d\epsilon \, 
n(\epsilon) \sum_{\{n_{lm}\}} 
\delta(\omega-[\epsilon+\sum_{lm} n_{lm} \omega_{lm}]) = 1/T_g 
\sum_{\{n_{lm}\}}  \theta(\omega-\sum_{lm} n_{lm} \omega_{lm}) 
e^{-\beta_g (\omega-\sum_{lm} n_{lm} \omega_{lm})},
\label{rho_setup}
\end{equation}
where we sum over all occupation numbers of the ripplons with
quantum numbers $l,m$ ($m=-l..l$). Using an integral representation
of the step function $\theta$, this can be rewritten as
\begin{equation}
\rho(\omega) = \frac{1}{T_g}\sum_{\{n_{lm}\}} 
\lim_{\epsilon_1 \rightarrow 0^+} \int_{-\infty}^\infty 
\frac{d k}{2 \pi (i k + \epsilon_1)} 
e^{i k (\omega-\sum_{lm} n_{lm} \omega_{lm})}
e^{-\beta_g (\omega-\sum_{lm} n_{lm} \omega_{lm})}.
\label{rho_om}
\end{equation}
\end{widetext}
The integral in Eq.(\ref{rho_om}) will be taken by the steepest
descent method (SDM). The reason why we do not apply an analogous
technique directly to the $\delta$-function in Eq.(\ref{rho_setup}) is
not only because we want to get rid of the $\epsilon$ integration, but
also because that the SDM proved more forgiving in terms of accuracy
when used to approximate the $\theta$-function, rather than the
$\delta$-function.
 
\begin{figure}[!htb]
\includegraphics[width=.8\columnwidth]{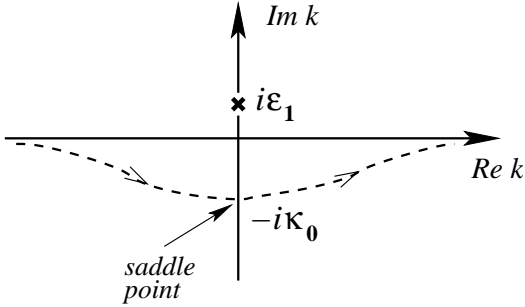}
\caption{\label{saddle_p} The integration contour from
Eq.(\ref{rho_om}) is distorted as explained in text. The pole $k= i
\epsilon_1$ is shown by a cross.  Note the contour does not cross the
pole when being shifted off the real axis.}
\end{figure}

For each $k$ on the real axis, the sum over the occupation numbers
$n_{lm}$ diverges, so each integral should be taken before the
summation. However, in the vicinity of the point that will turn out to
be the saddle point $k_0$ ($\Im k_0 < -\beta_g$) all the sums are
finite, so we reverse the order of summation and integration.  The
integration contour is shifted as shown in see Fig.\ref{saddle_p}.

Hence, the saddle-point approximation yields for the value of
the integral in Eq.(\ref{rho_om}) ($\kappa \equiv i k$):
\begin{widetext}
\begin{equation}
\rho(\omega) = \frac{1}{T_g} \frac{1}{\sqrt{2 \pi |f''(\kappa_0)|}}
\frac{1}{\kappa_0} \exp\{(\kappa_0 - \beta_g) \omega - \sum_{lm}
\log[1-e^{-(\kappa_0-\beta_g)\omega_{lm}}] \},
\label{SDM_value}
\end{equation}
\end{widetext} 
where the saddle point $\kappa_0$ is determined from
\begin{equation}
\omega = \sum_{lm} \frac{\omega_{lm}}{e^{(\kappa_0-\beta_g)\omega_{lm}}-1}
+\frac{1}{\kappa_0}
\label{rho_om1}
\end{equation}
and the curvature at the saddle point is equal to
\begin{equation}
|f''(\kappa_0)| = \sum_{lm} 
\frac{\omega_{lm}^2}{4 \sinh^2[(\kappa_0-\beta_g)\omega_{lm}/2]}
+\frac{1}{\kappa_0^2}.
\label{curv}
\end{equation}
As is clear from (\ref{rho_om1}), the approximation amounts to finding
the effective temperature so as to populate the ripplonic states to
match the excitation energy $\omega$. The expression for the curvature
(\ref{curv}) appropriately involves the corresponding heat capacity of
the excitations.

\begin{figure}[!htb]
\includegraphics[width=.8\columnwidth]{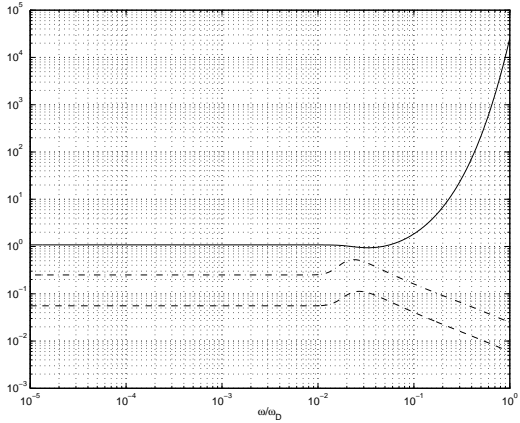}
\caption{\label{nE1_err} The solid line shows $\rho(\omega) T_g$ from
Eq.(\ref{SDM_value}). The dash-dotted line shows the value of the
fourth order term. The third order term, being purely imaginary,
contributes only in the sixth order; it is shown as the dashed line,
however there are other contributions to the 6-th order.}
\end{figure}

The $\omega \rightarrow 0$ and the barely relevant $\omega \rightarrow
\infty$ asymptotics are easily found. As luck has it, the $\omega
\rightarrow 0$ limit of Eq.(\ref{SDM_value}), apart from $1/T_g$
factor, gives $\exp(1)/\sqrt{2 \pi} \simeq 1.08$, only 8\% away from
the correct $1$. The $\omega \rightarrow \infty$ yields, on the other
hand, $\rho(\omega) \propto \prod_{lm} (\omega/\omega_{lm}) \propto
\omega^{96}$, as expected ($\sum_{l=2}^9 (2 l+1) = 96$).  The SDM is
thus reasonably accurate in this case, which could be at least
somewhat evaluated by computing the value of the fourth order term
under the exponent at ``one sigma'' distance from the extremal action
point. This turns out to be satisfactorily small, as demonstrated in
Fig.\ref{nE1_err}, along with the density of states itself as a
function of $\omega$.

\begin{figure}[htb]
\includegraphics[width=.8\columnwidth]{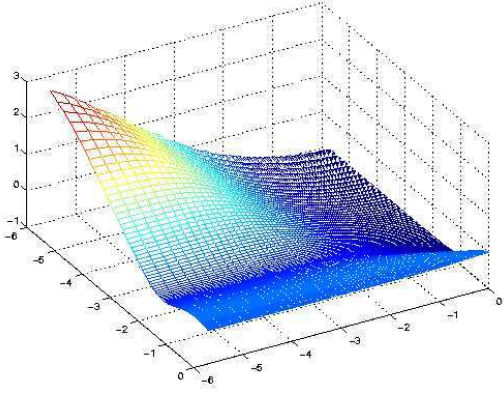}
\caption{\label{err4_2D} The value of the fourth order correction in
the exponent at the distance of one $\sigma$ from the saddle point,
used to evaluate the adequacy of the SDM, is shown here. Each axes is
shown in the base 10 logarithm scale.  The error at the low energies
($< 10^{-2} \omega_D$) is extremely large, which is a consequence of
the fact that the expression in Eq.(\ref{dbl_1}) gives an incorrect
asymptotics as $\omega, E \rightarrow 0$.  However it still gives a
density of states which is less than 1, which is all that matters to
us. The performance at the plateau energies, which is essential here,
is good. Actually, in order to save space, we have presented here the
data for the case with the phonon coupling effects on the ripplon
spectrum taken into account ($T_g/\omega_D=5$).  This case is more
interesting anyway, and the error here is slightly larger (but still
tolerable!), we thus have covered all the cases.}
\end{figure}

When estimating absorption from the ground state, we totally ignore
the depletion of ground state population at finite temperatures, when
the system spends some time in an excited state. This is fine because
by the relevant temperatures, the excited state absorption dominates
anyway (see Fig.\ref{en_levels} and note that $\omega_l - |\epsilon| <
\omega_l + |\epsilon|$).  This case, i.e. $\epsilon < 0$, is somewhat
less straightforward. Let us calculate
\begin{equation}
N_E (\omega) \equiv \int_0^{E} 
d\epsilon \, n(\epsilon) \sum_{\{n_{lm}\}} 
\delta(\omega-[\sum_{lm} n_{lm} \omega_{lm}-\epsilon]). 
\label{N_E(om)}
\end{equation}
This expression gives the {\em cumulative} density
of absorbing states between energies $0$ and $E$ (note
the change of sign in front of $\epsilon$). This 
expression can be used to estimate the total excited state
absorption by computing 
\begin{equation}
\rho_{exc}(\omega,T) 
\equiv \int_0^{\infty} 
d E \, f(E,T) \, \frac{\prtl N_E (\omega)}{\prtl E},
\label{rho_exc} 
\end{equation} 
where $f(E,T) \equiv 2/(e^{\beta E}+1)$ gives the appropriate
Boltzmann weights. The factor of $2$ is used in order to calibrate the
excited state absorption relative to the ground state case: $f(0,T) =
1$. We now have two $\theta$-functions and consequently two
integrations.  The SDM value for $N_E(\omega)$ is given by
\begin{widetext}
\begin{equation}
N_E (\omega) = \frac{1}{T_g} \frac{1}{2 \pi |\mbox{``Det''}|^{1/2}}
\frac{1}{\lambda_0 \mu_0 } \exp\{ (\beta_g +\lambda_0-\mu_0) \omega
+\lambda_0 E  - 
\sum_{lm} \log[1-e^{-(\beta_g+\lambda_0-\mu_0)\omega_{lm}}] \}.
\label{dbl_1}
\end{equation}
The corresponding saddle points are determined from
\begin{equation}
\omega + E = \sum_{lm} 
\frac{\omega_{lm}}{e^{(\beta_g +\lambda_0-\mu_0)\omega_{lm}}-1}
+\frac{1}{\lambda_0}
\end{equation}
and
\begin{equation}
\omega = \sum_{lm} 
\frac{\omega_{lm}}{e^{(\beta_g +\lambda_0-\mu_0)\omega_{lm}}-1}
-\frac{1}{\mu_0}.
\end{equation}
Here,
\begin{equation}
|\mbox{``Det''}| \equiv \sum_{lm} \frac{\omega_{lm}^2}{4
  \sinh^2[(\beta_g +\lambda_0-\mu_0)\omega_{lm}/2]}
\left(\frac{1}{\lambda_0^2} + \frac{1}{\mu_0^2} \right)
+\frac{1}{\lambda_0^2 \mu_0^2}
\label{dbl_last}
\end{equation}  
\end{widetext}
is the determinant of the curvature tensor in the direction (i.e. 2D
subset) of the fastest descent in the 4-dimensional (complex)
$\lambda, \mu$ space.  The steepest descent approximation turns out to
perform well, except at very low frequencies ($\omega < 10^{-2}
\omega_D$).  However, even though it overestimates the answer, it is
still very small compared to the $\rho(\omega)$ calculated earlier at
these frequencies, much as the complete result would be.  The
appropriate graph is shown in Fig.\ref{err4_2D}.

\begin{figure}[htb]
\includegraphics[width=.8\columnwidth]{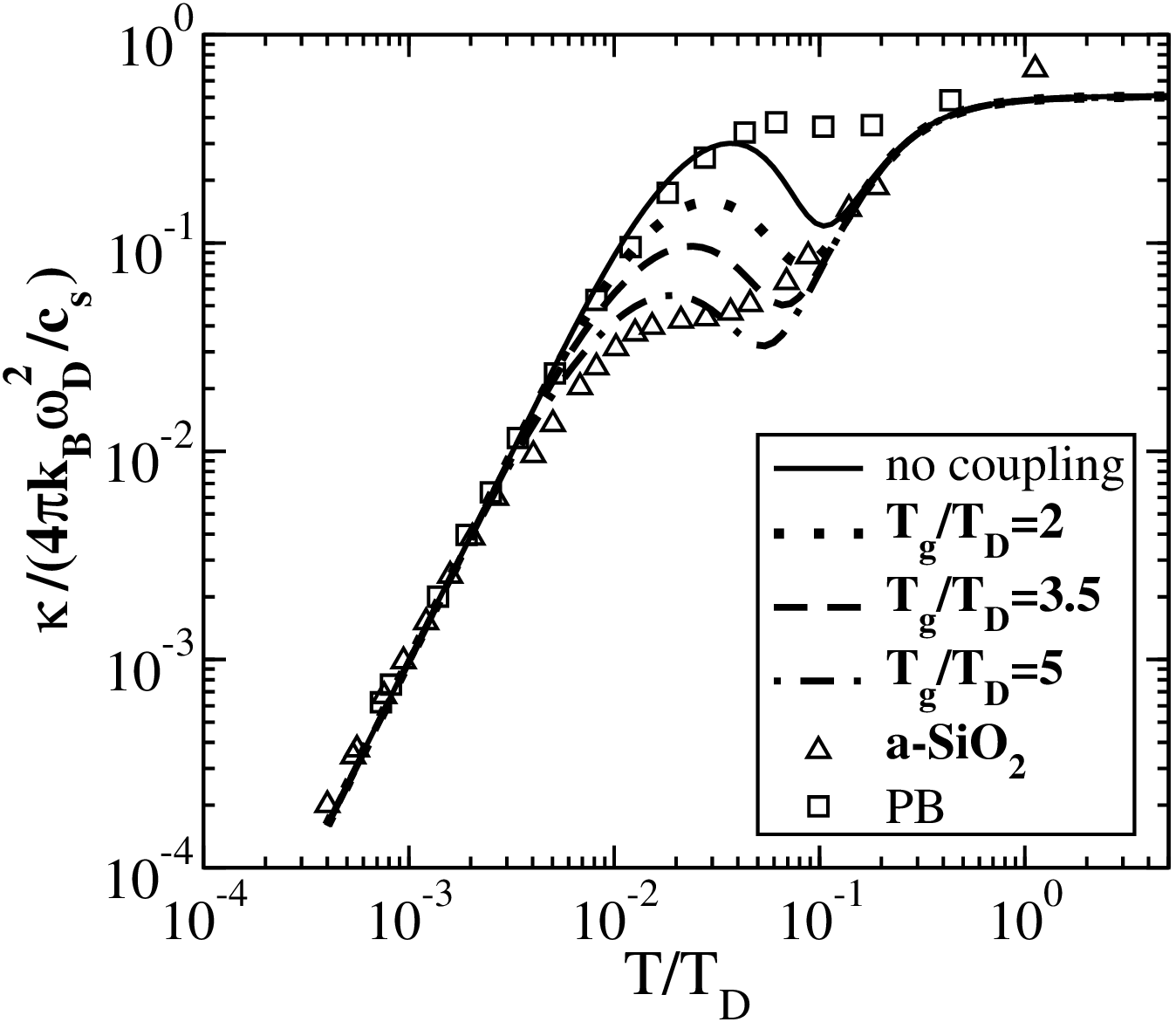}
\caption{\label{plateau_fig} The predicted low $T$ heat conductivity.
The ``no coupling'' case neglects phonon coupling effects on the
ripplon spectrum. The (scaled) experimental data are taken from
\protect\cite{smith_thesis} for a-SiO$_2$ ($k_B T_g/\hbar \omega_D
\simeq 4.4$) and \protect\cite{FreemanAnderson} for polybutadiene
($k_B T_g/\hbar \omega_D \simeq 2.5$). The empirical universal lower
$T$ ratio $l_\smfp/l \simeq 150$ \protect\cite{FreemanAnderson}, used
explicitly here to superimpose our results on the experiment, was
predicted by the present theory earlier within a factor of order
unity, as explained in Section \ref{universality_sec}.  The effects of
``non-universality'' due to the phonon coupling are explained in
Section \ref{friction}.}
\end{figure}

An accurate calculation of the heat conductivity requires solving a
kinetic equation for the phonons coupled with the multilevel systems,
which would account for thermal saturation effects etc.  We
encountered one example of such saturation in the expression
(\ref{l_l1}) for the scattering strength by a two-level system, where
the factor of $\tanh(\beta \omega/2)$ reflected the difference between
thermal populations of the two states. Neglecting these effects should
lead to an error of the order unity for the thermal frequencies.
Within this single relaxation time approximation for each phonon
frequency, the Fermi golden rule yields for the scattering rate of a
phonon with $\hbar \omega \sim k_B T$:
\begin{equation}
\tau_{\omega}^{-1} \sim   \, \omega \frac{\pi g^2}
{\rho c_s^2} \, [\rho(\omega) + \rho_{exc}(\omega,T)].
\end{equation}
The heat conductivity then equals $\kappa = \frac{1}{3} \sum_\omega
l_\smfp(\omega) C_\omega c_s$.  The mean free path cannot be less than
the phonon's wave-length $\lambda$ (which occurs at the Ioffe-Riegel
condition). Since our theory does not cover the phonon localization
regime we account for multiple scattering effects by simply putting
$l_\smfp = c_s \tau_{\omega} + \lambda$.  At high $T$, the heat is not
carried by ``ballistic'' phonons, but rather is transfered by a random
walk from site to site, as originally anticipated by Einstein
\cite{Einstein} for homogeneous isotropic solids. The resultant heat
conductivity is shown in Fig.\ref{plateau_fig}
\vspace{5mm}
We postpone further discussion of the results above until we include
the effects of coupling of the resonant transitions to the phonons on
the transitions' spectrum.

\subsection{The effects of friction and dispersion}
\label{friction}

A transition linearly coupled to the phonon field gradient will
experience, from the perturbation theory perspective, a frequency
shift and a drag force owing to phonon emission/absorption. Here we
resort to the simplest way to model these effects by assuming that our
degree of freedom behaves like a localized boson with frequency
$\omega_l$.  The corresponding Hamiltonian reads:
\begin{equation}
H =  \omega_l a^\dagger a +  
\sum_{\bf k} \omega_k b^\dagger_{\bf k} b_{\bf k} +
\sum_{\bf k} \frac{({\bf g k})}{\sqrt{2 \omega_k V \rho}} 
(a^\dagger b_{\bf k} + b^\dagger_{\bf k} a).
\end{equation}
The ensuing equations of motion are
\begin{eqnarray}
\dot{a}= -i \left[ \omega_l a + 
\sum_{\bf k} \frac{({\bf g k})}{\sqrt{2 \omega_k V \rho}} b_{\bf k} 
\right], \nonumber \\ 
\dot{b}_{\bf k} = -i \left[\omega_k b + 
\frac{({\bf g k})}{\sqrt{2 \omega_k V \rho}} a \right].
\label{eqn_mtn}
\end{eqnarray}
We next introduce the following (retarded) Green's functions $A(t)
\equiv - i \theta(t) \la [a(t), a^\dagger(0)] \ra$ and $B(t) \equiv -
i \theta(t) \la [b(t), a^\dagger(0)] \ra$.  The fourier transforms of
these Green's functions will consequently obey
\begin{eqnarray}
(\omega - \omega_l) \we{A} = \frac{1}{2\pi} + 
\sum_{\bf k} \frac{({\bf g k})}{\sqrt{2 \omega_k V \rho}} 
\we{B}_{\bf k} 
 \nonumber \\ 
(\omega - \omega_k) \we{B}_{\bf k} =  
\frac{({\bf g k})}{\sqrt{2 \omega_k V \rho}} \we{A}.
\label{Green_om}
\end{eqnarray} 
From Eqns.(\ref{Green_om}), one determines the real and imaginary
parts of the Green's functions self-consistently. We however can
disregard the phonons' dispersion and damping which introduces an
error in a higher order, in so far as the shifted frequencies
$\omega_l$'s are concerned. This yields
\begin{equation}
\we{A} = \frac{1}{2 \pi} \left( 
\omega -\omega_l - \frac{1}{3} \frac{g^2}{4 \pi^2 \rho c_s^2}
\lim_{\epsilon_1 \rightarrow 0^+}
\int_0^{k_c} \frac{k^3 \, d k}{\omega/c_s + i \epsilon_1 - k}
\right)^{-1},
\label{A_om}
\end{equation}
where $k_c$ is the cut-off wave-vector whose value will be discussed
shortly (we have also replaced $\sum_{\bf k} \rightarrow 
V \int \frac{d^3 {\bf k}}{(2\pi)^3} )$. Eqn.(\ref{A_om}) gives 
immediately for the inverse life-time of the internal resonance
\begin{equation}
\tau^{-1}_{\omega_l} = \frac{g^2}{4 \pi \rho c_s^2} 
\left(\frac{\omega}{c_s} \right)^3
\simeq \frac{3 \pi}{2 \hbar} T_g \left(\frac{\omega}{\omega_D}\right)^3
,  \hspace{10mm} \omega \le \omega_c
\label{fric}
\end{equation}
and its frequency shift
\begin{eqnarray}
\omega_l(\omega) & = & \omega_l - \frac{g^2}{4 \pi^2 \rho c_s^2}
\dashint_0^{\omega_c} 
\frac{d \omega' (\omega'/\omega_c)^3}{\omega'-\omega}
\nonumber \\
& \simeq & \omega_l - \frac{3}{2 \hbar} T_g
\left(\frac{\omega_c}{\omega_D}\right)^3 \dashint_0^{\omega_c} 
\frac{d \omega' (\omega'/\omega_c)^3}{\omega'-\omega},
\label{disp}
\end{eqnarray} 
where the factor of $1/3$ has disappeared because we have accounted
for the three phonon polarizations and also ignored the distinction
between the longitudinal and transverse branches. The singularity in
Eq.(\ref{disp}) at $\omega \rightarrow \omega_c$ is completely
artificial, as the cut-off is not supposed to be sharp. In our
numerical estimates, we use a cut-off smeared by $\delta \omega_c =
\omega_c/\sqrt{D}$, where $D$ is the glass' fragility (see Appendix
\ref{Rayleigh_app}); this is however totally unimportant as the
divergence is only logarithmic. According to Eq.(\ref{disp}), the
frequency shift scales roughly with $\omega_c^3$ and is thus rather
sensitive to its value. Due to the dispersion, the resonance in
Eq.(\ref{A_om}) is effectively broadened because the value of the
integral in Eq.(\ref{disp}) is positive for sufficiently small
$\omega$, but turns negative at a frequency which is a multiple of
$\omega_c$.

We approximate the phonon coupling effects by replacing in our spectral 
sums in Eqs.(\ref{SDM_value}-\ref{curv}), (\ref{dbl_1}-\ref{dbl_last}) 
the discrete summation over different ripplon harmonics 
by integration over ``lorentzian'' profiles: 
\begin{equation}
\sum_l \int d\omega \, \delta(\omega-\omega_l) \rightarrow
\sum_l \int d\omega \, 
\frac{\gamma_\omega/\pi}{[\omega-\omega_l(\omega)]^2+\gamma_\omega^2},
\label{replace}
\end{equation} 
where $\gamma_\omega \equiv \tau_\omega^{-1}$ is a (frequency
dependent) friction coefficient and $\omega_l(\omega)$ is the ripplon
frequency shifted due to the dispersion effects. This approximation
amounts to having the total inverse life-time of a transition
involving more than one mode being the sum of the inverse life-times
of the participating modes. This would be correct in the case of a
frequency independent $\gamma$, but should be still adequate at the
low $T$ end of the plateau, where the absorption is mostly due to
single ripplon mode processes.

The value of the cut-off frequency $\omega_c$ is close to but larger
than $(a/\xi) \omega_D$ (see Appendix \ref{wc_app}), as the phonons
whose wave-length is shorter than $\xi$ cause an increasingly smaller
effective gradient of the phonon field as sensed by a region of size
$\xi$. These shorter wave-length phonons will still strongly interact
with the droplets, however at this point we could only emulate that to
some extent by increasing $\omega_c$.  This also brings us back to the
radiation life-time's frequency dependence. It is now clear that for
$\omega_l(\omega) > \omega_c$, $\gamma_\omega$ will not follow the
simple cubic dependence cited above, the latter being probably still a
safe lower estimate. We will thus use the above expression as it makes
little difference computationally in the region of such intense
damping.  At the corresponding temperatures, the scattering is
probably better formally described by the stochastic resonance
\cite{stoch} methodology anyway.

We are now ready to discuss the non-universality of the plateau.  It
is evident from Eqs.(\ref{fric})-(\ref{disp}) that even though the
absorbers' frequencies are determined by the quantum energy scale
$\omega_D$, the overall effective frequency {\em shifts} scale with
$T_g$.  The ratio $T_g/\omega_D$ seems to vary within the range of
between 2 and 5 among different glasses, and the non-universality in
this number could have a substantial effect subject to the value of
$\omega_c$. As argued in Appendix \ref{wc_app}, a value of $\omega_c <
2.5 (a/\xi) \omega_D$ is justified. $\omega_c = 1.8 (a/\xi) \omega_D$
seems to yield the experimentally observed spread in the plateau's
position.  Our results for three values of $T_g/\omega_D$ are shown in
Fig.\ref{plateau_fig}. Since $\omega_c$ should be regarded as an
adjustable parameter we can claim to possess only circumstantial
evidence that the plateau's non-universality is caused by the spread
in the value of the ratio of the two main energy scales in the
problem: the classical $T_g$ and the quantum $\omega_D$. On a
speculative note, this phenomenon may be a sign of strong mixing (and
thus level repulsion) between the ripplons and the phonons, as
implicitly confirmed by a phonon localization transition at
frequencies just above those at the plateau. Indeed, the self-energy
of an internal resonance of dimensions $\xi$ coupled with strength $g$
to an elastic medium scales (within perturbation theory) as $g^2/\rho
c_s^2 \xi^3 \propto T_g$.  This can be viewed as lowering of an
impurity band edge due to the interaction with the phonons, yet
another way to express the existence of mixing between the resonant
transitions and the elastic waves. Within our theory, the
non-universality of the plateau is an internally consistent proof that
the degrees of freedom causing the Boson Peak are {\em inelastic}
ones, whose coupling with the phonons then must be equal to $g$
related to the value of $T_g$ through the stability requirement
explained in Section \ref{Intrinsic}.

We now comment on the plateau slopes in Fig.\ref{plateau_fig} being
noticeably more negative than the experimental value.  The
explaination is, we did not solve the full kinetic equation for the
interacting system, but used a simplistic single life-time
approximation.  We demonstrate this issue by briefly presenting a
slightly different way to estimate $\rho_{exc}(\omega,T)$ from
Eq.(\ref{rho_exc}). Here, we imagine we do not exactly know the
thermal weight function $f(E,T)$ due to the lack of knowledge of the
life-times in the multi-level system. On general grounds, however,
this function should decrease rapidly for $\omega > \alpha T$, where
the $\alpha$ is of order unity. This yields $\rho_{exc}(\omega,T)
\simeq N_{\alpha T} (\omega)$ (where $N_{E} (\omega)$ was defined in
Eq.(\ref{N_E(om)})). We show the result of this approximation for
reasonable $\alpha = 1$ and $\omega_c = 2 (a/\xi) \omega_D$.
\begin{figure}[htb]
\includegraphics[width=.8\columnwidth]{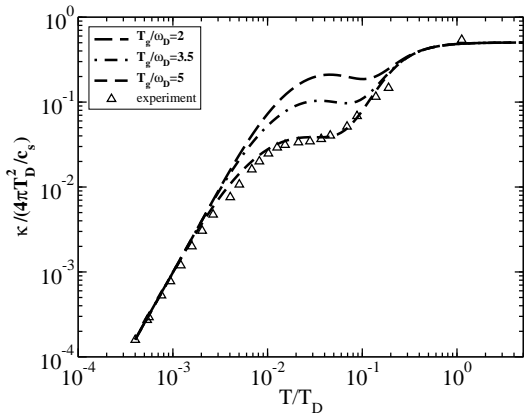}
\caption{\label{cond} The predicted low $T$ heat conductivity for 
three diferent values of $T_g/\omega_D=5, 3.5, 2$ 
for a simpler model of the scattering density as explained 
in text. The a-SiO$_2$ is the same as in Fig.\ref{plateau_fig}.}
\end{figure}
Even though these curves resemble the experimental data better than in
the previous figure, they do not really provide more material support
for the theory than the earlier method. This discussion simply
demonstrates that the basic estimates are robust enough to ``survive''
different levels of treatment.  Also, curiously, these curves reflect
the experimental tendency that the higher $T$ plateaux seem to have a
more negative slope as compared to the low $T$ ones (see
Fig.\ref{l_lambda}), which was less obvious in Fig.\ref{plateau_fig}.

Finally in this subsection, we return to the specific heat. The
effects of the phonon coupling on the ripplon spectrum can be taken
into account in the same fashion as in the conductivity case. Here, we
replace the discrete summation in Eq.(\ref{c_eps}) by integration over
the broadened resonances, as prescribed by Eq.(\ref{replace}).  The
bump, as shown in Fig.\ref{bump}, is also predicted to be
non-universal depending on $T_g/\omega_D$. The predicted bump for
$T_g/\omega_D =2 $ seems to match the best the available data for
a-SiO$_2$, whereas the more appropriate $T_g/\omega_D \sim 4$ is about
a factor of $3$ lower in temperature. It is somewhat unsatisfying that
the plateau's and the bump's position can not be thus both made to
exactly match the experiment at the same time say by adjusting
$\omega_l$, which is certainly allowed given the qualitative character
of some of the estimates. However, since we had to employ an
approximation when calculating the scattering density of states, the
discrepancy does not warrant too much concern, in our opinion.

\subsection{The Relaxational Absorption}
\label{rel_abs}

In addition to the resonant absorption, an internal resonance will
also provide a so called ``relaxational'' scattering mechanism.  Since
a cross-over to the multilevel behavior of the tunneling centers leads
to an increased resonant scattering, we must check whether the
relaxation mechanism is enhanced as well.  This latter mechanism
arises because a passing phonon modifies the energy bias of a
particular pair of internal states. This causes irreversible thermal
equilibration processes within each pair, resulting in energy
dissipation \cite{relax,Maynard}.  This phenomenon is sometimes
referred to as the bulk viscosity \cite{LLhydro}. One important
difference between the relaxational and resonant absorption is that
the former does not saturate and can easily exceed the latter at low
enough temperature and high enough sound intensity, which is what is
usually observed in ultrasonic experiments unless special care is
taken \cite{HunklingerRaychaudhuri} (this saturation is not an issue
in heat conductance, owing to the rather low sound intensities in
these experiments).  Applying the notion of the relaxational
absorption to the two-level systems explained well the shape of the
maximum in the temperature dependence of the sound speed at very low
frequencies at $\sim 1K$ \cite{HunklingerRaychaudhuri}, which is one
of the impressive achievements of the TLS model.  In
\cite{HunklingerRaychaudhuri}, the relation between the slopes of the
logarithmic temperature profiles around the maximum was explained.  At
higher $T$, the logarithmic decrease in $c_s$ is followed by what has
been viewed by others as a mysterious linear law \cite{Belessa}.  At
higher frequencies still, the logarithmic decrease is outweighed by
the just mentioned linear $T$ dependence. We have argued that the
increase in the density of the scattering states is due to thermal
activation of the vibrational states of the domain walls, or matching
of the thermal phonon frequency with that of a ripplon on a mobile
domain wall. Does the existence of the vibrational modes modify the
relaxational scattering as compared to a bare underlying two-level
system? The answer is: not significantly, for the following
reason. The magnitude of the dissipation due to the bulk viscosity
depends on the number of local distinct molecular configurations,
populated according to the Boltzmann statistics.  A shift in this
population results in relaxational dissipation. While having a domain
wall excited may modify the energy scale in the Boltzmann
distribution, which may produce some effect, it does not change the
number of the intrinsic (``inelastic'') glassy states, and thus will
not on average enhance the relaxational scattering. This is to be
compared to the resonant scattering, which depends on the degeneracy
of the ripplon states and will thus intensify at higher $T$, subject
to the degree of the ripplon's linearity. While the relaxational
mechanism thus seems to play only a minor role in the phonon
absorption at the plateau temperatures, its effects are observable and
can explain, as we will argue below, the temperature independent $\log
\omega$ part in the sound speed variation as measured in
\cite{Belessa}.  According to \cite{relax}, the variation in the speed
of sound due to a collection of two-level systems is
\begin{equation}
\left. \frac{\delta c_s}{c_s} \right|_\omega = \dla  
\frac{g^2}{2 \rho c_s^2} 
\left( \frac{\epsilon_i}{E_i} \right)^2 \, 
\frac{\beta}{\cosh^2 \beta E_i} \, 
\frac{1}{1 + \omega^2 \tau_i^2} \dra,
\label{del_cs_rel}
\end{equation}
where 
\begin{equation}
\tau_i \simeq \frac{3 g^2 \Delta_i^2 E_i}{2 \pi c_s^5} \coth(\beta E_i/2)
\label{tau_TLS}
\end{equation}
is the radiative life-time of the $i$th TLS \cite{Jackle} (see also
Eq.(\ref{fric})), and the double angular brackets denote averaging
with respect to $E_i$, $\Delta_i$ and $\tau_i$.  While it would seem
that detailed information on the relevant parameters' distribution is
necessary to use Eq.(\ref{del_cs_rel}), some qualitative conclusions
can be made on general grounds. First, for small $\omega$ the average
is dominated by the long life-time systems, i.e. those with $\Delta
\ll E$ and thus $\epsilon \sim E$.  As a result, the averaging over
these systems is not very sensitive to the possible correlation
between $E_i$ and $\tau_i$, and thus the summation over the two-level
system (nearly flat!)  spectral density $\sim \int d \epsilon (1/T_g
\xi^3)$ introduces, within order unity, only a numerical factor
proportional to $T/T_g$ (and eliminates the explicit temperature
dependence). As just argued, the $(\epsilon/E)^2$ factor should only
give a correction factor of order unity, and we are left with
averaging expression $1/(1 + \omega^2 \tau_i^2)$ with respect to the
life-time distribution. At low frequencies $\omega$, this averaging
will be dominated by the TLS with the long life-times.  Quite
generally, for large $\tau$, $P(\tau) d \tau \propto d \tau/\tau$
because $\tau^{-1}$ scales algebraically with $\Delta$, and the
distribution of $\log \Delta$ is flat (at least for small $\Delta$),
or, almost flat, up to a weak power law, as argued earlier. More
specifically, for a two-level system coupled linearly to the elastic
strain, $\tau^{-1} \propto \Delta^2 E$ (Eq.\ref{tau_TLS}).  Therefore
at each $E$ (which is incidentally only weakly dependent on $\Delta$
in the relevant long life-time case $\Delta/E \ll 1$), obviously
$d(\log \Delta) = const \Rightarrow d(\log \tau) = const$.  Thus the
averaging w.r.t. $\tau$ will produce a term of the order $\sim
(g^2/\rho c_s^2) (1/T_g \xi^3) \log \omega$, which is of the right
order of magnitude (and sign!). Since the dimensionless factor in
front of the $\log \omega$ term has been shown to be universal
($\propto (a/\xi)^3$), the present theory predicts that it should not
vary significantly among the insulating glasses; in fact, according to
our argument, it is proportional to the coefficient $\alpha$ at the
logarithmic temperature dependence of the sound speed variance in the
TLS regime, a rather universal quantity indeed \cite{Leggett}. We
stress however that the just predicted TLS-like property should be
observed in the {\em plateau} regime. A deviation would be a sign of
more than two inelastic states playing a role in the transition.  We
finally mention that the lower limit in the integral over the
life-time distribution should produce a $\log T$ term, which would be
however masked by the stronger linear dependence.

\section{Quantum Effects beyond the strict Semi-Classical Picture}
\label{LevelRepulsion}

\subsection{Quantum Mixing of a Tunneling Center and the Black-Halperin
Paradox}  
\label{Mixing}

The preceding sections have shown that structural transitions,
accompanied at high enough temperatures by vibrational excitations of
the mosaic, account for the most conspicuous departures of the low
temperature behavior of glasses from the prescriptions of a standard
harmonic lattice theory - namely the existence of multilevel intrinsic
resonances in a amorphous sample made by quenching a supecooled
liquid.  At the lowest temperatures these resonances behave for the
most part as if they were two-level systems, while at higher $T$ the
density of states of these intrinsic excitations grows considerably
and leads to the Boson Peak phenomena. While we have computed the
density of states accessible by {\em tunneling} even at the lowest
temperatures, we have assumed, within a semi-classical approach, that
having a small tunneling barrier between alternative local structural
states does not affect significantly the corresponding spectrum
$n(\epsilon)$ of the lowest energy transitions from its classically
defined value. Likewise, we have assumed that the vibrational spectrum
of moving domain walls is unaffected by the presence of tunneling,
that would in principle mix those vibrations quantum
mechanically. Clearly, the transitions that are active at low $T$ must
have some significant (even if small) overlap between the
wave-functions corresponding to the alternative structural
states. This overlap would lead to the familiar effects of repulsion
between the semi-classically determined energy levels. This could be
described as partial quantum melting of some tunneling centers, but it
is probably better to use term ``quantum mixing''.

In this section we estimate the magnitude of these quantum mixing
effects. Even though the strictly semiclassical theory agrees well
with experiment as is, making such estimates that go beyond it is
useful for two distinct reasons. First, we must check to what extent
the semi-classical picture, tacitly assumed earlier, is a consistent
zeroth order approximation to a more complete treatment. Second, it is
important to ask whether the expected corrections to the strict
semiclassical theory lead to observable consequences. In what follows,
we provide approximate arguments that indeed such corrections are
discernible and may even potentially answer some long-standing puzzles
in this field.

\begin{figure}[htb]
\includegraphics[width=.9\columnwidth]{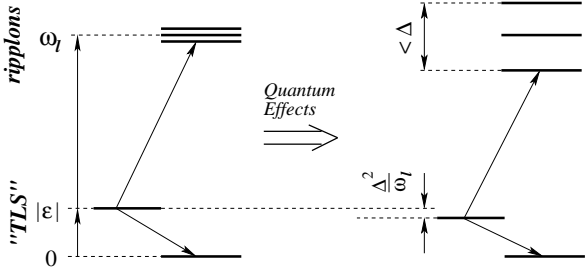}
\caption{\label{en_levels_simple} A low energy portion of the energy
level structure of a tunneling center is shown. Here, $\epsilon < 0$,
which means that the reference, liquid, state structure is {\em
higher} in energy than the alternative configuration available to this
local region. A transition to the latter configuration may be
accompanied by a distortion of the domain wall, as reflected by the
band of higher energy states, denoted as ``ripplon'' states.}
\end{figure}

{\em Quantum Mixing.} As the starting point in the discusion, we
consider a simplified version of the diagram of a tunneling center's
energy states from Fig.\ref{en_levels} with $\epsilon < 0$, as shown
on the left hand side of Fig.\ref{en_levels_simple}.
We remind the reader that the $\epsilon < 0$ situation, explicitly
depicted in Fig.\ref{en_levels_simple}, implies lower transition
energies than when the semiclassical energy difference $\epsilon > 0$
and thus dominates the low temperature onset of the Boson Peak and the
plateau.

Accounting for tunneling, the low energy portion of the Hamiltonian
that corresponds to Fig.\ref{en_levels_simple} is as follows:
\begin{equation}
H_{i} = \left(\begin{array}{cccc}
0        & \Delta/2     &  0  &  0\\
\Delta/2 & |\epsilon|   & \Delta_{i_1} & \Delta_{i_2} \\
0        & \Delta_{i_1} & \hbar \omega_{i_1} & 0            \\
0        & \Delta_{i_2} &  0           & \hbar \omega_{i_2}
\end{array}
\right),
\label{H_class}
\end{equation}
where the semi-classical values of the ripplonic energies are denoted
as $\hbar \omega_i$ and the transition amplitudes to those levels are
$\Delta_i$ respectively (only two lowest of those ripplonic states are
shown in Eq.(\ref{H_class})).  As argued in detail in Section
\ref{plateau_chapter}, only one of the lowest two energy levels in a
tunneling center (the top one in this case) is directly coupled to the
higher, ripplonic, energy states.  Obviously, virtual transitions to
those high energy states will result in lowering the energy of the
higher level.  There are no direct transitions from the bottom state
of energy $0$, as explained in the previous Section, and therefore its
position is unaffected by the presence of the ripplons. Consequently,
the effective energy splitting of the two-level system (with $\epsilon
< 0$) will be lower than the classical value obtained earlier, and the
smaller the original value of $\epsilon$ was, the more pronounced the
effect will be.  In what follows we estimate the consquences of this
effect on the apparent energy spectrum of the lower excitations,
i.e. the empirical two-level systems. In the limit of inifinitely
small tunneling amplitude $\Delta$, the decrease in $\epsilon$ could
be estimated using a perturbative expansion:
\begin{equation}
|\tilde \epsilon| = |\epsilon| - \sum_i \frac{\Delta_i^2}{\hbar \omega_i -
  |\tilde \epsilon|}.
\label{eps_rnrm}
\end{equation}
Here, $\tilde \epsilon$ is the new value of the energy splitting,
$\omega_i$'s are the ripplon frequencies and $\Delta_i$'s are
tunneling amplitudes of transitions that excite the corresponding
vibrational mode of the domain wall. Those amplitudes will be
discussed in due time; for now, we repeat, the expression above will
be correct in the limit $\Delta_i/\hbar \omega_i \rightarrow 0$.
Finally, the renormalized value $\tilde{\epsilon}$ was used in the
denominator.  While, according to Feenberg's expansion
\cite{Feenberg}, including $\tilde{\epsilon}$ in the resolvent is
actually more accurate, we do it here mostly for convenience.

\begin{figure}[htb]
\includegraphics[width=.8\columnwidth]{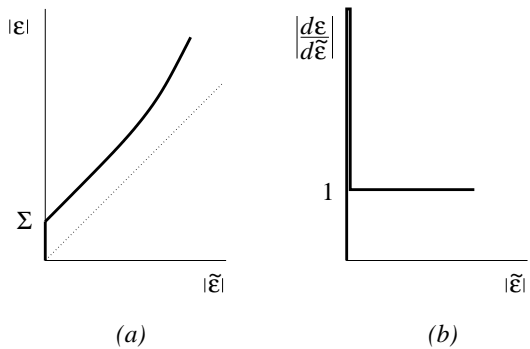}
\caption{\label{eps_eps} Shown in panel (a) is the relation between
the bare energy difference $\epsilon$ between frozen-in structural
states in a glass and the effective splitting $\tilde{\epsilon}$ that
is smaller due the level repulsion in the tunneling center.  Panel (b)
depicts schematically the derivative of $\epsilon$ with respect to
$\tilde{\epsilon}$, which is used to compute the new effective
distribution $P(\tilde{\epsilon})$ of the transition energies.}
\end{figure}

Given that the semi-classical values of eigen-values $\hbar \omega_i$
are known, the low energy portion of the energy level structure of the
tunneling center, as shown in Fig.\ref{en_levels_simple}, gives a
quantitative idea of the eigen-energies of the full Hamiltonian only
in the limit of a very small tunneling splitting $\Delta$. In a
complete treatment, all transition amplitudes must be included and the
Hamiltonian diagonalized. In general, such diagonalization (and, in
our case, the system's ``quantization'') is difficult, however could
still be conducted approximately in some cases of interest.  Consider,
for the sake of argument, the following situation, where $\Delta$ is
not necessarily smaller than $\epsilon$ but $\sum_i \Delta^2_i/\hbar
\omega_i$ is. In this arrangement, the energy shift due to the higher
lying states can be computed using perturbation theory and yields a
``renormalized'' value of the classical energy difference that we have
called $\tilde{\epsilon}$.  This procedure also modifies the tunneling
amplitude $\Delta$ of the underlying TLS by a {\em multiplicative}
factor according to
\begin{equation}
\widetilde{\Delta} = \Delta \left( 1- \frac{1}{2} \sum_i
\frac{\Delta_i^2}{(\hbar \omega_i)^2} \right).
\label{d_d1}
\end{equation}
Following this, the full energy splitting of the TLS tunneling
transition is computed using $E = \sqrt{\Delta^2 +
\tilde{\epsilon}^2}$.  The important feature of the argument is that
$\Delta$ (or $\widetilde{\Delta}$) is allowed to take arbitrarily
large values relative to $\epsilon$ and the ratio of the two
parameters is not treated perturbatively. The lowering of $\epsilon$
due to virtual transitions among the higher energy states changes
somewhat the effective density of transition energies $E$ that
directly enters into the heat capacity and conductivity
calculations. While Eq.(\ref{eps_rnrm}) is perturbative, it should
accurately give finite effects in the mean-field limit of infinitely
many transitions $\hbar \omega_i$ coupled infinitely weakly to one of
the two bottom states of the tunneling center.  We will analyze the
physical consequences assuming the accuracy of Eq.(\ref{eps_rnrm}).
We must, of course, bear in mind that while a tunneling center is
nearly a meanfield entity, owing to the strong correlations, it is
actually of finite, albeit molecularly large size. Let us plot
$|\epsilon|$ as a function of $|\tilde{\epsilon}|$ (see
Fig.\ref{eps_eps}(a)).
Clearly, the smallest allowed value of the effective classical
splitting $|\tilde{\epsilon}|$ of zero corresponds to a finite value
of $|\epsilon|$ equal to 
\begin{equation}
\Sigma \equiv \sum_i \frac{\Delta_i^2}{\hbar \omega_i}.
\label{sigma1}
\end{equation}
Therefore, smaller values of $|\epsilon|$ do not correspond to
physically realizable systems, according to Eq.(\ref{eps_rnrm}).  The
overall excitation spectrum of the structural transitions with those
(small) values of $\epsilon < \Sigma$ is strongly affected by the ease
of tunneling. As a consequence, the $\epsilon$ and $\Delta$
distributions become correlated. The quantity $\sum \Delta_i^2/\hbar
\omega_i$ is of central importance for this section of the article,
therefore we will discuss it now in some detail. The wave-functions of
the highly quantum tunneling centers are heavily mixed combinations of
the classical states corresponding to potential energy
minima. Transitions between such states are strongly coupled to
lattice vibrations and, among other things, would strongly scatter
phonons. This result is expected on rather general grounds and was
exploited earlier when we noted that the eigen-states of the classical
potential energy are largely unrelated to the eigenstates of the
vibrational modes of the domain wall, hence one expects that
transition amplitudes of exciting a ripplon (which, loosely speaking,
is a highly anharmonic combination of both structural and vibrational
modes of the lattice) are expected to be comparable to the ripplon
energy itself. If this is the case, then number $\Sigma$ is actually
very large relative to $\epsilon$ and one, strictly speaking, should
not use a {\em perturbative} expression, such as in
Eq.(\ref{eps_rnrm}) in order to assess the lowering of $\epsilon$ due
to quantum effects. We note, however, it is more likely that the
seeming discrepancy simply stems from our ``quantization'' procedure
being so far rather naive, so let us slow down a bit and attempt to
outline briefly a more careful way to quantize the tunneling centers's
dynamics. First of all, recall that we actually know the values
$\omega_i$ in a strongly quantum regime, because they were computed
assuming a freely moving membrane. On the other hand, we know that in
the classical limit domain wall vibrations are indistinguishible from
the lattice vibrations. Remarkably, the vibrational eigen-frequencies
of a ``box'' of dimensions $\xi$ - $\omega_D$ times a number from
$a/\xi$ to $1$ - span roughly the same range of energies that the
$\omega_i$'s do. Therefore, even though the quantization procedure
will ``reshuffle'' all the ripplonic states, it will not significantly
shift their position as a whole. Next, since static lattice
inhomogeneities scatter phonons only elastically, the coupling of the
ripplonic excitations to the phonons must scale with a positive power
of $\Delta$ so as to vanish in the classical limit. Consequently, in
the limit of small tunneling matrix element $\Delta$, the transition
amplitudes $\Delta_i$'s must scale with a positive power of
$\Delta$. On the other hand, as mentioned many times, in the quantum
regime, the $\Delta_i$'s are of the order $\hbar \omega_i$ and are not
directly related to $\Delta$.  Therefore, a careful quantization
procedure of introducing quantum tunneling in the system must combine
and rationalize both of those seemingly conflicting notions. It will
turn out that in the quantum regime, the $\Delta_i$'s are related to
the strength of the underlying tunneling transition $\Delta$ only in a
certain renormalized sense.  We will now outline this
renormalization/quantization procedure. This procedure imposes certain
restrictions on how adding the possibility of tunneling to a local
structural transition can be performed, so that the structure of the
energy levels of the transition, that we know from general arguments,
is {\em preserved}. Recall that ``switching on'' tunneling to the
higher energy states $\epsilon + \hbar \omega_i$ not only lowered
$\epsilon$, but also made $\Delta$ smaller by a factor of $[1-\sum_i
\Delta_i^2/(\hbar \omega_i)^2]$. This expression is only valid if the
sum is a small number, so that the whole correction factor is
necessarily positive and only changes the effective {\em magnitude} of
the matrix tunneling element, but not its sign. We must require that
$\Delta$ not change its sign in the course of the ``quantization'',
but only its absolute value, because the (ordinarily small) value
$\Delta$ only reflects the (ordinarily small) {\em configurational}
overlap between two local structural states, while the sign (or
complex phase, in general) bears no special meaning here because no
particular {\em spatial} symmetry is involved in the problem. (Such
spatial symmetry is important, for instance, when computing overlaps
between eigenstates, or near-eigenstates of orbital momentum centered
around close locations in space.) Thus, the final answer should depend
only on $|\Delta|^2$. This becomes especially evident after the
following realization.  Note that the expression in the brackets in
Eq.(\ref{d_d1}) is a small coupling limit of what can be considered a
Franck-Condon factor. The appearance of such a factor after the
introduction of non-zero transition amplitudes is natural: the degrees
of freedom that used to be classical and static, can now follow to
some extent a selected motion in the system. The Franck-Condon (FC)
factor is the overlap between the the initial and final wave-functions
of these other (``ripplonic'') degrees of freedom, corresponding to
the initial and final configuration of that selected motion.  (A well
known example of a FC factor arising in an analogous, dynamical
fashion is the tunneling matrix element renormalization in the
spin-boson problem \cite{SB_review}.)  Let us suppose, in a simplified
manner, that the effective renormalization of {\em all} of the newly
introduced tunneling amplitudes occurs in a similar fashion. This
allows one to self-consistently close Eq.(\ref{d_d1}) and rewrite it
for a representative amplitude $\Delta_Q$:
\begin{equation}
\widetilde{\Delta}_Q = \Delta_Q \left[ 1- \frac{B}{2} 
\frac{\widetilde{\Delta}_Q^2}{(\hbar \omega_D)^2} \right],
\label{d_d2}
\end{equation}
where replacing $\Delta_Q$ by $\widetilde{\Delta}_Q$ inside the
brackets preserves the approximation's order. $B$ is a numerical
constant, reflecting the sum over ripplon states with their
vibrational frequencies $\omega_i$, and we have replaced $\omega_i$ by
the Debye frequency $\omega_D$. The two must be related at the end of
the renormalization, as we already know.  Identifying the expression
in brackets with a Franck-Condon factor reminds us that $\Delta_Q$ (or
$\widetilde{\Delta}_Q$) is not only a (generic) tunneling amplitude
but also can be considered a coupling constant, therefore only its
absolute value is physically relevant in the present context, not its
sign. The smallness of $\Delta_Q$ and $\widetilde{\Delta}_Q$ lets us
recast Eq.(\ref{d_d2}) as
\begin{equation}
d \log \left(\frac{\widetilde{\Delta}_Q}{\Delta_Q} \right) = -
\frac{B}{2} d \left[ \frac{\widetilde{\Delta}_Q^2}{(\hbar \omega_D)^2}
\right],
\label{d_d3}
\end{equation}
where the reference state is $\widetilde{\Delta}_Q = \Delta_Q = 0$,
$\lim_{\Delta_Q \rightarrow 0} (\widetilde{\Delta}_Q/\Delta_Q) =1$
(that is ``no tunneling'' $\Rightarrow$ ``no renormalization'').
Notice that the r.h.s. of Eq.(\ref{d_d3}) depends explicitly only on
the effective tunneling element $\widetilde{\Delta}_Q$, but not on the
original (tunable) perturbation strength $\Delta_Q$. We therefore can
use the differential relation in Eq.(\ref{d_d3}) to extend the
perturbative construction from Eq.(\ref{d_d2}) into the region of
arbitrarily large values of the bare coupling $\Delta_Q$ by using the
outcome of the previous (infinitesimal) change in $\Delta_Q$ as the
initial input in the subsequent increment $d \Delta_Q$. Each of these
increments is a small perturbation around a new, self-consistently
determined, value of $\widetilde{\Delta}_Q$. One gets the following
self-consistent equation for the effective tunneling amplitude as a
result:
\begin{equation}
\widetilde{\Delta}_Q = \Delta_Q e^{- \frac{B}{2}
\frac{\widetilde{\Delta}_Q^2}{(\hbar \omega)^2}}.
\label{d_d4}
\end{equation}
Our renormalization procedure is internally consistent in that the
physical value of the tunneling amplitude depends on the scaling
variable - the bare coupling $\Delta_Q$ - only logarithmically.  This
bare coupling must scale with the only quantum scale in the problem -
the Debye frequency, as pointed out yet in the first section.

In a more complete treatment, the $\Delta$ renormalization would not
be characterized by a single ``Franck-Condon'' parameter, but by a
distribution of Franck-Condon factors. Therefore, the exponential form
in Eq.(\ref{d_d4}) might be possibly replaced by a different, perhaps
a polynomial expression. In fact, one may think minimally of
Eq.(\ref{d_d4}) as of one of the possible Pad\'{e} extensions of the
perturbative formula (\ref{d_d2}). At any rate, such a Pad\'{e}
approximant will retain the main feature of Eq.(\ref{d_d4}) in that
the value of the observable tunneling matrix element
$\widetilde{\Delta}_Q$ is bounded from above and depends strongly on
the (semi-)classical energies $\hbar \omega_i$.  According to the
discussion above, this restriction stems from a self-consistency
condition, namely that the leading term in the exponent in
Eq.(\ref{d_d4}) must scale with $\widetilde{\Delta}_k^2$, if this same
number $\widetilde{\Delta}_k$ is on the l.h.s. in that equation. An
important corollary of this is that the perturbative term inside the
brackets of Eq.(\ref{d_d1}) must scale with $\widetilde{\Delta}^2$
itself, hence the perturbation correction $\Sigma$ will scale with
$\widetilde{\Delta}^2$ too (from now on, we will drop tildes from the
symbols denoting the physical tunneling amplitudes, but retain them
for the effective $\epsilon$'s).  That is, with our definition of $B$,
it is roughly true that
\begin{equation}
\Sigma = B \frac{\Delta^2}{\hbar \omega_D}.
\label{sigma2}
\end{equation}
We have thus demonstrated explicitly that the magnitude of quantum
effects on the classical energy splitting $\epsilon$ on a particular
site should depend on the facility of tunneling at that same site. We
have therefore established that the fact of quantum $\Delta_i$ being
close in value to a rather large energy scale $\omega_i$ is consistent
with a relatively small value of the correction in Eq.(\ref{eps_rnrm})
and its scaling with $\Delta^2$. As another dividend from the
argument, we obtain a ballpark estimate of the constant $B$. A
structural transition that is thermally active at a plateau
temeperature $k_B T \sim \hbar \omega_i$ and being an efficient phonon
scatterer, will have $\Delta \sim \hbar \omega_i$. Therefore, $B =
\sum_i \omega_D/\omega_i$, will be a number on the order of several
hundreds, since the total number of the ripplonic modes (at the
laboratory glass transition, at which $\xi/a \simeq 5..6$) is
approximately a hundred, and $\omega_i$ is proportional to, but
somewhat smaller than the Debye frequency $\omega_D$.

Eq.(\ref{sigma2}) implies that while the distributions of the clasical
energy splittings $\epsilon$ and the bare semiclassical tunneling
amplitude $\Delta$ may be uncorrelated, quantum corrections require
that $\tilde{\epsilon}$ and $\Delta$ be correlated for systems with a
sufficiently low barriers and which simultaneously have small energy
difference between the initial and final structural state.
Conversely, the independence approximation is valid when, roughly,
$|\epsilon| > B \frac{\Delta^2}{\hbar \omega_D}$. Since $\Delta$ is
proportional to $\hbar \omega_D$, this criterion is a formal
restatement of an earlier comment that the theory is strictly valid in
the classical limit. (Note that there is also a (much stronger)
$\hbar$ dependence in the exponent of the tunneling element $\Delta$
(see Eq.(\ref{D_D0})).  While, obviously, only a negligible fraction
of the {\em total} number of structural rearrangements in the liquid
at $T_g$ would not satisfy the classicality criterion, these
particularly facile transitions do actually comprise a significant
portion of those transitions that are thermally active at {\em
cryogenic} temperatures. We will now indicate what the observable
consequences of this deviation for the strict semiclassical limit are.
In order to do this, let us discuss first the difference between the
strongly quantum and the bare ``classical'' structural transitions.

According to Eq.(\ref{eps_rnrm}), for all transitions, whose diagonal
energy difference would be $|\epsilon| < \Sigma$ in the classical
limit, the effective diagonal splitting $\tilde{\epsilon}$ is actually
zero, meaning that the full energy splitting $E$ is entirely comprised
of the originally {\em off-diagonal} energy scale $\Delta$. This
implies that the energy eigen-states of such highly mixed tunneling
centers are heavy superpositions of the original classical structural
states and would not be easily interpreted in terms of the atomic
coordinates of the potential minima alone, but must include the
kinetic energy term as well.  This is directly related to the well
known ambiguity in separating the energy of such systems into
potential and kinetic components even at conditions that are entirely
classical, such as at a high temperature.  Of course, in such cases
{\em free energy} formulations must be employed that allow one to
count the number of configurational states unambiguously, while using
``inherent'' structures based on {\em potential} energy stationary
points alone is of limited utility.  The strongly quantum case can be
loosely understood by transcribing the complex multiparticle
rearrangements onto a single collective ``reaction'' coordinate (as in
the soft potential model \cite{soft}). In fact, this analogy to a
single coordinate soft potential model is quite loose because of the
much higher density of states of the ripplons (that give rise to the
Boson Peak and correspond to the vibrations of the membrane) compared
with the density of states of the soft potential model, which is one
dimensional so that only one coordinate is vibrationally excited.
Nevertheless, following this analogy, consider a two-well potential
(with very steep outer walls) with a barrier high enough so that the
physical coordinate eigen-states corresponding to the particle being
in the left or the right well are unambiguously definable and the
diagonal component of the transition's energy is equal to the
difference in the potential energy of the two well with high
accuracy. Imagine next lowering the barrier. In the limit of zero
barrier the system is simply a particle in a square box, whose energy
scale is determined by the quantum energy scale in the problem - that
is the particle's kinetic energy alone. This analogy reminds us that
just like the transition from a largely classical to quantum behavior
in a double well potential, the transition at $|\epsilon|=\Sigma$ is
not sharp (note, however, that unlike in a one dimensional
soft-potential model, the density of excited states of a tunneling
center is very high thus possibly leading to a sharper
cross-over). Put another way, this ``phase transition'' clearly
corresponds to term-crossing and therefore would be gradual in a
finite system.  From a mean-field perspective, the transition at
$\epsilon = \Sigma$ resembles a de-localization phase transition (see
e.g. \cite{AAT}) which we may think of as quantum depinning of the
domain wall. Alternatively, one could say that the local structure of
classical energy levels melts out locally in that the energy variation
on the mostly classical landscape (determined by $T_g$) happens
locally to be smaller than the confinement kinetic energy of the
domain wall motion. Of course, this is occuring for only small parts
of an otherwise rigid matrix. Again, since the system is finite, one
expects a soft cross-over rather than a sharp transition when such
``melting'' occurs.  Both ways of interpreting the quantum
mixing/melting described above are consistent with our view of the
tunneling process leading to the expression (\ref{D_D0}) for the
tunneling amplitude $\Delta$.  The action exponent in Eq.(\ref{D_D0})
scales as the height of the barrier relative to the under-barrier
frequency. The former quantity, while distributed, scales with the
classical energy scale in the problem - $T_g$, while the latter is
proportional to the Debye frequency (and, most likely, is somewhat
distributed too).  The quantum limit of large $\hbar \omega_D$
corresponds to a narrow barrier and a short tunneling path. This would
imply the relative unimportance of the classical energy landscape
modulation during the tunneling process.  Finally, in order to avoid
ambiguity, we stress that the structural transitions of both types of
tunneling centers, that we have called ``classical'' (in that the wave
functions are well localized near minima and are well defined
structurally, i.e. in position space) and ``quantum'' (i.e. in a
superposition of structural states), at low temperature occur in a
purely quantum mechanical fashion, that is by tunneling.
 
We now show that the presence of a somewhat distinct class of such low
barrier, or ``fast'', two-level systems, whose effective diagonal
splitting is zero, leads to additional phonon scattering in comparison
with the strictly semiclassical analysis, which neglects the
renormalization from quantum mixing effects. This additional
scattering at low energy is consistent with the apparent subquadratic
temperature dependence of the heat conductivity in the TLS regime. The
mixing also leads to a super-linear addition to the heat capacity at
subKelvin temperatures. These highly quantum tunneling centers in
strongly mixed superpositions of structural states, therefore, give a
mechanism to resolve a quantitative deviation from the standard
tunneling model, which was brought up by Black and Halperin
\cite{BlackHalperin} in 1977. They noted that the short time heat
capacity of a-SiO$_2$ is larger than would be predicted by the
logarithmic dependence obtained in the STM, if one uses the TLS
parameters extracted from ultra-sonic measurements. The quantitative
mismatch appears to be as if there were two kinds of two-level
systems: one set obeying the distribution postulated in STM, and
another set of ``fast'' tunneling centers responsible for the short
time value of the heat capacity. We can see our analysis of mode
mixing leading to the existence of a finite number of two-level
systems with $\tilde{\epsilon}$ very nearly $0$, as suggested by
Eq.(\ref{eps_rnrm}) is quite consistent with this empirical
notion\footnote{We must stress however that the Black-Halperin
analysis has been conducted only for a single substance, namely
amorphous silica, and systematic studies on other materials should be
done. The discovered numerical inconsistency may well turn out to be
within the deviations of the heat capacity and conductivity from the
strict linear and quadratic laws repsectively.  Finally, a
controllable kinetic treatment of a time-dependent experiment would be
necessary.}.

To see this more explicitly we note that Eq.(\ref{eps_rnrm}) allows
one to formulate the effects of quantum mode mixing as a change in the
apparent distribution of the diagonal energy splitting. Whatever the
old distribution of classical energy difference $n(\epsilon)$, the new
distribution of the effective classical component of the transition
energy can be found using $n(\tilde{\epsilon}) |d \tilde{\epsilon}| =
n(\epsilon)|d \epsilon|$. For $\tilde{\epsilon}$'s not too close to
$\hbar \omega_i$ (case $\tilde{\epsilon} \sim \hbar \omega_i$ will be
discussed later), which is appropriate in the TLS regime, the function
$|\prtl \epsilon/\prtl \tilde{\epsilon}|$ that describes the relative
probability distribution of the two quantities, is given by
\begin{equation}
\left| \frac{\prtl \epsilon}{\prtl \tilde{\epsilon}} \right| = \Sigma
\delta(\tilde{\epsilon}) + 1,
\end{equation}
where the $\delta$-function is positioned to the right of the origin:
$\int^{0^+}_0 d \epsilon \delta(\epsilon) = 1$ (see also
Fig.\ref{eps_eps}b).  Consequently, the distribution of the effective
diagonal splitting is:
\begin{equation}
n_\Delta(\tilde{\epsilon}) = \frac{1}{T_g \xi^3} \left[B
\frac{\Delta^2}{\hbar \omega_D} \delta(\tilde{\epsilon}) +
e^{-|\tilde{\epsilon}|/T_g}\right].
\label{P_eps}
\end{equation}
The coefficient of the $\delta$-function reflects the ``pile-up'' of
the two-level systems that would have had a value of $|\epsilon| <
\Sigma$ were it not for quantum effects. These fast two level systems
will contribute to short time value of the heat capacity in
glasses. The precise distribution in Eq.(\ref{P_eps}) was only derived
within perturbation theory and so is expected to provide only a crude
description of the interplay of clasical and quantum effects in
forming low barrier TLS. Quantitative discrepancies from the simple
perturbative distribution may be expected owing to the finite size of
a tunneling mosaic cell, as mentioned earlier, and the finite
life-times of each energy state due to phonon emission. These effects
would also smoothen the local quantum melting transition as
$\tilde{\epsilon} \rightarrow 0$. While various improvements of the
functional form of $n(\tilde{\epsilon})$ might be suggested, it seems
unwarranted, at present, to use any more complicated expressions for
this function. Thus, to see the main consequences of the quantum
mixing effect, we will proceed with the perturbative
expression. Assuming a particular value of the coefficient $B$ allows
one to derive the contribution of the fast two-level systems to the
heat capacity and scattering of the thermal phonons. Before we start,
let us note that since we now have to deal with a specific coupled
distribution of $\epsilon$ and $\Delta$, the generic two-level system
model that only specifies the distribution of the total splitting $E$
is not sufficient. We must use the full tunneling model where the
tunneling elements $\Delta$'s are distributed according to
Eq.(\ref{P_D}). The exact value of constant $A$ in equation
(\ref{P_D}) depends (weakly!)  on the (possibly $\epsilon$-dependent)
cut-off value of the $P(\Delta)$ distribution.  {\em Both} the heat
capacity and the phonon scattering strength depend on the coefficient
$A$, therefore it is possible to check the {\em relative} contribution
of the ``quantum'' centers to both of those quantities, regardless of
$A$'s value. The $n(\epsilon,\Delta)$ distribution obtained in this
way is now a product of the $P(\Delta)$ distribution from
Eq.(\ref{P_D}) and the density of states from Eq.(\ref{P_eps}). The
new normalization coefficient $A_1$ is found from the requirement that
$\int d\epsilon d\Delta n(\epsilon,\Delta) = 1/\xi^3$. This gives $A_1
= \left[\frac{B}{T_g \hbar \omega_D} \Delta^{2-c} + \frac{1}{c}
\left(\frac{1}{\Delta_{min}^{c}} -
\frac{1}{\Delta_{max}^{c}}\right)\right]^{-1}$). In order to compute
the life-time of a phonon of energy $E$, one averages the Golden Rule
scattering rate $\frac{\pi g^2 \Delta^2}{\rho c_s^2 E}
\tanh\frac{\beta E}{2}$ with respect to $n(\epsilon,\Delta)$, subject
to the resonance condition $E=\sqrt{\epsilon^2 + \Delta^2}$ \cite{AHV,
Jackle,LowTProp}.  This yields two contributions to the decay rate:
\begin{eqnarray}
\tau_{E}^{-1} & = & \frac{\pi}{3} A_1 \left(\frac{a}{\xi}\right)^3 E
\left(\frac{\Delta_{max}}{E}\right)^c \nonumber \\ &\times&
\left[\frac{B E}{\hbar \omega_D} + \int^{1}_{\Delta_{min}/E} d x
\frac{x^{1-c}}{\sqrt{1-x^2}} \right].
\label{tau_phononE}
\end{eqnarray}
The first term in the square brackets is the contribution owing to the
fast, or highly quantum, two-level systems. Note that this term scales
faster with $E$ than the other term. Provided the magnitude of this
first term is comparable to the other term, the fast modes will
somewhat modify the overall scaling of the heat conductivity
$\kappa$. Without the first term, $\kappa$ scales superquadratically
according to $T^{2+c}$ (recall that the heat conductivity is {\em
inversely} proportional to the scattering rate from
Eq.(\ref{tau_phononE})). If we use a numerical value of $B$ of the
order 100, this leads to a {\em subquadratic} $T$ dependence of
$\kappa$: Experimentally, $\kappa(T)$ scales like $T^{1.9 \pm .1}$ as
extracted from a decade and a half of data (see
Fig.\ref{l_lambda}). Without the fast TLS, one, again, would have
$\kappa \propto T^{2+c}$. Using the theoretical approximation for $c$,
this differs from the empirically observed value at least by a factor
of $(10^{1.5})^{c+.1} \sim 2$ at $T \simeq 10^{-2} T_D$. Obviously,
this is a very crude estimate because, first, we do not know how far
down in temperature the power law scaling of $\kappa$ goes; second,
our correction, while going in the right direction, summed with the
older result, is not strictly a power law. Since the integral in the
square brackets of Eq.(\ref{tau_phononE}) varies between 1 and $\pi/2$
for $0 < c < 1$ ($\Delta_{min}/E \ll 1$, surely at $E \simeq 10^{-2}
T_D$), we conclude that the first term must be between $10^0$ and
$10^1$ in order to make a sizable contribution to the phonon
scattering and modify its functional form. Since $E \simeq 10^{-2}
T_D$, this shows that $B$ indeed must be of the order of several
hundreds, consistent with our expectations based on the number of
vibrational modes in the Boson Peak.

Does this mixing induced correction to the density of states with the
value of $B$ around a hundred make an appreciable contribution to the
time-dependent heat capacity? Following the calculation from
subsection \ref{Barr_Dist}, but now using the new distribution
$n(\epsilon,\Delta)$, one finds:
\begin{widetext}
\begin{equation}
C(t) = \frac{A_1}{T_g \xi^3} \int_0^\infty dE \left(\frac{\beta E}{2
\cosh(\beta E/2)} \right)^2 \left( \frac{\Delta_{max}}{E} \right)^c
\left[\frac{B E}{\hbar \omega_D} \theta(t - \tau_{min}) +
\int_0^{\log(t/\tau_{min}(E))} d z \, \frac{e^{\frac{c}{2} z}}{2
\sqrt{1-e^{-z}}} \right],
\label{hcap3}
\end{equation}
\end{widetext}
where $\theta(t)$ is the usual step-function and $\tau_{min}$ is the
fastest possible relaxation time of a TLS with the total energy
splitting $E$, defined in Eq.(\ref{tau_min}). Again, the first term in
the square brackets gives the contribution of the ``fast'' TLS.  Using
the same numbers as given in subsection \ref{Barr_Dist}, it is
straightforward to show that the second, regular, term is of the order
a hundred at temperatures $T \sim 10^{-2} T_D$ when measured on the
time scale of minutes. At the same time, the first term is at most of
order ten. Note that at the shortest times $t \sim \tau_{min}$, when
the regular two-level system only begin to contribute to the heat
capacity, the theory with quantum corrections says the actual heat
capacity is {\em finite} and is at the most one tenth of the long-time
value. At the same time, the fast tunneling centers do not seem to
contribute significantly to the long-time heat capacity. We note
however that the result obtained $c(T) \propto T^{1+c/2}$ with $c=.1$
gives a somewhat slower rise with temperature than seen in
experiment. The quantum correction again goes in the right direction
of {\em increasing} the rate of the heat capacity growth with
temperature relative to the $T^{1+c/2}$ law. 

We have established that effects beyond the strict semi-classical
analysis give rise to a subset of tunneling centers that undergo
faster tunneling than the rest. Nevertheless, there are some
quantitative issues in the heat capacity magnitude that remain to be
understood, namely that the computed contribution of the ``fast''
centers seems somewhat lower than what is necessary to explain the
deviation of the experimental $T$ dependence from the supelinear
dependence $T^{1+c/2}$ predicted by the present (approximate)
argument.  It is posssible that ultimately a broader view of the
time-dependence of the heat capacity needs to be taken. Since, in
fact, the system will clearly be aging by tunneling at those low
temperatures, the notion of fixed frozen-in ``defects'' may no longer
be adequate - essentially interactions between defects play a role.
``Aging'' by definition implies irreversible {\em structural}
changes. More work on understanding the long time evolution of the
tunneling centers is necessary.

We have concentrated on the quantum corrections to the low lying
tunneling states with low barriers. Quantum mixing applies to the
higher energy states too.  Energy shifts and quantum melting occur
within sub-bands of the ripplonic states of order $l$ and respective
degeneracy $(2 l+1)$, thus mixing these states. As tunneling can take
place on a given time scale and the vibrationally excited levels
become observable, their apparent energies can not be degenerate
because the levels are coupled through those same tunneling
transitions. The magnitude of energy level repulsion from the quantum
mixing can be assessed qualitatively.  In the limit of weak coupling,
the deviation of a ripplonic frequency from its classical value scales
$\sum_i \Delta_i^2/\hbar \omega_i$. The width of the ripplonic band of
order $l$ is probably limited from above by the tunneling amplitude
$\Delta_i$ itself. Does this band broadening affect our previous
results on the Boson Peak phenomena?  Not very much. Since the
observables depend mostly on the number of new excitations and the
{\em number} of the ripplonic modes is not changed by these mixing
effects, the essential core of our conclusions from Section
\ref{plateau_chapter} remains intact. Nevertheless, some quantitative
modifications are to be expected. For example, the lowest ripplonic
energies may be lowered to the extent so as to cause a cross-over to a
multi-level behavior in some of the internal resonances, thus possibly
modifying the derived magnitude of the heat capacity and phonon
scattering at sub-plateau temperatures. This effect will further
contribute to the phonon interaction induced broadening of the
ripplonic transitions, as estimated in Section \ref{plateau_chapter}.

\subsection{Mosaic Stiffening and Temperature Evolution of the Boson
Peak} 
\label{stiffening}

Eq.(\ref{eps_rnrm}) raises another interesting point.  According to
that equation, the values of both the bare and the effective classical
energy bias of a transition - $\epsilon$ and $\tilde{\epsilon}$
respectively - are limited from above by the lowest ripplon frequency
($\omega_2$). (Note that this is only realized in the $\epsilon < 0$
case, discussed in this section.) This is unimportant at low
temperatures. But what happens at higher $T$, near this limit?  Unlike
in the low energy situation just discussed, one simply cannot ignore
here that all the energy states have a rather short
life-time. Therefore the singularity in Eq.(\ref{eps_rnrm}) does not
occur, but will be rounded.  This observation does not completely
answer the question that one should have asked in the first place on
general grounds alone: what happens to the structure of the energy
spectrum of a tunneling center, when the energy of the transition
becomes comparable to a vibrational eigen-frequency of the domain
wall\footnote{We remind the reader that the tunneling transition
energy could be also thought of as an eigen-energy of the wall's
motion, but of a lower, $l=1$ order, associated with the translational
motion of the shell's center of mass}?

\begin{figure}[htb]
\vskip 2mm
\includegraphics[width=.8\columnwidth]{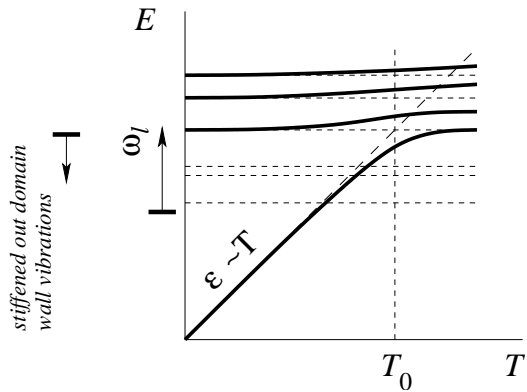}
\caption{\label{level_crossing} This caricature demonstrates the
predicted phenomena of energy level crossing in domains whose energy
bias is comparable or larger than the vibronic frequency of the domain
wall distortions. The vertical axis is the energy measured from the
bottom state; the horisontal axis denotes temperature.  The diagonal
dashed line denotes roughly the thermal energies.  A tunneling center,
that would become thermally active at some temperature $T_0$ will not
possess ripplons whose frequency is less than $T_0$.}
\end{figure}

When attempting to answer this question, a general multi-level
perspective on each tunneling center is somewhat easier to use than
the very mechanical view of the wall's excitations that we have mostly
employed so far, in which the ripplonic energy states are obtained by
quantizing vibrations of a freely moving classical mambrane. The
``singularity'' at $|\tilde{\epsilon}| \sim \hbar \omega_i$ is
actually a term-crossing phenomenon that, again, would not take place
in the strict classical limit. Let us go back to our argument on the
density of states, but consider a case when $\epsilon$ is larger than
a ripplonic frequency. As mentioned many times already, vibrational
excitations of a domain wall can be defined meaningfully only when a
structural {\em transition} takes place in a given region of the
material. The energy of the transition must be the {\em lowest}
excited state of a mosaic cell. On the other hand, the values of the
ripplon frequencies are determined by a (fixed) surface tension
coefficient and the wall's mass density. They have fixed values. The
necessary conclusion from this is that the tunneling centers will not
have ripplons whose frequency is lower than the transition frequency.
We provide a cartoon illustrating this idea in
Fig.\ref{level_crossing}.
We see the quantum mixing reduces the number of the lower frequency
vibrational modes. The mosaic appears stiffer than expected. This
effect may contribute to the temperature evolution of the Boson Peak
as observed in inelastic scattering experiments. Wischnewski et
al. find \cite{bos_peak} that at temperatures between 51 K
(numerically close to silica's $\hbar \omega_2$) and above the glass
transition, the left hand side of the Boson Peak decreases in size as
the temperature was raised. At the same time, the high frequency side
remained relatively unchanged. Note that, as temperature is raised,
the total area of the peak in Fig.2 of Ref.\cite{bos_peak} does not
increase.  In this temperature range mosaic cell motion loses
oscillating character and becomes a rather featureless activated
relaxation process.

To summarize this section, we have seen that the possibility of
quantum tunneling between structurally close states in glass does have
a predictable effect on the spectrum and must be taken into account
when computing the density of low (and not so low) energy structural
excitations in these materials. At the same time, the main conclusions
of the original semi-classical argument remain valid: each structural
transition may be thought of as a rearrangement of about 200 molecules
accompanied by distortion of the domain wall that separates the two
alternative local atomic arragements.

\section{The Negative Gr\"{u}neisen Parameter: an Elastic 
Casimir Effect?}
\label{Gruneisen}

With the exception of the plateau's position and the quantum mixing
effects, we have so far dealt with those anomalies in low temperature
glasses that are more or less universal. These universal patterns are
of particular interest because they cannot be easily blamed on
chemical peculiarities of each substance. Indeed, given the flatness
of the low energy excitation spectrum in glasses, the apparent
universal ratio $l_\smfp/\lambda \simeq 150$ is the dimensionless
quantity that seems to express the general, intrinsic character of
those low energy excitations, as arising from the non-equilibrium
nature of the glass transition. The number $150$ reflects the size of
nearly independent fragments into which a supecooled liquid is broken
up at the laboratory glass transition.  Yet, there is another
dimensionless quantity, namely the Gr\"{u}neisen parameter $\gamma$,
that also reflects the necessity of going beyond a harmonic picture
for amorphous solids. This parameter is always a positive number of
order one for simple cubic crystals (at low enough $T$), but varies
wildly among amorphous materials \cite{Ackerman} (see also a
discussion in \cite{Leggett}). $\gamma$ in glasses has been reported
to be as large as several tens and often negative in sign! A negative
$\gamma$ implies a negative thermal expansion coefficient
$\frac{1}{V}\left(\frac{\prtl V}{\prtl T}\right)_p$ (the {\em linear}
expansion coefficient $\alpha \equiv \frac{1}{L}\left(\frac{\prtl
L}{\prtl T}\right)_p = \frac{1}{3} \frac{1}{V}\left(\frac{\prtl
V}{\prtl T}\right)_p$ is a commonly used quantity chracterizing
anharmonicity too).  Contraction with heating is observed in some
crystals at {\em not too low} temperatures, owing to the details of
the anharmonic couplings in a specific substance that may result in
the negativity of the Gr\"{u}neisen parameter of a lattice mode of a
finite frequency (see e.g.  \cite{cryst_Gru}). Thermal contraction
along a single direction in anisotropic materials is even more
common. Nonetheless, as the temperature is lowered, the thermal
expansion coefficient in an insulating crystal eventually becomes
positive and approaches the cubic $T$ dependence predicted by standard
thermodynamics.  In contrast, an isotropic negative thermal
expansivity is observed in many amorphous substances even at the
lowest temperatures. In addition, the expansivity is not cubic in $T$.
The most widely known example of a substance with a negative $\alpha$
is rubber. Rubber owes this property to the largely entropic nature of
its elasticity.  Here, we will see that a distinct mechanism of
thermal contraction in glasses in the TLS temperature range arises,
which is a direct consequence of the existence of the spatially
extended tunneling centers that give rise to the universal phenomena
considered earlier.

As shown above, the excitation spectrum of the tunneling centers may
be represented as a combination of the two lowest energy levels,
corresponding to the structural transition and a set of higher energy
states involving vibrations of the moving domain wall. By the exchange
of phonons, these local (quantum) fluctuations in the elastic stress
will be attracted to each other much like in the Van der Waals
interaction between neutral molecules. The elastic Casimir effect
seems a more appropriate name for this phenomenon, since the moving
domain walls are not point-like but, instead, resemble fluctuating
membranes.  While we do not claim this attraction is solely
responsible for the negative expansion coefficient, it turns out to
provide a large contribution to the thermal contraction in glasses. We
will see how this effect arising from interaction of amorphous state
excitations depends on the material constants and the preparation
speed of the glass is derived and, therefore, is not universal!

We note first that not all amorphous substances actually exhibit a
negative $\alpha$ in the experimentally probed temperature range. In
such cases, it is likely that the contraction coming from those
interactions in these materials is simply weaker than the regular,
anharmonic lattice thermal expansion. Other contributions to the
Gr\"{u}neisen parameter will be discussed below as well.

Coupling the motion of the mosaic cell (TLS and Boson Peak) to phonons
is necesssary to explain thermal conductivity, therefore the
interaction effects discussed below follow from our identification of
the origin of amorphous state excitations.  The emission of a phonon
followed by its absorption by another cell will give an effective
interaction, in the same way that photon exchange leads to
inter-particle interactions in QED. The longest range coupling between
local degrees of freedom coupled linearly to the elastic stress has
the form of a dipole-dipole interaction.  Since the structural
transitions are of finite size, the dipole assumption is only
approximate for the closer centers. For the time being, we take for
granted that there is no {\em first} order, static, interaction
between the vibrating domain walls, which, if non-zero, could be {\it
\`{a} priori} of either sign.  The next, second order interaction is
always negative in sign and is proportional to $- \sum_{ij}
\frac{1}{r^{6}_{ij}} \propto - \left(1 - \frac{\delta V}{V} \right)^2
\simeq - 1 + 2 \frac{\delta V}{V}$. This favors a sample's contraction
($V$ is the volume). This attractive force, which will be temperature
dependent, is balanced by the regular temperature independent elastic
energy of the lattice: $F_{elast}/V = \frac{K}{2} \left(\frac{\delta
V}{V} \right)^2$.  Calculating the equilibrium volume from this
balance allows us to estimate the thermal expansion coefficient
$\alpha$.  More specifically, the simplest Hamiltonian describing two
local resonances that interact off-diagonally is $H =
\frac{\omega_i}{2} \sigma_x^i + \frac{\omega_j}{2} \sigma_x^j + J_{ij}
\sigma_z^i \sigma_z^j$, where $\omega_i$ and $\omega_j$ would be the
frequencies of ripplons on sites $i$ and $j$ and
\begin{equation}
J_{ij} \equiv \frac{3}{4\pi \rho c_s^2} \frac{({\bf g}_i {\bf g}_j) - 
3 ({\bf g}_i {\bf r}_{ij}) ({\bf g}_j {\bf r}_{ij})/r_{ij}^2}{r_{ij}^3}
\end{equation}
is the dipole-dipole interaction following from Eqs.(\ref{H_ph}) and
(\ref{H_int}). (Having the interaction be off-diagonal automatically
removes the first order term in $J_{ij}$.) The factor $3$ accounts in
our usual simplistic way for all three acoustic phonon branches.  This
ignores a distinction between the longitudinal and transverse speed of
sound. This simplification is however accurate enough for our
purposes.  Since $g \simeq \sqrt{\rho c_s^2 a^3 k_B T_g}$, the
$J_{ij}$'s turn out to scale in a very simple way with the glass
transition temperature and the molecular size $a$, giving $J_{ij} \sim
k_B T_g \left(\frac{a}{r}\right)^3$.

Since only mobile domain walls give rise to local dynamic
heterogeneities, one may conclude intuitively that only the sites of
thermally active structural transitions can contribute to
$\alpha$. Therefore one expects that as temperature in increased, more
tunneling centers will contribute to the Van der Waals attraction thus
leading to negative expansivity. As already mentioned, the excitations
of a tunneling centers are conveniently subdivided into a low energy
TLS-like pair of states, and higher energy, ``ripplonic'' excitations
corresponding to distortions of an active center's domain wall. Hence
we may view the total Van der Waals attraction as having three
somewhat distinct contributions: ``TLS-TLS'', ``ripplon-ripplon'' and
``TLS-ripplon'' attractions. In this section, we focus on the
relatively low, subplateau temperature regime, for reasons that will
be explained later. At these low temperatures, transitions to the
ripplonic states are only virtual, whereas the TLS structural may well
be thermally active.  This, in addition to the differences in the
respective spectra of these excitations, will lead to some difference
in the dependence of the mutual interactions between those excitations
on temperature and other parameters.  In order to assess the magnitude
of those interactions let us consider the following, very simple,
three-level Hamiltonian that is designed to model a transition of
energy $\epsilon_i$ between two different structures that may also be
accompanied by a wall vibration of frequency $\omega_i$:
\begin{equation}
H_{i} = \left(\begin{array}{ccc}
0 & 0           & 0 \\
0 & \epsilon_i  & 0 \\
0 & 0           & \epsilon_i + \omega_i 
\end{array}
\right)
\label{H_rippl}
\end{equation}
Note that, even though, for simplicity's sake, we use the
semiclassical energy $\epsilon$ in the Hamiltonian above, the latter
is meant as (the lowest energy portion of) the full, diagonalized
Hamiltonian with quantum corrections included. This corresponds to the
plain two-level system formalism that does not specify a distribution
of the tunneling matrix element $\Delta$. Also, in comparison with the
general case of Eq.(\ref{part_fun}), we only include an excitation by
a {\em single} quantum of a {\em single} ripplon. The latter
simplification is obviously justified in the lowest perturbation
order, where all pairs of excitations contribute to the total in an
additive fashion. Considering only single-quantum excitations is a low
temperature approximation, made mostly to avoid adopting extra
modelling assumptions necessary to embody the mixed spin/boson
statistics on each site.  This simplification is nevertheless
adequate, as will become clear later in the discussion.

Since the contributions of the three constituents of the Van der Waals
attraction are additive, one can consider each contribution
separately. This indeed proves to be convenient not only because all
the contributions exhibit distinct scaling with the parameters, but
each contribution comes to dominate the expansivity at somewhat
distinct temperatures. We consider first the ripplon-ripplon
attraction. This contribution appears to dominate the most studied
region around 1 K.  The off-diagonal (flip-flop) interaction between
the ripplons has the form:
\begin{equation}
H_{ij}^{int} = J_{ij} | 2_i 3_j \rangle \langle 3_i 2_j | + H.C.,
\label{H_int23}
\end{equation}
where the rows and columns in the unperturbed Hamiltonian from
Eq.(\ref{H_rippl}) are numbered in the conventional way from the upper
left corner. The ``ripplon-ripplon'' case appears the simplest of the
three because here, the issue of how many tunneling centers contribute
to the effect is more or less separate from the strength of the
interaction. The former is (qualitatively) determined by the number of
thermally active two-level systems, that scales roughly with the heat
capacity. The latter is nothing but the ground state lowering of a
pair of resonances after interaction is switched on, which scales as
$-J_{ij}^2/(\omega_i + \omega_j)$ and is $T$-independent at these low
temperatures. This contribution to the negative thermal expansion is
therefore expected to be roughly quadratic in temperature (this
corresponds to linear {\em expansivity}), which is similar, if not
somewhat slower than observed in amorphous silica around 1 degree K.

Calculating the correction to the system's free energy in the lowest
order in $J_{ij}$, that corresponds to the interaction term from
Eq.(\ref{H_int23}) is entirely straightforward and yields:
\begin{eqnarray}
\delta F_{rr} & = & - \sum_{ij} J_{ij}^2
\frac{e^{-\beta(\epsilon_i+\epsilon_j)}(1+e^{-\beta \omega_i})
(1+e^{-\beta \omega_j})}{Z_i Z_j} \nonumber \\ & \times &
\frac{\omega_i \tanh(\beta \omega_i/2) - \omega_j \tanh(\beta
\omega_j/2)} {\omega_i^2 - \omega_j^2},
\label{sumVdW}
\end{eqnarray}  
Here, $Z_i \equiv 1+e^{-\beta \epsilon_i}+e^{-\beta
(\epsilon_i + \omega_i)}$ is the unperturbed on-site
partition function, corresponding to Eq.(\ref{H_rippl}).
Here, subscript ``$rr$'' signifies the ``ripplon-ripplon''
contribution.

At low - subplateau - temperatures $T < \omega_i$, that we are
primarily interested in here, the expression above reduces to the
following Van der Waals energy:
\begin{equation}
\delta F_{rr} = - \sum_{ij} \sum_{l_1 l_2} \frac{J_{ij}^2}{\omega_{l_1}^i +
\omega_{l_2}^j} \frac{1}{(1+e^{\beta \epsilon_i})(1+e^{\beta
\epsilon_j})},
\label{sumVdW1}
\end{equation}  
where we have explicitly written out summation over distinct ripplon
harmonics $l_1$ and $l_2$ at sites $i$ and $j$.

A few intermediate calculations are needed to compute the sum in
Eq.(\ref{sumVdW1}). First, averaging of $J_{ij}^2$ with respect to
different mutual orientations of ${\bf g}_i$, ${\bf g}_j$ and ${\bf
r}_{ij}$ yields an effective isotropic attractive interaction
$\frac{2}{3} \left(\frac{3}{4 \pi}\right)^2 T_g^2
\left(\frac{a}{r}\right)^6$.  Second, the sum over all harmonics
amounts to $\sum_{l_1,l_2=2}^{l_{max}}
\frac{(2l_1+1)(2l_2+1)}{\omega_{l_1}+\omega_{l_2}}$, where $\omega_l$
is found using the dispersion relation from
Eq.(\ref{om_l_numer}). Here we assume that $\omega_i$'s are not
correlated with $J_{ij}$ and $\epsilon_i$. As we already know from the
discussion in the previous section, the latter assumption is adequate
for values $\epsilon$ smaller than the Boson Peak frequency. Now,
recall that $l_{max}$ actually depends on the droplet's perimeter,
thus introducing an additional (cubic!)  scaling with $\xi/a$. In the
end, the sum over the $l$'s is equal, within sufficient accuracy, to
$1.5 \omega_D^{-1} (3/4\pi) \pi^3 (\xi/a)^{5/4} (\xi/a)^3$.  Finally,
assuming $J_{ij}$'s and $\epsilon$'s to be uncorrelated enables one to
present the double sum over $\epsilon_i$ as a product of two identical
sums: $\left(\sum_i (1+\beta \epsilon_i)^{-1}\right)^2$.  Each sum is
the effective concentration of thermally active tunneling centers:
$k_B (\ln 2) \frac{T}{T_g \xi^3}$ as computed by integrating
$1/(1+e^{\beta \epsilon})$ with the density of states from
Eq.(\ref{n_eps}). Note that here we use the simple $1/T_g \xi^3$
expression for density of the tunneling transitions, in keeping with
the assumption $E \sim \epsilon_i$ of the plain two-level system model
adopted in this section. This is reasonable, given the qualitative
character of this calculation.  Finally, the summation over the
ripplon sites can now be reduced to an integration with the lower
limit equal to $\xi (3/4\pi)^{1/3}$.

As a result of the previous discussion, one recovers the following
expression for the energy gain (per volume) due to a volume change
$\delta V$: $\delta F_{rr}/V \simeq 1.5 \left(\ln 2\right)^2 \pi^2
\frac{k_B T^2}{\xi^3 T_D} \left(\frac{a}{\xi}\right)^{7/4}
\left(\frac{\delta V}{V}\right)$.  This works against the regular
elastic energy $\delta F_{elast}/V = \frac{K}{2} \left(\frac{\delta
V}{V} \right)^2$, introduced earlier. The equilibrium relative change
$\delta V/V$ as a function of $T$ is obtained by setting $\prtl
F/\prtl V = 0$.  Differentiating the equilibrium value of $\delta V$
with respect to temperature yields the following estimate for the
thermal (volume) expansion coefficient:
\begin{equation}
\frac{1}{V}\left(\frac{\prtl V}{\prtl T}\right)_p \simeq - 3.0
\left(\ln 2\right)^2 \pi^2 \frac{1}{K} \frac{k_B T}{\xi^3 T_D}
\left(\frac{a}{\xi}\right)^{7/4}.
\label{th_exp}
\end{equation}
This can already be used to estimate the magnitude of the
ripplon-ripplon contribution to the ``Casimir'' effect
numerically. One can do it in several ways. The simplest thing to do
that does not require knowing $K$, is simply to use Eq.(\ref{th_exp})
to calculate the Gr\"{u}neisen parameter $\gamma$ itself according to
$\gamma = (\prtl p/\prtl T)_V/c_V$ \cite{Kittel}, also using $(\prtl
p/\prtl T)_V = - (\prtl p/\prtl V)_T/(\prtl T/\prtl V)_p$.  This
yields a temperature independent Gr\"{u}neisen parameter:
\begin{equation}
\gamma_{rr} \simeq - 3.0 \left(\ln 2\right)^2 \pi^2 \frac{T_g}{T_D}
\left(\frac{a}{\xi}\right)^{7/4}.
\label{grun}
\end{equation} 
 Using $(\xi/a)^3 \simeq 200$ and silica's $T_g/T_D \simeq 1500/350$
one obtains $\gamma \simeq - 3.$, within an order of magnitude of what
is observed in amorphous silica at low temperatures (that experimental
number varies between $-5$ and $-20$ among different kinds of silica
at $1$ K and seems to grow larger with lowering the temperature, see
Fig.3 from \cite{Ackerman}).  We will argue shortly that this growth
may be explained by other contributions to the attraction between
local resonances.

We can also directly compare the conribution in Eq.(\ref{th_exp}) to
the {\em linear} thermal expansion coefficient $\alpha = \frac{1}{3
V}\left(\frac{\prtl V}{\prtl T}\right)_p$ for silica as measured in
\cite{Ackerman}. According to the Fig.2 from \cite{Ackerman}, the
$\alpha$ of silica is linear (possibly slightly sub-linear) in
temperature and equals $- 1.0 \cdot 10^{-9} K^{-1}$ at 1K.  The
compressibility $K$ was obtained in \cite{Ackerman} from measured
speed of sound and density. For internal consistency, we use the
scalar elasticity to estimate $K$ in this way. Summing up three single
polarization phonon Hamiltonians from Eq.(\ref{H_ph}) yields $K \simeq
\rho c_s^2/3$ (remember, $\Delta V/V = 3 \Delta \phi$).  Using
silica's constants, given in Fig.\ref{bump} and the earlier obtained
$\xi = 20 \AA$ and recalling that $\delta l/l = \delta V/3V$,
Eq.(\ref{th_exp}) gives linear expansion coefficient $\alpha \simeq -
0.4 \cdot 10^{-9} K^{-1}$ at 1K, indeed strongly suggesting that
attraction between the tunneling centers is a significant contributor
to the negativity of the expansion coefficient. The numbers just
obtained are also a convenient benchmark in assessing other
contributions to the negative thermal expansivity.

Next, we estimate the magnitude of the attraction between virtual
transition and the direct, lowest energy transitions on different
sites. The corresponding coupling term - $J_{ij} | 2_i 2_j \rangle
\langle 3_i 1_j | + H.C.$ - leads to the following contribution to the
free energy in the lowest order:
\begin{eqnarray}
\delta F_{rT} & = & - \sum_{ij} J_{ij}^2 \frac{e^{-\beta \epsilon_i} (1+
e^{-\beta \omega_i})}{Z_i} \nonumber \\ & \times & \frac{\omega_i
\tanh(\beta \omega_i/2) - \epsilon_j \tanh(\beta \epsilon_j/2)}
{\omega_i^2 - \epsilon_j^2},
\label{sumVdW_Tr}
\end{eqnarray}  
At subplateau temperatures, when $\beta\omega_i \gg 1$, $\tanh(\beta
\omega_i/2)$ can replaced by unity. Furthemore, the summation with
respect to $\epsilon_j$ is no longer cut off by the temperature and
the respective integral (weighted by $n(\epsilon) =
\frac{1}{T_g}e^{-|\epsilon|/T_g}$) picks up most of its value at
$\epsilon \gg T$. Therefore $\tanh(\beta \epsilon_j/2)$ may be
replaced by unity as well. (Actually, both of those replacements must
be made simultaneously lest the sum becomes potentially ill-behaved
when $\omega_i \sim \epsilon$.) As a result, the expression in
Eq.(\ref{sumVdW_Tr}) simplifies:
\begin{equation}
\delta F_{rT} = - \sum_{ij} J_{ij}^2 \frac{1}{1+e^{\beta \epsilon_i}}
\frac{1}{\omega_i + \epsilon_j}.
\label{sumVdW_Tr1}
\end{equation}
The $\epsilon_j$ integral is related to an exponential integral $E_1$
and yields in the two lowest orders: $(\ln(T_g/\omega_i)-\gamma_E)$,
where $\gamma_E = 0.577...$ is the Euler constant. As in the previous
calculation, we regard $\epsilon_i$, $\omega_i$ and $J_{ij}$ as
uncorrelated. The summation over $\omega_i$ can be approximately
represented as a continuous integral between 0 and $l_{max}$ and leads
to a quantity that scales as the area of the domain wall with a
logarithmic correction.  The final result is $\delta F_{rT}/V = 0.5
\frac{T}{\xi^3} (a/\xi)^4 \ln\left[2.0\frac{T_g}{\omega_D}
\left(\frac{\xi}{a}\right)^{1/4}\right] (-1 + 2 \frac{\delta V}{V})$.
Up to a logarithmic correction, the expression is independent of the
energy parameters in the problem and thus must scale linearly in
$T$. Note that we have written out the full expression of $\delta
F_{rT}/V$ that includes the bigger, $\delta V$ independent term
``$-1$'', for the following reason: This larger negative term is
linear in temperature, which apparently would lead to a non-zero
(positive) entropy at $T=0$.  This observation signals a breakdown of
a perturbative picture of largely non-interacting two-level systems.
For the sake of argument, let us estimate at what temperature this
breakdown occurs we compare the magnitude of the $\delta F_{rT}/V$
term, {\em assuming} it is correct, to the free energy of
non-interacting two-level systems per unit volume: $\int \frac{d
\epsilon}{T_g \xi^3} e^{-\epsilon/T_g} [-T \ln(1+e^{-\beta
\epsilon})]$, where we have appropriately chosen $E=0$ as the
reference energy. The latter expression is equal to $(\pi^2/12)
T^2/T_g \xi^3$ and becomes smaller (in absolute value) than the
$\delta F_{rT}/V$ term at temperatures below $10^{-3} T_g$. This
temperature is actually less, but still within an order of magnitude
from the lower end of the plateau, which is well within the {\em
empirical} validity of the non-interacting two-level systems
regime. Let us recall, however, that a perturbative expansion is an
asymptotic one and therefore always {\em over}estimates the {\em
magnitude} of a correction (we suspect that most of the error comes
from the low $\epsilon$ two-level systems). Therefore, a more accurate
estimate would probably yield a break-down temperature lower in value
than the estimate above. There is a reason to believe the
``break-down'' temperature is just at the edge of the lowest
temperatures routinely accessed in the experiments. This is suggested
by several experiments such as on internal friction where deviations
from the standard non-interacting two-level system picture have been
seen (see, for example, a recent review by Pohl et
al. \cite{Pohl_review}). In general, the effect of interaction between
two-level systems could exhibit itself under several guises. One of
those is an apparent gap in the excitation spectrum of the effective
individual TLS. Such effects may have in fact been observed
\cite{Pohl2,LMV}. The estimates above show thise effects are more
likely to be observed in substances with a higher glass transition
temperature, such amorphous silica, or, germania (GeO$_2$).  Note,
however, that the effects of interaction on the apparent TLS spectrum
must be separated from quantum effects of level repulsion on each
sites, that we have considered in Section \ref{LevelRepulsion}. At any
rate, the volume expansion coefficient, corresponding to the computed
value of the ripplon-TLS term, is approximately equal to
\begin{equation}
\frac{1}{V}\left(\frac{\prtl V}{\prtl T}\right)_p \simeq - 1.0
\frac{1}{\xi^3 K} \left(\frac{a}{\xi}\right)^4
\ln\left[2.0\frac{T_g}{\omega_D}
\left(\frac{\xi}{a}\right)^{1/4}\right].
\label{th_exp1}
\end{equation} 
Substituting the numerical values for a-SiO$_2$ in Eqs.(\ref{th_exp1})
and (\ref{th_exp}) shows that at 1 K, the ratio of the ripplon-TLS
contribution to the ripplon-ripplon term is about 1.2 - that is they
contribute comparably to the ``contraction'' free energy at this
temperature. However, since the ripplon-TLS $\alpha$ is temperature
independent, it will dominate at {\em sub}Kelvin temperatures. The
Gr\"uneisen parameter's value corresponding to Eq.(\ref{th_exp1}) is
\begin{equation}
\gamma_{rT} \simeq - 1.0 \frac{T_g}{T} \left(\frac{a}{\xi}\right)^4
\ln\left[2.0\frac{T_g}{\omega_D}
\left(\frac{\xi}{a}\right)^{1/4}\right].
\label{grun1}
\end{equation}
The ripplon-TLS term, as estimated here, therefore seems somewhat
larger relative to the ripplon-ripplon term than seen in experiment,
consistent with our earlier notion that it is somewhat overestimated.
Still, qualitatively our estimates are consistent with the observed
tendency of $\gamma$ to increase in magnitude, when the temperature is
lowered.  We point out that the results obtained above disregard
possible effects of a specific distribution of $\Delta$ that will
influence the precise value of the coupling between phonons and
tunneling centers.

Note that the heat capacity like expression reflecting the number of
thermally active sites $i$ enters into the expressions from
Eqs.(\ref{th_exp1}) and (\ref{grun1}) in a linear fashion. Therefore,
in contrast to Eq.(\ref{grun}), the temperature dependence of
expression (\ref{grun1}) is expected to be largely independent of the
exact $T$-scaling of the heat capacity. Therefore, according to
Eq.(\ref{grun1}), the Gr\"{u}neisen parameter should eventually scale
as $1/T$ at low enough temperatures in all substances (however,
unrealistically long observation times may be required to verify this
prediction; see the discussion at the end of this section). And again,
the apparent density of states of the tunneling centers may be
modified at those low temperatures due to interaction effects (such as
the Burin-Kagan \cite{BurinKagan} effect).

According to Eqs.(\ref{grun}) and (\ref{grun1}), the {\em
dimensionless} contribution of the attractive forces between the
tunneling centers can be expressed in a simple manner through the
$T_g/T_D$ and $T_g/T$ ratios, as well as the relative size of the
mosaic. Note that the effects of varying the quenching speed of the
liquid on the number in Eqs.(\ref{grun}) and (\ref{grun1}) add up. For
instance, making the quenching faster will increase $T_g$ and decrease
$\xi$.  The ripplon-TLS term is especially convenient with regard to
testing our results, because it is nearly insensitive to changes in
the Debye temperature potentially induced by altering the speed of
glass preparation.

Finally, we show that the second order coupling between direct
tunneling transitions is subdominant to the already computed
quantities.  Consider an interaction of the form $J_{ij} | 1_i 2_j
\rangle \langle 2_i 1_j | + H.C.$. If one repeats simple-mindedly the
steps leading to Eq.(\ref{sumVdW}), one obtains the following simple
expression for the free energy correction due to interaction between
the underlying structural transitions:
\begin{equation}
\delta F_{TT} = - \sum_{ij} J_{ij}^2 \frac{\epsilon_i \tanh(\beta
\epsilon_i/2) - \epsilon_j \tanh(\beta \epsilon_j/2)} {\epsilon_i^2 -
\epsilon_j^2}.
\label{F_int}
\end{equation}  
Assuming, again, that $J_{ij}$ and $\epsilon_i$'s are uncorrelated,
the $\epsilon$ summation can be performed via averaging with respect
to the distribution from Eq.(\ref{n_eps}). One can show that the low
temperature expansion of the expression above yields, within two
leading terms, $\delta F_{TT}/V = - (2T_g/3\xi^3)
\left(\frac{a}{\xi}\right)^6 [1+(\pi T/T_g)^2 \ln(T_g/T)/3]$. The
$T$-independent term in itself is curious in that it is a contribution
to the ``vacuum energy'' of the lattice that is of purely glassy
origin and is entirely due to the locality of the free energy
landscape of a liquid. Indeed, as attested by its scaling with
$T_g/\xi^3$, this ``vacuum energy'' contribution would disappear at
the {\em ideal} glass transition at which the whole space is occupied
by a {\em non-extensive} number of distinct aperiodic solutions of the
free energy functional. However, this constant term will have no
effect on the {\em thermal} expansion. The lowest order $T$-dependent
term - $T^2 \ln T$ - actually has a slightly stronger temperature
dependence than the ripplon-ripplon contribution, however the latter
is larger by at least three orders of magnitude, mostly owing to the
large number of ripplon modes. Apropos, we would like to stress again
that the presence of vibrational modes of the (extended) mosaic walls
is essential to the existence of the negative thermal expansivity
effect that we just estimated.  Therefore, while the present theory
predicts that many (and most conspicuous) effects that distinguish
amorphous lattices from crystals should be described well by a set of
non-interacting two-level-like entities at cryogenic temperatures, the
intrinsic {\em multilevel} character of the structural transitions,
that follows from the present theory, in glasses exhibits itself even
at these low energies in higher order perturbation theory.

To complete the discussion of the second order interaction between
tunneling centers we note that the corresponding contribution to the
heat capacity in the leading low $T$ term comes from the
``ripplon-TLS'' term and scales as $T^{1+2\alpha}$, where $\alpha$ is
the anomalous exponent of the specific law. Within the approximation
adopted in this section, $\alpha = 0$.  However it is easily seen that
the magnitude of the interaction induced specific heat is down from
the two-level system value by a factor of $10.(a/\xi)^5 (d_L/a)^2 \sim
10^{-4..5}$ and therefore may be safely neglected.

We have so far considered the second order part of the induced
interactions (square in $J_{ij}^2$, but forth order in $g$).  There
could be also, {\it \`{a} priori}, lower order contributions - first
order in $g$, and first order in $J_{ij}$.  First, let us consider the
term linear in $J_{ij}$, which {\em also} has to do with interaction,
mediated by the phonons. If non-zero, it could be of either sign. In
our case, it is identically zero for the following reason. It is known
\cite{YuLeggett,Silbey}, that the apparent TLS's are only weakly
interacting (one could also infer this implicitly from the smallness
of the second order term that we have already estimated. The first
order term, if non-zero, is comparable to the second order in a
mean-field disordered system. The dipole-dipole $1/r^3$ interaction is
long range and is indeed well described by the mean field).  But we
are dealing here with a non-polarized state, for which the first order
term, linear in the average on-site magnetization, vanishes. In any
case, even if the system were in a ``ferromagnetic'' state, the first
order term would still be only very weakly temperature dependent and
thus would not contribute to the thermal contraction. Whether to
consider such first order term non-zero or not is, to some degree, a
matter of choice. If non-zero, it must be simply thought of as the
effective Weiss-like field that is part of molecular field at each
site. That field implies a {\em hard} gap of the order $T_g$ and
indeed is negligible at low $T$. Yet, at low enough temperatures -
microKelvins or so \cite{Silbey}, the {\em phonon-mediated} first
order interaction between the tunneling centers may become important
and one can no longer use the bare frozen-in values of the on-site TLS
energies, but those determined by the interaction. In this regime an
independent two-level system picture breaks down and more complicated
renormalized excitations may begin to play a role \cite{BurinKagan}.

On the other hand, the other possible contribution to $\alpha$, a term
linear in $g$ does not have to do with interactions between the
anharmonic amorphous solid excitations but is due to the direct
coupling of the tunneling centers with the phonons. This direct
TLS-phonon interaction has so far been the main suspect
\cite{Phillips_Gru,Papoular,GGP_Gru} behind the anomalous thermal
expansion properties of the glassses. This mechanism requires however
the existence of a correlation \cite{Phillips_Gru} (in our notation)
between the on-site values $g$ and $\epsilon$, or else between
$\Delta$ and $\prtl \Delta/\prtl \phi_{ii}$. In other words, the value
of either classical or quantum splitting of a two-level system must be
correlated with the way its energy changes when elastic stress is
applied locally.  The $\Delta$ with $\prtl \Delta/\prtl \phi_{ii}$
correlation has been argued to make a small contribution relative to
the $g$ versus $\epsilon$ correlation because of the smallness of the
value of the $\Delta$'s for the majority of the thermally active TLS
\cite{Phillips_Gru}. On the other hand, a correlation between $g$ and
$\epsilon$ could produce, in principle, both a negative or positive
Gr\"{u}neisen parameter and therefore could explain, by itself, the
observed variety of expansion anomalies in the low $T$
glasses. However, the degree of correlation between $g$ and $\epsilon$
and its temperature dependence is not really known and has to be
parametrized. The soft-potential model offers enough richness in
behavior to accomodate two possible contributions - one dilating and
the other contracting - to the sample's volume. In fact, Galperin et
al. \cite{GGP_Gru} suggest that those two types of the TLS may well be
the two types of the tunneling centers that were postulated early on
by Black and Halperin \cite{BlackHalperin} in order to resolve the
apparent discrepancy between the value of the TLS density $\bar{P}$ as
deduced from the phonon scattering experiments and the equilibrium and
time dependent heat capacity measurements. This, of course, could be
checked experimentally by comparing the degree of the discrepancy in
$\bar{P}$ and the sign of the thermal expansion coefficient in
different substances. (We have shown in the previous section how the
Black-Halperin paradox is, at least partially, explained by quantum
corrections to the semi-classical landscape picture of structural
transitions in glass.)  With regard to the linear in $g$ effect, we
suggest here a modification to the original argument of Phillips from
\cite{Phillips_Gru}. According to Phillips (note some notational
differences), $|\gamma| = 12 g \alpha_0 \ln 2/\pi^2 k_B T$ (he {\em
also} assumed a linear heat capacity linear in $T$).  Here,
$|\alpha_0| \le 1$ is an (unknown) coefficient that reflects the
degree of correlation between $g$ and $\epsilon$: $\dla \epsilon_i
g_i(\epsilon_i) \dra = \alpha_0 \epsilon_i g$. $\alpha_0 = \pm 1$
means complete correlation and $\alpha_0 = 0$ means no correlation.
Now, due to symmetry, $\alpha_0$ must be odd power in $\epsilon$, the
dominant term being therefore linear (see the form of
$\alpha_0(\epsilon)$, somewhat cryptically mentioned as a remark of
B. Halperin, at the end of Phillips' article). We must note, that
although we have pretended, within our one-polarization phonon theory,
that $g$ is a vector quantity, it is in reality a tensor, if the
phonons are treated properly.  The off-diagonal terms, corresponding
to interaction with shear, will indeed be uncorrelated with $\epsilon$
due to symmetry. However, the {\em trace} of the tensor, corresponding
to coupling of the TLS to a uniform volume change could be, in
principle, correlated with the energy of the transition. For example,
it may happen that when the sample is locally dilated, the structural
transitions in that region will require less energy to occur.  At
present, we do not have an argument in favor of or against such a
correlation.  Note, however that at the glass transition temperature,
when the current arrangement of the defects freezes in, most
structural transitions involve a thermal energy around $T_g$. On the
other hand, the energy spliting $\epsilon$ of the tunneling centers
relevant at the cryogenic temperatures is significantly
smaller. Informally speaking, relative to the thermal energy scale at
$T_g$ all two-level systems with low splitting will feel the same to
the phonons. Therefore, qualitatively, the correlation factor
$\alpha_0$ should be at least a factor of $\epsilon/T_g$ down from the
largest value of one. Note, this coincides with the form
$\alpha_0(\epsilon) \propto \epsilon$, suggested by
Halperin. Therefore, the contribution of TLS-phonon coupling to the
thermal expansivity of Phillips (who left the issue of the degree of
correlation open at the time) should be multiplied by a factor of
$T/T_g$. This takes into account, in a very naive way, both the
symmetry and our knowledge of the energy scales relevant at the moment
of the tunneling centers formation. This modifies Phillips' result to
yield
\begin{equation} 
|\gamma| < 12 g \ln 2/\pi^2 k_B T_g = \frac{12 \ln 2}{\pi^2}
\sqrt{\frac{\rho c_s^2 a^3}{T_g}} = \frac{12 \ln 2}{\pi^2}
\frac{a}{d_L}.
\label{gamma_Phil}
\end{equation}
The temperature independence of this contribution to the Gr\"{u}neisen
constant is the main difference between Eq.(\ref{gamma_Phil}) and the
original calculation of Phillips.  The numerical value of the
expression should be nearly the same for all substances and is about
8. This suggests that the direct coupling to phonons is a potential
contributor to the elastic Casimir effect at temperatures around 1 K.
Remember, however, the sign of the expression in Eq.(\ref{gamma_Phil})
is unknown and its numerical value of $10^1$ only provides an estimate
from the above.

From the qualitative analysis in this section, we conclude,
tentatively, that there are several contributions of comparable
magnitude to the thermal expansion at low temperatures. Higher order
effects may also be present. In this case, it may be more
straightforward to estimate the interaction between ripplons as
extended membranes without using a multipole expansion, as indeed is
done when computing the regular Casimir force between extended plates.

The qualitative treatment above of the second-order interaction
between the ripplons on different sites can be extended to higher
temperatures as well. It is easily seen from Eq.(\ref{sumVdW}) that an
excitation of energy $\omega_l$ will contribute only $\beta J_{ij}^2$
at temperatures comparable to $\omega_l$ and above. Therefore one
might expect that at the temperatures near the end of the plateau the
ripplonic transitions become thermally saturated and this attractive
mechanism becomes increasingly less important. The expression in
Eq.(\ref{sumVdW_Tr}), in contrast, is subject to thermal saturation to
a lesser degree. Still, we have seen that its scaling with temperature
is subdominant to the ripplon-ripplon term at temperatures above 1
K. Finally, we remind the reader about the effect of mosaic stiffening
explained in the previous sections. This should also diminish the
attraction between the tunneling centers, owing to a smaller number of
resonant modes at the sites of centers thermally active at these
higher temperatures. On the other hand, the usual anharmonic effects
also become more significant at a higher $T$ leading to a turn-over in
the temperatrure dependence of $\alpha$, as circumstantially supported
by the old data on several materials cited in
Ref.\cite{Krause}. However, in order to assess this ``cross-over''
temperature, one needs to know the magnitude of the regular thermal
expansion due to the non-linearities of the lattice. This is something
that would be extremely difficult to measure independently, because
even a crystal with the same stoichiometry as the respective glass, is
not guaranteed to have the same non-linearity. Direct computer
simulation estimates of the Gr\"{u}neisen parameter, on the other
hand, may be problematic due to the current difficulty of generating
amorphous structures corresponding to realistic quenching rates. This
is the main reason we have confined ourselves here to sub-plateau
temperatures.

Finally, we note again that even at the low temperatures we have been
discussing, not all glasses have been shown to exhibit a negative
$\alpha$.  According to our theory, however, the ``Casimir''
contribution to $\alpha$ is negative and sub-linear in $T$, whereas
the regular non-linear expansion coefficient is positive but only
cubic in temperature. Therefore, there should be a (perhaps very low)
temperature at which the Casimir force should dominate.  Data for many
substances, although still positive at the achieved degree of cooling,
do extrapolate to negative values of $\alpha$ at finite
temperatures. This is not the case, however, for all substances
\cite{Ackerman}.  Even excluding the possibility of error in these
difficult experiments, this is not necessarily inconsistent with our
theory for the following reasons. As the temperature is lowered, it
takes a long time (proportional to $T^{-3}$) for the tunneling
transitions to occur and appear thermally activated. For these same
reasons, like the amorphous heat capacity, the direct interaction
effect is time dependent at low temperatures.  It may therefore take
an excessively long time to actually observe the effects, discussed in
this section, at very low temperatures, thus making it difficult to
see a sign change in $\alpha$ for lattices with relatively large
anharmonicity. Incidentally, this analysis predicts that the response
of the length of an amorphous sample to a temperature change at
sub-plateau temperatures must be time dependent (such time-dependence,
acompanied by heat release, has been observed in polycrystalline NbTi
\cite{Escher}). Since the interaction effect is quadratic in
concentration, one expects qualitatively that the relative rate of the
expansion's time dependence should be twice that of the specific heat.

\section{Conclusions}

In summary, this work elucidates the origin of the thermal phenomena
observed in the amorphous materials at temperatures $\sim T_D/3$ and
below, down to the so far reached milliKelvins. The nature of these
phenomena can be boiled down to the existence of excitations other
than elastic strains of a stable lattice.  The peculiarity of these
excitations is exhibited most conspicuously in the following
phenomena: The specific heat obeys a nearly linear dependence on the
temperature at the lowest $T$, greatly exceeding the Debye
contribution. At the same temperatures, the heat conductivity is
nearly quadratic in $T$ and is universal if scaled in terms of the
elastic constants.  At higher temperatures ($\sim T_D/30$), the
density of these mysterious excitations grows considerably leading to
enhanced phonon scattering and thus a plateau in the temperature
dependence of the heat conductance.  This increase in the density of
states is also directly observed as the so called Boson Peak in the
heat capacity data, as well as inelastic scattering experiments.

We have argued that the origin of these excitations is a necessary
consequence of the non-equilibrium nature of the structural glass
transition. This transition, not strictly being a phase transition at
all in a regular equilibrium sense, occurs if the barriers for
molecular motions in a supercooled liquid become so high as to
prohibit any macroscopic shape changes in the material on the scales
of hours and longer \cite{XW}.  The origin of these high barriers lies
in a cooperative character of the molecular motions, which involve
around $200$ molecules at the glass transition temperature.  Unlike
regular crystals, where the correlation between the molecular motions
is rather long range, thus leading to the emergence of translational
symmetry below solidification, the motions within the cooperative
regions in a supercooled liquid, or entropic droplets, are only weakly
correlated with their surrounding.  In the language of the energy
landscape paradigm, a crystal is a (possibly non-unique) ground state
of the sample (thus the long-range correlation!), whereas a glass is
caught in a high energy state, not being able to reach the true ground
state for kinetic reasons. The respective dense energy spectrum at
these energies exhibits itself in the existence of alternative
mutually accessible conformational states of regions, or domains, of
about $200$ molecules in size. It was argued that quantum transitions
between these alternative states are the additional excitations
observed in glasses at low temperatures. The knowledge of the spectral
and spatial density of these excitations allowed us to estimate from
first principles the magnitude of the observed linear specific heat.
The relevant energy scale here is the glass transition temperature
$T_g$ itself.

Stability requirements for the existence of these alternative
conformational states at $T_g$ allowed us also to estimate the
strength of their coupling to the regular lattice vibrations, which is
determined by $T_g$, the material mass density and the speed of
sound. This enabled us to understand the universality of the phonon
scattering at the low temperatures.

The novelty of this picture is that we have established rather
generally a {\em multiparticle} character of the tunneling
events. This is counter-intuitive because, naively, the larger the
number of particles involved in a tunneling event, the larger the
tunneling mass is, and the harder the tunneling becomes. This is
indeed the case for systems like disordered crystals or crystals with
substitutional impurities, where the tunneling mass is that of an
atom, and the barrier heights are determined by the energy of
stretching a chemical bond by a molecular distance; this virtually
excludes the possibility of tunneling. The existence of structural
rearrangements in a macroscopically rigid system is a sign of the
system being in a high energy state in which the available phase space
is potentially macroscopically large. However, a decrease in this
density of states for glass transitions occuring at a slower pace of
quenching would result in the necessity to engage a larger number of
atoms in these structural rearrangements.  Transitions between the
internal states of a domain involve only a very minor length change of
each individual bond and atomic displacements not exceeding the
Lindemann length, which is of the order one-tenth of the atomic length
scale. It is not particularly beneficial to picture the tunneling
events as individual atomic motions but rather as the motion of an
interface between the alternative states of the domain. This domain
wall is a quasi-particle of a sort, which has a low mass indeed: per
molecule in the domain, it is only about {\em one-hundredth} of the
atomic mass. The contributes to the ease of the tunneling events that
are thermally relevant at cryogenic temperatures: These events are
subject to only very {\em mild} potential variations and are possible,
again, because the lattice is frozen-in in a high energy state.

The spatially extended character of the domain wall excitations along
with their strongly anharmonic nature explains also higher temperature
phenomena, such as the Boson peak and the plateau in the heat
conductivity. By using our knowledge of the surface tension and the
mass density of the domain wall we were able to calculate the energy
spectrum of vibrational excitations of the active domain walls, or
ripplons. This spectrum is in good agreement with the observed
frequency of the Boson peak.  The ripplonic excitations accompany the
transitions between the domain's internal states and thus are strongly
coupled to the phonons. This has enabled us to understand the
experimentally observed rapid drop in the phonon mean free path at the
plateau temperatures. In addition, we have investigated the effects of
phonon coupling on the spectrum of the ripplons. These spectral shifts
scale with $T_g$ and seem to be the cause of the non-universal
position of the plateau.

We have carrried out an analysis of the multi-level structure of the
tunneling centers that goes beyond a semi-classical picture of the
formation of those centers at the glass transition, that was primarily
employed in this work. These effects exhibit themselves in a deviation
of the heat capacity and conductivity from the nearly linear and
quadratic laws respectively, that are predicted by the semi-classical
theory.

A Van der Waals attraction between the domain walls
undergoing tunneling motions was argued to contribute to the puzzling
negative expansivity, observed in a number of low $T$ glasses.

Finally, we note that the conclusions of this work strictly apply only
to glasses made by quenching a supercooled liquid. One may ask,
nevertherless, to what extent the present results are pertinent to
{\em other} types of disordered solids, such as ``amorphous'' films
made chemically or by vapor depositions, or, say, disordered crystals.
Indeed, phenomena, reminiscent of real glasses, such as an excess
density of states, are observed in many types of disordered materials,
although they do not appear to be as universal as in true glasses
(see, for example \cite{Pohl_review}).  In this regard, we note that
most of the phenomena discussed in the present work should indeed take
place in other types of aperiodic structures. What makes quenched
glasses special is the {\em intrinsic} character of their additional
degrees of freedom that stems from the non-equilibrium nature of the
glass transition. Since the characteristics of this transition (while
not being a transition in a strict thermodynamic sense!) are nearly
universal from substance to substance, many low (and not so low)
temperature properties of all those substances can be understood
within a unified approach.

\section*{Acknowledgments} We thank J.Schmalian, A.Leggett and
A.C.Anderson for helpful discussions.  This work was supported by NSF
grant CHE 0317017.

\appendix

\section{Rayleigh Scattering of the Phonons due to the Elastic 
Component of Ripplon-Phonon Interaction}
\label{Rayleigh_app}

In this Appendix, we present an argument on the strength of the phonon
scattering due to the direct coupling with the ripplons via lattice
distortions, but not due to the inelastic momentum absorbing
transition in which the internal state of the domain changes. We thus
consider phonon scattering processes which do obey selection rules and
couple to the lattice strain only in the second and higher order. This
scattering is of the Rayleigh type (and higher order) and occurs off
the domain walls as localized modes. Importantly, we will use only
derived quantities and no adjustable parameters in this estimate. We
show here that, indeed, this absorption mechanism is not significant
compared to the resonant scattering by the inelastic transitions
between the internal states of a thermally active domain.

First, it proves handy to rederive the ripplon spectrum from
Eq.(\ref{spec_final}) in the less general case $\rho_g = 0$ (but
non-zero pressure!).  As argued in Section \ref{plateau_chapter}, the
droplet wall is at equilibrium pressure $p = \frac{3}{2}
\frac{\sigma}{R} = \frac{3}{2} \frac{\sigma_0 a^{1/2}}{R^{3/2}}$. If
the surface is distorted locally by $\Omega$, this results in an extra
force on this portion of the wall due to a changed curvature
\cite{Morse}.  The second Newton's law (as applied per unit area)
yields then:
\begin{equation} \frac{9}{8} \frac{\sigma}{R^2}
\left[2 +  \frac{1}{ \sin \theta} 
\frac{\prtl}{\prtl \theta}
\left(\sin \theta \frac{\prtl \Omega}{\prtl \theta} \right)
+ \frac{1}{\sin^2 \theta} \frac{\prtl^2 \Omega}{\prtl \phi^2} 
\right] = \rho_W \frac{\prtl^2 \Omega}{\prtl^2 t},
\label{eq_mot}
\end{equation}
where $\theta$ and $\phi$ are the usual polar and asimuth angular
coordinates on the surface and we took into account the $r$ dependence
of pressure. The equation above can be solved by a linear combination
of the eigen-functions of angular momentum in 3D:
\begin{equation}
\chi \equiv \sum_{lm} \Omega_{lm} (t) 
Y_{lm} (\theta, \phi),
\end{equation}
$Y_{lm} (\theta, \phi)$ are the spherical Laplace functions
($m=-l..1$). Substituting a harmonic of $l$-th order in
Eq.(\ref{eq_mot}) yields the equation for $\omega_l$ derived in text
as Eq.(\ref{spec_final}).  We will absorb the $9/8$ factor into the
definition of $\sigma$ in the rest of the Appendix.

A (fake) potential energy, yielding the equation of motion
(\ref{eq_mot}), is (c.f. the discussion of surface waves on a
spherical liquid droplet in \cite{LLhydro}):
\begin{widetext}
\begin{equation}
f_{surf} = \sigma \int d \phi \int d(\cos \theta) \left\{(R+\Omega)^2 + 
\frac{1}{2} \left[ 
\left(\frac{\prtl \Omega}{\prtl \theta} \right)^2
+ \frac{1}{\sin^2 \theta} 
\left(\frac{\prtl \Omega}{\prtl \phi} \right)^2
\right] \right\}.
\label{f}
\end{equation}
\end{widetext}
Although varying Eq.(\ref{f}) w.r.t. $\Omega$ does produce the
Eq.(\ref{eq_mot}), note that it differs (by a factor of $9/8$!) from
the original surface energy $\sigma 4 \pi r^2$.  The resulting error
is sufficiently small for our purposes, however this subtlety may be
worth thinking about as this could reveal an extra friction mechanism
due to the wetting phenomenon and surface tension renormalization
mentioned in our discussion of the random first order transition in
Section \ref{RFOT}.

While the domain wall positions are not strictly tied to the atomic
locations, they {\em are} tied to the lattice as a continuum and
follow the lattice distortions. Let us employ our usual ``scalar''
phonons descibed by Hamiltonian
\begin{equation}
H\sph=\int d^3 {\bf r} \left[ \frac{\pi^2}{2\rho}
+\frac{\rho c_s^2 (\nabla \psi)^2}{2} \right], 
\end{equation} 
where $[\psi({\bf r}_1),\pi({\bf r}_2)] = 
i \hbar \delta({\bf r}_1-{\bf r}_2)$.
The surface energy due to
the presence of both $\Omega$ and $\psi$ is:
\begin{widetext}
\begin{equation}
H_{surf} = \sigma \int d \phi \int d(\cos \theta) 
\left\{(R+[\psi-\psi(r_i)]+ \Omega)^2 + 
\frac{1}{2} \left[ 
\left(\frac{\prtl (\psi+\Omega)}{\prtl \theta} \right)^2
+ \frac{1}{\sin^2 \theta} 
\left(\frac{\prtl (\psi+\Omega)}{\prtl \phi} \right)^2
\right] \right\},
\label{f1}
\end{equation}  
\end{widetext}
where $\psi$ is taken on the sphere of radius $R$ with the center
located at ${\bf r}_i$. The potential energy in Eq.(\ref{f1})
thus provides an explicit form of phonon-ripplon interaction
due to the liquid free energy functional solutions being
imbedded in the real space.

If we expand the value of the displacement field $\phi$ in terms of
spherical harmonics according to $\psi_{lm} \equiv \int d \phi d (\cos
\theta) \psi (r=R) Y^*_{lm}(\phi,\theta)$, it is then possible to
write down equations of motion for the ($l,m$)-components of both
ripplon and phonon displacements:
\begin{equation}
\frac{\prtl^2 \Omega_{lm}}{\prtl t^2}  +  \omega_l^2
\left(\Omega_{lm} + \psi_{lm} \right) = 0.
\label{eqmot_O}
\end{equation}

The equation of motion for the phonon field can be obtained e.g. 
from $\ddot{\psi} = i [H\sph+H_{surf}, \pi/\rho]$ to yield:
\begin{widetext}
\begin{eqnarray}
\ddot{\psi} -c_s^2 \Delta \psi & = & - \frac{\sigma}{\rho} 
\int_0^{2 \pi} d \phi' 
\int_{-1}^1 d(\cos \theta') \int d r' \delta(r'-R) 
\left\{ 2(R + [\psi({\bf r}')-\psi({\bf r}_i)])- \frac{1}{\sin \theta'}
\left(\sin \theta' \frac{\prtl \psi}{\prtl \theta'} \right) \right. 
\nonumber \\ 
& - & \left.
 \frac{1}{\sin^2 \theta'} \left( \frac{\prtl^2 \psi}{\prtl \phi'^2} 
\right)  
+  \sum_{lm} \Omega_{lm} [2+l(l+1)] 
Y_{lm} (\theta',\phi') \right\} \delta({\bf r}-{\bf r}').
\label{eqmot_P}
\end{eqnarray} 
\end{widetext} 
The terms with $\psi$ on the r.h.s. serve only to modify the local
elastic constants, and therefore give rise to the regular
Rayleigh scattering, so we will ignore them from now on.

Equations (\ref{eqmot_O}-\ref{eqmot_P}) can be used to write down
equations of motion for the retarded Green's functions, which are
preferable due to their convenient analytical properties (see
\cite{Zubarev} for our conventions). We are interested in the system's
response to ``plucking'' the latice at site ${\bf r} = {\bf 0}$ at
time zero, hence the choice of the Green's function corresponding to
an operator $X$: $-i \theta(t-t') \la [X(t), \psi({\bf r} = {\bf
0},t'=0)] \ra$.  Eqs.(\ref{eqmot_O}-\ref{eqmot_P}), if rewritten for
the corresponding Green's functions, will preserve except there will
be an additional term $-\frac{1}{\rho} \, \delta(t) \delta^3({\bf
r})$, corresponding to the ``plucking'' event, in the r.h.s. of
Eq.(\ref{eqmot_P}) (note also a change in units). Thus obtained
equations are possible to rewrite in the Fourier space:
\begin{equation}
-\omega^2 \we{\Omega}_{lm}^i+\omega^2_l 
\left[\we{\Omega}_{lm}^i + \we{\psi}_{lm}^i \right] = 0
\label{om_k}
\end{equation}
and
\begin{widetext}
\begin{equation}
-\omega^2  \we{\psi}_{\bf k} + c_s^2 k^2 \we{\psi}_{\bf k} =
- \sum_i \frac{\sigma}{\rho} \sum_{lm} \we{\Omega}_{lm}^i  [2+l(l+1)] 
\frac{e^{-i {\bf k}{\bf r_i}}}{2 \pi^2} Y_{lm}(-{\bf k}/k)
i^l j_l(kR) - \frac{1}{(2 \pi)^4 \rho},
\label{psi_k}
\end{equation}
\end{widetext}
where 
$\we{\psi}_{lm}^i \equiv \int d^3 {\bf k} \, \we{\psi}_{\bf k}
e^{i {\bf k}{\bf r}_i} (4 \pi) i^l j_l(kR) Y_{lm}^*({\bf k}/k)$
and we used the expansion of a plane wave in terms of the spherical
harmonics: $e^{i {\bf k}{\bf r}}=4\pi \sum_{l=0}^\infty \sum_{m=-l}^l
i^l j_l(kr) Y_{lm}^*({\bf k}/k) Y_{lm}({\bf r}/r)$. Here, $j_l(x)
\equiv \sqrt{\pi/2 x} \, J_{l+1/2}(x)$
is the spherical Bessel function, which scales as $x^l$ for small
$x$, hence we see that the ripplons' coupling with the phonons
is quadratic or higher order in ${\bf k}$ as the second harmonic
is the lowest order term allowed. Modes $l=0$ and $l=1$ have
the meaning of the droplet's growth and translation respectively, 
as was discussed in Section \ref{spectrum}. These modes are not
covered by this Section's formalism. Even though the theory
as a whole could be thought of as a multipole expansion 
of a molecular cluster interacting with the rest of the lattice, 
the modes of different orders end up being described by 
different theories. 

The system of Eqs. (\ref{om_k}) and (\ref{psi_k}) can now be used to
determine the sound dissipation due to the interaction with the
ripplons. Since the system is infinite and has a continuous spectrum,
all excitations will have finite life-times, which can be, in
principle, obtained self-consistently by using e.g. the Feenberg's
perturbative expansion \cite{Feenberg, AAT} (one in the end arrives at
Green's functions that are well behaved at infinity, as implied in the
thus greatly simplified derivation). We do not have to do this
self-consistent self-energy determination as long as we are interested
in the lowest order estimate, as justified in the end by the smallness
of the obtained value of the perturbation.  Substituting
Eq.(\ref{om_k}) into Eq.(\ref{psi_k}) yields
\begin{widetext}
\begin{eqnarray}
-\omega^2  \we{\psi}_{\bf k} + c_s^2 k^2 \we{\psi}_{\bf k} & = &
(4 \pi)^2 \frac{\sigma}{\rho}
 \sum_i \sum_{lm} \frac{[2+l(l+1)] \omega_l^2}{\omega_l^2-\omega^2}
\nonumber \\ 
& \times &\int \frac{d^3 {\bf k}_1}{(2 \pi)^3} e^{i ({\bf k}_1 - {\bf k}){\bf r_i}}
(-1)^l j_l(k_1 R) j_l(k R) Y_{lm}(-{\bf k}/k) Y_{lm}^*({\bf k}_1/k_1)
\we{\psi}_{{\bf k}_1}.
\label{psi_k1}
\end{eqnarray}
\end{widetext}
Since the spatial locations ${\bf r}_i$ of active droplets are not
correlated \footnote{This is not strictly true - they, of course, can
not be on top of each other.}, we can replace the summation over the
droplets by a continuous integral, assuming at the same time that the
ripplon frequency corresponding to $\omega_l$ varies from droplet to
droplet within a (normalized) distribution ${\cal P}_l(\omega)$
centered around $\omega_l$ and having a characteristic width $\delta
\omega_l$, whose value will be discussed shortly. There is no reason
to believe that the frequency and location of the tunneling centers
are correlated, therefore one obtains
\begin{widetext}
\begin{eqnarray}
-\omega^2  \we{\psi}_{\bf k} + c_s^2 k^2 \we{\psi}_{\bf k} & = &
 n \frac{\sigma}{\rho}
\sum_{l} \int d \omega' {\cal P}_l(\omega') 
\frac{4 \pi [2+l(l+1)](2l+1) \omega'^2}{\omega'^2-(\omega+i\epsilon)^2}
j_l^2(k R) \we{\psi}_{\bf k},
\label{psi_k2}
\end{eqnarray}
\end{widetext}
where $n$ is the concentration of the active domain walls to be
estimated shortly and we have displaced $\omega$ by $\epsilon$ into
the upper half-plane because we are looking for the {\em retarded}
Green's function.  Also, in order to derive Eq.(\ref{psi_k2}), we have
used the summation theorem for the spherical functions $P_l({\bf n}
{\bf n}') =\frac{4 \pi}{2l+1} \sum_{m=-1}^l Y_{lm}^*({\bf n}')
Y_{lm}({\bf n})$, as well as $P_l(-1)=(-1)^l$, where $P_l$ is the
Legendre polynomial.  If we ignore the real part of the r.h.s. of
Eq.(\ref{psi_k2}), responsible only for the dispersion, the poles of
the resultant phonon Green's function are found by solving $\omega^2 -
c_s^2 k^2 + i \, 2 \omega \tau_\omega^{-1} = 0$, where
$\tau_\omega^{-1}$ clearly has the meaning of the inverse life-time of
a phonon of frequency $\omega$ and is given by
\begin{equation}
\tau^{-1}_\omega =  n \frac{\sigma}{\rho}
\sum_{l=2}^{9} \pi^2 [2+l(l+1)](2l+1)
j_l^2(k R) {\cal P}_l(\omega),
\label{tau_1w}
\end{equation}
where we have ignored the contribution of the peaks centered around
$(-\omega_l)$. We remind the reader that $l_{max} \simeq 9$ is 
dictated by the finite size of a droplet.

One can find the value of $\delta \omega_l$ from an argument identical
to the one used in \cite{XWbeta} to obtain the width of the
distribution of the barriers for the droplet growth free energy
profile.  At the glass transition, a liquid breaks up into dynamically
cooperative regions, so that a translation of one atom involves moving
about $200$ atoms around it, which involves overcoming a large (on
average) barrier. This barrier's height is determined, together with
the domain surface tension coefficient, by the configurational entropy
density, which in its turn reflects the number of metastable states
available to a particular volume of liquid at this temperature. Even
though a good description of freezing is achieved by assuming that
this number of available states does not strongly depend on where
exactly on the free energy surface a particular molecular cluster is
\cite{XW}, it should vary from domain to domain. The size of the
variation can be estimated from the known magnitude of the entropy
fluctuations at constant energy, so that the ratio of the variance to
the mean is related to the jump in the heat capacity at $T_g$ and
subsequently turns out to be $1/2 \sqrt{D}$ \cite{XWbeta}, where $D$
is the liquid's fragility, entering the Vogel-Fulcher law for
relaxation times in a supercooled liquid $\tau_{relaxation} \propto
e^{\frac{D T_K}{T-T_K}}$.  We conclude then that the lower bound on
the fluctuations of the ripplon frequency $\omega_l$ is given by
$\delta \omega_l \simeq \omega_l/2 \sqrt{D}$.

Lastly, in order to use Eq.(\ref{tau_1w}) to compute the phonon
absorption due to this particular mechanism, we need to estimate the
density of the active domain walls. It will suffice for our purposes
here to consider as active the defects that contribute to the specific
heat, that is, roughly, $n \simeq \frac{1}{\xi^3} T/T_g$. A more
accurate estimate would be similar to the one we made when calculating
the bump in the heat capacity in Section \ref{spectrum}.

We are now ready to give a numerical estimate of the expression in
Eq.(\ref{tau_1w}). We will compute here the contribution of the $l=2$
term in the plateau region. It is convenient to represent $kR$ from
Eq.(\ref{tau_1w}) as $kR \sim \frac{\omega}{0.4 \, (a/\xi)\omega_D}$.
For the reference, $(a/\xi)T_D \sim 0.2 T_D$ is at the high
temperature end of the plateau, whereas its middle is about an order
of magnitude lower depending on the substance (see $\kappa$ vs.
$T/T_D$ plot in Fig.\ref{l_lambda}). We can now use our usual
expressions connecting $\sigma, T_g, \omega_D, c_s, \rho, a$ etc to
obtain a numerical estimate of Eq.(\ref{tau_1w}) at the plateau
frequencies $\omega_{plateau} \sim 10^{-1.5} \omega_D$. Even if one
favorably assumes that $\omega_2 \sim \omega_{plateau}$ (it is
somewhat larger according to Section \ref{plateau_sec}), one still
gets $l_\smfp/\lambda > \sim 10^4$ at the plateau frequency, whereas
the {\em resonant} absorption by the TLS would give $l_\smfp/\lambda
\sim 10^2$. The amplitude of this type of absorption is small due to
the weakness of direct coupling to the ripplons for the processes not
accompanied by a change in the domain's internal state.

\section{Frequency Cutoff in the Interaction Between the
Tunneling Centers and the Linear Strain} 
\label{wc_app}

\begin{figure}[htb]
\includegraphics[width=.6\columnwidth]{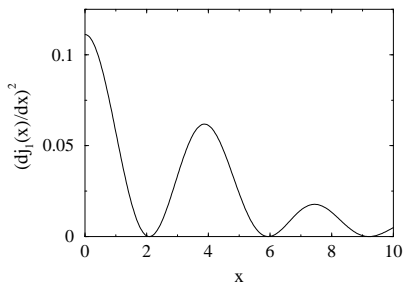}
\caption{\label{bessel} Shown is the derivative of the 1st order spherical
Bessel function determining the effective decrease in the
elastic field gradient produced by a phonon of wave-length $k$
($x = kR$).}
\end{figure}

As argued in Section \ref{universality_sec}, the coupling of the
tunneling transition to a phonon can be found from an additional
energy cost of moving the molecules within the domain in the presence
of a strain and is given by an integral over the droplet's volume (we
consider only longitudinal strain for simplicity):
\begin{equation}
g =  \rho c_s^2 \int_V d^3 {\bf r} ({\bf \nabla} \vec{\phi}) 
({\bf \nabla d}),
\label{g_est}
\end{equation}
where $\rho c_s^2$ is basically the elastic modulus, $\vec{\phi}$ and
${\bf d}$ are elastic and inelastic components of the atomic
displacements respectively.  If the phonon's wave-length is much
larger than $\xi$, the elastic component is constant throughout the
integration region and the integral reduces to one over the droplet's
surface and thus the $g$ estimate obtained in text.  Otherwise, one
obtains:
\begin{equation}
g =  \rho c_s^2 \left\{ \int_S d{\bf S} \, {\bf d} \, 
({\bf \nabla} \vec{\phi}) \, 
- \int_V d^3 {\bf r} \, ({\bf d \nabla}) ({\bf \nabla} \vec{\phi})  \right\}.
\label{g_est1}
\end{equation}
The volume integral will give a higher order term in $k$, so for now,
we focus on the surface integral. The displacement due to the phonon
is conveniently expanded it terms of the spherical waves: $e^{i {\bf
k}{\bf r}}=4\pi \sum_{l=0}^\infty \sum_{m=-l}^l i^l j_l(kr)
Y_{lm}^*({\bf k}/k) Y_{lm}({\bf r}/r)$. Since it is the first
derivative with respect to ${\bf r}$ that we are interested in, we
only need the $l=1$ term from this expansion. The angular part
contributes only to the overall constant, but it is the spherical
function $j_1 (kr)$ that sets the cut-off value of the wave-vector,
above which the phonons do not produce significant linear uniform
stress on the domain. In Fig.\ref{bessel}, we plot the derivative
$\prtl j_1(x)/\prtl x$ (or, rather, we plot the square of it, which
enters into all the final expressions).

We see that it is not unreasonable to assume that only the phonons
with $kR < \sim 6$ will exert an appreciable linear strain on the
domain. $kR = 6$ translates into $\omega_c \sim 2.5 (a/\xi) \omega_D$.

While we are at it, we estimate the interaction of the domain with the
higher order strain, at least due to the term (\ref{g_est}), in the
frequency region of interest. The next order term in the $k$ expansion
in the surface integral from Eq.(\ref{g_est1}) has the same structure,
but is scaled down from the linear term by a factor of $kR$.  At the
plateau frequencies $\sim \omega_D/30$, $kR < 0.5$ as immediately
follows from the previous paragraph. While this is not a large number,
it is not very small either. Therefore, this interaction term is of
potential importance.

The volume integral in Eq.(\ref{g_est1}) produces a quadratic term,
which is roughly equal to \\ $({\bf \nabla} \vec{\phi}) \int_V d^3
{\bf r} \, ({\bf d \, k})$. We then proceed in a completely identical
fashion to our earlier estimate of $g$.  Assuming the diplacements
within the droplet are random, one gets for the integral $\frac{1}{4}
\sqrt{N^*} \, a^3 \, d_L k$, where factor of $1/4$ comes about because
the displacement is assumed do decrease from $d_L$ in the center of
the droplet to zero at the edge \cite{LW}. This yields that this term
becomes comparable to the linear one at frequencies $\omega \simeq
\omega_D \sqrt{(a/\xi)} \, 4/(6\pi^2)^{1/2} \simeq 0.4 \, \omega_D$ -
well beyond the high $T$ end of the plateu.

We must note, there are other sources of non-linearity in the system,
such as the intrinsic anharmonicity of the molecular interactions
present also in the corresponding crystals. While these issues are of
potential importance to other problems, such as the Gr\"{u}neisen
parameter, expression (\ref{g_est}) only considers the lowest order
harmonic interactions and thus does not account for this non-linear
effect.  We must note that if this non-linearity is significant, it
could contribute to the non-univrsality of the plateau, in addition to
the variation in $T_g/\omega_D$ ratio. It would be thus helpful to
conduct an experiment comparing the thermal expansion of different
glasses and see whether there is any correlation with the plateau's
location.

\bibliography{lowT}
\end{document}